\documentclass[a4paper,twocolumn,10pt,unpublished]{quantumarticle}
\pdfoutput=1

\usepackage[utf8]{inputenc}
\usepackage[english]{babel}
\usepackage[T1]{fontenc}

\usepackage{graphicx}
\usepackage{subcaption}

\usepackage{hyperref}
\hypersetup{
	colorlinks=true,
	linkcolor=blue,
	citecolor=blue,
	urlcolor=blue,
}

\usepackage{physics}
\usepackage{siunitx}
\usepackage{amsmath}
\usepackage{amssymb}
\usepackage{bm}  % bold symbols (for vectors, etc.)
\usepackage{bbm} % bbold symbols (for sets, etc.)
\usepackage{mathtools}

\usepackage[numbers,square,comma,sort&compress]{natbib}

\usepackage{xspace}
\usepackage[dvipsnames]{xcolor}
\usepackage{orcidlink}

\usepackage{tabularx} % fit table to linewidth
\usepackage{enumitem}

\usepackage[normalem]{ulem}

\newcommand{\pennylanedataset}{%
    \href{https://pennylane.ai/datasets/collection/low-depth-image-circuits}%
    {\texttt{pennylane.ai/\allowbreak datasets/\allowbreak collection/\allowbreak low-\allowbreak depth-\allowbreak image-\allowbreak circuits}}%
    \xspace%
}

\newcommand{\gitrepo}{%
    \href{https://github.com/fkiwit/low-depth-image-circuits}%
    {\texttt{github.com/\allowbreak fkiwit/\allowbreak low-\allowbreak depth-\allowbreak image-\allowbreak circuits}}%
    \xspace%
}

\newcommand{\bigO}[1]{\mathcal{O}\!\left(#1\right)}

\renewcommand{\vec}[1]{\bm{#1}}
\newcommand{\Id}{\mathbbm{1}}
\newcommand{\transpose}{T}
\newcommand{\E}{\mathbb{E}}
\newcommand{\Var}{\text{Var}}

\begin{document}

\title{Typical Machine Learning Datasets as Low-Depth Quantum Circuits}

\author{Florian J. Kiwit\,\orcidlink{0009-0000-4065-1535}\,}
\affiliation{Ludwig Maximilian University, Munich, Germany}
\affiliation{BMW Group, Munich, Germany}
\orcid{0009-0000-4065-1535}
\email{florian.kiwit@ifi.lmu.de}
\author{Bernhard Jobst\,\orcidlink{0000-0001-7027-3918}\,}
\affiliation{Technical University of Munich, TUM School of Natural Sciences, Physics Department, 85748 Garching, Germany}
\affiliation{Munich Center for Quantum Science and Technology (MCQST), Schellingstr. 4, 80799 München, Germany}
\orcid{0000-0001-7027-3918}
\email{bernhard.jobst@tum.de}
\author{Andre Luckow\,\orcidlink{0000-0002-1225-4062}\,}
\affiliation{Ludwig Maximilian University, Munich, Germany}
\affiliation{BMW Group, Munich, Germany}
\orcid{0000-0002-1225-4062}
\author{Frank Pollmann\,\orcidlink{0000-0003-0320-9304}\,}
\affiliation{Technical University of Munich, TUM School of Natural Sciences, Physics Department, 85748 Garching, Germany}
\affiliation{Munich Center for Quantum Science and Technology (MCQST), Schellingstr. 4, 80799 München, Germany}
\orcid{0000-0003-0320-9304}
\author{Carlos A. Riofrío\,\orcidlink{0000-0002-7346-9198}\,}
\affiliation{BMW Group, Munich, Germany}
\orcid{0000-0002-7346-9198}

\maketitle

\begin{abstract}
Quantum machine learning (QML) is an emerging field that investigates the capabilities of quantum computers for learning tasks.
While QML models can theoretically offer advantages such as exponential speed-ups, challenges in data loading and the ability to scale to relevant problem sizes have prevented demonstrations of such advantages on practical problems.
In particular, the encoding of arbitrary classical data into quantum states usually comes at a high computational cost, either in terms of qubits or gate count. However, real-world data typically exhibits some inherent structure (such as image data) which can be leveraged to load them with a much smaller cost on a quantum computer.
This work
% (i)
further develops an efficient algorithm for finding low-depth quantum circuits to load classical image data as quantum states. To evaluate its effectiveness, we
% (ii)
conduct systematic studies on the MNIST, Fashion-MNIST, CIFAR-10, and Imagenette datasets. The corresponding circuits for loading the full large-scale datasets are available publicly as PennyLane datasets and can be used by the community for their own benchmarks. We further
% (iii)
analyze the performance of various quantum classifiers, such as quantum kernel methods, parameterized quantum circuits, and tensor-network classifiers, and we compare them to convolutional neural networks.
In particular, we focus on the performance of the quantum classifiers as we introduce nonlinear functions of the input state, e.g., by letting the circuit parameters depend on the input state.

\end{abstract}

\section{Introduction}

% Intro QML
Machine learning is a powerful framework that has facilitated tremendous progress in problems such as image recognition and text generation, which would have been considered intractable using conventional programming. Quantum computers promise to enhance the capabilities of machine learning models further, generating a significant amount of interest in the field of quantum machine learning (QML)~\cite{schuld2022machine, Biamonte_2017}. Indeed, several theoretical works suggest an exponential advantage of QML over classical machine learning methods for certain specific tasks~\cite{Liu2021, Sweke2021quantumversus, Huang_2021, PhysRevA.107.042416, Gyurik2023}, such as solving a synthetic classification task constructed from partitioning the data based on the discrete logarithm~\cite{Liu2021}. These tasks tend to be quite contrived from an application-based view, owing to the difficulty of making more natural tasks amenable to mathematical proofs~\cite{Gil-Fuster2024}. Hence, the demonstration or existence of the advantages mentioned above is still an important open question~\cite{zimboras2025mythsquantumcomputationfault}.

% Data encoding challenges
A key challenge is the loading of classical datasets as quantum states. On the one hand, popular data encoding methods include \emph{rotation encoding}~\cite{schuld2022machine, Stoudenmire2017}, which uses a single qubit per data component and rotates the state of the corresponding qubit proportional to the data component. The encoding is popular because it is very efficient to implement as a circuit and allows to prove universality results based on data reuploading techniques~\cite{P_rez_Salinas_2020, Goto_2021, Schuld_2021}---however, the number of qubits for relevant datasets quickly becomes of the order of millions, which are not available in the near term.

On the other hand, there are popular techniques like \emph{amplitude encoding}~\cite{schuld2022machine, latorre2005imagecompressionentanglement}, which only need a number of qubits logarithmic in the number of data components by storing the (normalized) values in the amplitudes of the wave function. In general, however, preparing such a state requires a number of gates that scales exponentially with the number of qubits, rendering the method inefficient and unrealistic for the gate fidelities and coherence times of the currently available quantum computers~\cite{Plesch2011, Iten2016}.

This data-loading problem has so far limited many demonstrations of QML to small-scale datasets in both experiments and numerical simulations, often utilizing either artificial or simple datasets, or applying dimensionality reduction methods like the principal component analysis or coarse-graining~\cite{Grant2018, Havl_ek_2019, Schuld_2019b, Schuld_2020, Bartkiewicz2020, Kerenidis2020, Chalumuri2021, Johri2021, Peters2021, Bokhan2022, Hur2022, Lu2025, bowles2024betterclassicalsubtleart}. While many QML models achieve promising results on simple datasets like MNIST and Fashion-MNIST, the same models tend to fail badly on more complex and realistic datasets like CIFAR-10~\cite{Zhou2021, Riaz2023, Baek2023, Khatun2024, Monbroussou2024}. This highlights the need to test QML models on relevant realistic datasets as a good performance on simple datasets does not necessarily translate to more complicated datasets.

Further, the use of many different datasets---unlike the common benchmarking datasets widely used in classical machine learning---makes it difficult to compare the performance of different models directly. Such standard datasets have been highly successful in classical machine learning. One example from the early development of modern neural networks is the `ImageNet Large Scale Visual Recognition Challenge'~\cite{russakovsky2015imagenetlargescalevisual}, in which AlexNet~\cite{NIPS2012_c399862d} first demonstrated the success of deep-learning architectures.

In the field of QML, efforts are now being undertaken to compare the performance of QML models on common datasets or using standardized benchmarks~\cite{Moussa2022, Kashif2023, bowles2024betterclassicalsubtleart, Finzgar_2022, kiwit2023applicationoriented, Kiwit_2024, Quetschlich_2023}. In a recent study, Ref.~\cite{bowles2024betterclassicalsubtleart} compared many popular QML models on several small-scale and mostly artificial datasets, and concluded that classical models consistently outperform their quantum counterparts. As already raised by the study's authors, an interesting open question is whether this observation also holds true for more complicated or real-world datasets. 

Moreover, a step towards real-world application-oriented benchmarking is taken by the creation of software tools that provide a modular and standardized pipeline to evaluate models~\cite{Finzgar_2022, kiwit2023applicationoriented, Kiwit_2024, Quetschlich_2023}---however, examples like the QUARK framework~\cite{Finzgar_2022, kiwit2023applicationoriented, Kiwit_2024} focus on optimization problems and generative quantum machine learning, while the MQT benchmark library~\cite{Quetschlich_2023} focuses on general quantum computing tasks. Nevertheless, there is still a lack of options for assessing the scalability and the capability of supervised QML models to deal with complex real-world data.

In this work, our main goal is to address the lack of real-world, large-scale datasets for benchmarking QML models, with a focus on keeping the hardware requirements to a minimum so that the datasets can be relevant for near-term experiments and numerical simulations. Inspired by the recent success of approximating amplitude-encoding-type states with tensor networks and shallow quantum circuits~\cite{Dilip2022, jobst2023efficientmpsrepresentationsquantum, iaconis2023tensornetworkbasedefficient, shen2024classification,maxwell_2024}, we devise and implement an algorithm for efficiently optimizing such quantum circuits on a classical computer. This yields quantum circuits whose number of qubits scales only logarithmically in the input data size and whose depth only scales linearly in the number of qubits. First, we apply this algorithm to image data, specifically the MNIST~\cite{Lecun1998, deng2012mnist}, Fashion-MNIST~\cite{FashionMNIST}, CIFAR-10~\cite{CIFAR10} and Imagenette~\cite{imagenette} datasets, and provide the encoded images as PennyLane datasets~\cite{bergholm2022pennylaneautomaticdifferentiationhybrid} at \pennylanedataset.

While our main focus is on obtaining these datasets, we also benchmark variational quantum circuit classifiers, support vector machines, and tensor-network classifiers on them and find that these models show promising results on the simple MNIST and Fashion-MNIST datasets but struggle on the harder and more realistic CIFAR-10 and Imagenette datasets. We further propose strategies to turn the quantum circuit classifiers from a linear into a nonlinear function of the input density matrix. While these strategies in their current form only enhance the accuracy of the classifier by several percentage points, which is still not enough to make them competitive with current classical state-of-the-art methods, we view them as potentially fruitful directions for further exploration. We invite researchers to try to reach state-of-the-art performance by using their own ideas for quantum classifiers on these datasets.

% Structure
This paper is structured as follows: In Sec.~\ref{sec:encoding}, we review the methods we use for encoding classical image data into quantum states. Building on this encoding, in Sec.~\ref{sec:efficient_encoding} we propose a low-depth quantum circuit architecture inspired by matrix-product states and present an algorithm that efficiently optimizes these circuits to approximate the encoded states on a classical computer. We show results on the MNIST, Fashion-MNIST, CIFAR-10, and Imagenette datasets, demonstrating the approach's efficacy. In Sec.~\ref{sec:classifiers}, we benchmark various quantum classifiers for image classification on these datasets, including variational quantum circuits (VQCs), a nonlinear variant of VQCs, support vector machines (SVMs) with a quantum kernel and tensor-network classifiers. First, we consider the exact dataset without the circuit approximation, then we assess the effect of the circuit approximation on the classification accuracy by also training the classifiers on the compressed dataset. We further compare the performance on the compressed dataset to that of a classical convolutional neural network (CNN) trained on the same data. In Sec.~\ref{sec:conclusion}, we conclude with a summary of our results.
\section{Quantum Image Representations}
\label{sec:encoding}

Our main goal is to construct quantum states encoding classical data from realistic and large-scale datasets, that also efficiently use available quantum resources---both in terms of the number of qubits used and the depth of the circuit needed to prepare the corresponding state. 

An encoding that minimizes the circuit depth is, for example, the \emph{rotation encoding}~\cite{schuld2022machine, Stoudenmire2017}, where each qubit encodes one component of the input data vector by rotating its state proportional to this data component, e.g., by applying a Pauli rotation gate $e^{i \sigma^{\alpha}_j x_j}$ where $\sigma^{\alpha}_j$ is a Pauli matrix acting on the $j$th qubit and $x_j$ is the $j$th data component. The single-qubit gates can all be applied in parallel and no entangling operations are needed, making the state preparation very efficient. However, this comes at the cost of requiring a large number of qubits: to encode moderate-sized images with $512\times512$ pixels and the three RGB color channels, we need over $\num{750000}$ qubits. Suppose a subsequent QML model needs to correlate all pixel values to make a correct prediction, then the total gate count will reach a similar order of magnitude. This makes the approach infeasible for experimental realizations for a long while.

On the other hand, approaches like \emph{amplitude encoding} minimize the number of qubits needed~\cite{schuld2022machine, latorre2005imagecompressionentanglement}. Encoding data in the amplitudes of a superposition of basis states allows, in principle, to encode an exponential number of data components in a linear number of qubits. In the previous example, this reduces the qubit requirements from over $\num{750000}$ to just about $20$ qubits. The caveat now being that preparing arbitrary data vectors using this encoding becomes exponentially costly. However, actual data of interest is often not arbitrary but follows some internal structure, e.g., how images differ from random pixel values. In fact, for typical images there have been several recent results (both numerical and theoretical) showing that their underlying structure leads to quantum states after the encoding that are well-captured by tensor-network states and by tensor-network-inspired quantum circuits~\cite{Dilip2022, iaconis2023tensornetworkbasedefficient, jobst2023efficientmpsrepresentationsquantum, shen2024classification,maxwell_2024}. They can thus be prepared in a depth linear in the number of qubits required for the encoding. While we focus on image data in this paper, there are other types of data which result in lowly entangled states after amplitude encoding, for which we expect the methods of this paper to work similarly well~\cite{Lubasch2018, Lubasch2020, Gourianov2022, Hoelscher2025, Ritter2024, jobst2023efficientmpsrepresentationsquantum}.

Motivated by these results, we consider similar encodings in the following. We will also consider a way to interpolate between the amplitude-type encoding and the rotation encoding~\cite{Dilip2022}---as the rotation encoding is often considered a more powerful nonlinear feature map for a subsequent machine learning model~\cite{schuld2022machine, Stoudenmire2017}. Specifically, we consider two encodings for grayscale and color images that result in quantum states of the form
\begin{equation}
    \label{eq:target_state}
    \ket{\psi(\vec{x})} = \frac{1}{\sqrt{2^n}} \sum_{j=0}^{2^{n}-1} \ket{c(\vec{x}_j)} \otimes \ket{j}
\end{equation}
for images with $2^n$ pixels~\cite{Amankwah2022}. The state $\ket{j}$ of the $n$ so-called \emph{address qubits} tracks the position index $j$ of the pixel in the image. The state $\ket{c(\vec{x}_j)}$ encodes the data value $\vec{x}_j$ of the $j$th pixel. For grayscale images this encodes the grayscale value in a single \emph{color qubit}, for color images this encodes the vector of RGB values using three color qubits. Similarly to amplitude encoding, these image encodings only require a number of qubits that scales logarithmically with the number of pixels, but they have some further advantages: whereas amplitude encoding can lose contrast information due to forced normalization, the states in Eq.~\eqref{eq:target_state} circumvent that issue by directly constructing correctly normalized states; empirically, they have a slightly smaller entanglement entropy than amplitude-encoded states making them more compressible as tensor-network states~\cite{jobst2023efficientmpsrepresentationsquantum}; and, finally, in image classification tasks states using such encodings could be classified with a slightly higher accuracy than their amplitude-encoded counterparts~\cite{shen2024classification}.

\begin{figure}[t]
    % Subfigure (a)
    \begin{subfigure}[b]{\linewidth}
        \begin{tikzpicture}
            \node[anchor=north west,inner sep=0] at (0.1, 0)
                {\includegraphics[
                    width=\linewidth-0.1cm
                ]{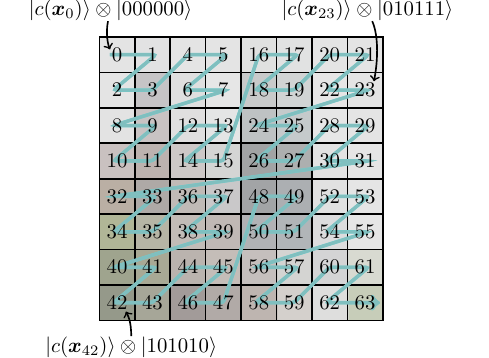}};
                % ]{figures_visualizations/image_encoding/hierarchical_order.pdf}};
            \node[anchor=north west] at (0, 0.1) {{\sffamily(a)}};
        \end{tikzpicture}
        \phantomcaption
        \label{fig:hierarchical_order}
    \end{subfigure}
    % Subfigure (b)
    \begin{subfigure}[b]{\linewidth}
        \begin{tikzpicture}
            \node[anchor=north west,inner sep=0] at (0.1, 0)
                {\includegraphics[
                    width=\linewidth-0.1cm
                ]{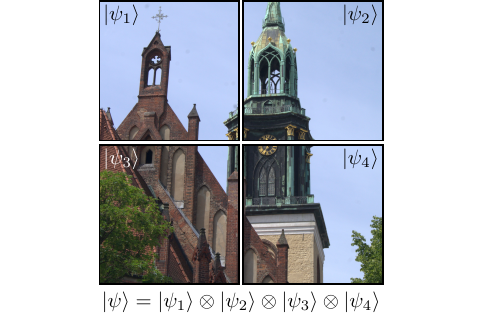}};
            \node[anchor=north west] at (0, 0.1) {{\sffamily(b)}};
        \end{tikzpicture}
        \phantomcaption
        \label{fig:patching}
    \end{subfigure}
    \vspace{-2em}
    
    \caption{\textbf{Hierarchical ordering of the pixels and encoding patches of an image.} (a) We choose a hierarchical ordering for enumerating the pixels when encoding an image according to Eq.~\eqref{eq:target_state}~\cite{latorre2005imagecompressionentanglement, Le2011, Le2011_2, jobst2023efficientmpsrepresentationsquantum}. In this order, the first two bits of the index labeling the position of the pixel denote the quadrant it lies in, the next two bits label its subquadrant, and so on. (b) To interpolate between the FRQI- or MCRQI-based encodings and the rotation encoding, we can encode patches of the image and take a product state of the individually encoded patches as an input state~\cite{Dilip2022}. This is shown here for four patches.}
    \label{fig:order_and_patching}
\end{figure}

The order in which the pixels are indexed can change the entanglement entropy of the resulting state~\cite{jobst2023efficientmpsrepresentationsquantum}. Here, we choose hierarchical indexing (shown in Fig.~\ref{fig:hierarchical_order}) based on the so-called $Z$- or Morton order~\cite{latorre2005imagecompressionentanglement, Le2011, Le2011_2, jobst2023efficientmpsrepresentationsquantum}---where the first two bits of the index $j$ label the quadrant of the image the pixel is in, the next two bits label the subquadrant, and so on. This tends to further decrease the entanglement entropy compared to other orderings, and thus should make the resulting states more compressible (see Ref.~\cite{jobst2023efficientmpsrepresentationsquantum} for grayscale images and App.~\ref{app:other_image_encodings} for color images).

\textbf{Flexible Representation of Quantum Images.}
For grayscale images, we use the \emph{flexible representation of quantum images (FRQI)}~\cite{Le2011, Le2011_2} as an encoding. In this case, the data value ${x}_j\equiv\vec{x}_j$ of each pixel is just a single number corresponding to the grayscale value of that pixel. We can encode this information in the $z$-polarization of an additional color qubit as
\begin{equation}
    \ket{c({x}_j)}
    = \cos({\textstyle\frac{\pi}{2}} {x}_j) \ket{0}
    + \sin({\textstyle\frac{\pi}{2}} {x}_j) \ket{1},
\label{eq:frqi_state}
\end{equation}
with the pixel value normalized to ${x}_j \in [0,1]$. Thus, a grayscale image with $2^n$ pixels is encoded into a quantum state with $n+1$ qubits---cf. Eq.~\eqref{eq:target_state}. In general, the complexity of preparing the resulting state \emph{exactly} scales exponentially with the number of qubits. Known constructions (without auxiliary qubits) use $\bigO{4^n}$ gates~\cite{Le2011, Le2011_2} or, allowing for classical preprocessing with $\bigO{n2^n}$ steps, construct a circuit with $\bigO{2^n}$ gates~\cite{Amankwah2022}. However, encoding typical images this way leads to lowly entangled quantum states that are well approximated by tensor-network states such as matrix-product states (MPSs)~\cite{Dilip2022, iaconis2023tensornetworkbasedefficient, jobst2023efficientmpsrepresentationsquantum, shen2024classification,maxwell_2024} whose bond dimension $\chi$ does not need to scale with the image resolution~\cite{jobst2023efficientmpsrepresentationsquantum}. Thus, preparing the state \emph{approximately} with a small error is possible with a number of gates that scales only as $\bigO{\chi^2n}$, i.e., linearly with the number of qubits. Note that the cost of classical preprocessing to obtain the MPS via successive SVDs is $\bigO{\chi2^n}$. While this is technically exponential in the number of qubits, this is not a problem in practice: the images we consider must still be stored classically. The cost of computing the MPS is only linear in the number of pixels, so the cost of compressing the image is not substantially more than storing it. In certain cases, this cost can even be further reduced to $\bigO{\chi^3 n}$ by making use of tensor cross interpolation algorithms~\cite{Savostyanov2011, Savostyanov2014, Dolgov2020, Nunez-Fernandez2022, Nunez-Fernandez2024}. In any case, while the cost of the classical preprocessing may be similar to the exact state preparation, the resulting quantum circuits are exponentially more efficient.

\textbf{Multi-Channel Representation of Quantum Images.}
For color images, we use the \emph{multi-channel representation of quantum images (MCRQI)}~\cite{Sun2011, Sun2013} as an encoding. For each pixel, the data value now has several components, $\vec{x}_j = \left(\!\begin{array}{cccc} x^R_j, & x^G_j, & x^B_j, & x^{\alpha}_j \end{array}\!\right)^{\transpose}$, corresponding to the three RGB color channels and a possible fourth $\alpha$ channel indicating the opacity of the image. If only the three RGB channels are available for a given image (as is the case for all color image datasets considered in this work), the image is at full opacity and we can simply set the $\alpha$ channel to zero~\cite{Sun2013}. The color information of a pixel is then encoded in a three-qubit state as
\begin{equation}
\begin{aligned}
    \ket{c(\bm{x}_j)} = \frac{1}{2}\Big(
     &\cos({\textstyle\frac{\pi}{2}} x^R_j)\ket{000}
    + \sin({\textstyle\frac{\pi}{2}} x^R_j)\ket{100}\\
    +&\cos({\textstyle\frac{\pi}{2}} x^G_j)\ket{001}
    + \sin({\textstyle\frac{\pi}{2}} x^G_j)\ket{101}\\
    +&\cos({\textstyle\frac{\pi}{2}} x^B_j)\ket{010}
    + \sin({\textstyle\frac{\pi}{2}} x^B_j)\ket{110}\\
    +&\cos({\textstyle\frac{\pi}{2}} x^{\alpha}_j)\ket{011}
    + \sin({\textstyle\frac{\pi}{2}} x^{\alpha}_j)\ket{111}\Big),
    \label{eq:RGB_color_state}
\end{aligned}
\end{equation}
with normalized values $x^R_j, x^G_j, x^B_j, x^{\alpha}_j \in [0,1]$. Thus, a color image with $2^n$ pixels is encoded into a quantum state with $n+3$ qubits---cf. Eq.~\eqref{eq:target_state}. To prepare the state \emph{exactly} on a quantum computer, we can essentially reuse the same circuit that prepares an FRQI state and run it for each color channel separately. Thus, known circuits for exact state preparation (without auxiliary qubits) use $\bigO{4^n}$ gates~\cite{Sun2011, Sun2013} or, after classical preprocessing with $\bigO{n2^n}$ steps, use $\bigO{2^n}$ gates~\cite{Amankwah2022}. However, just as for grayscale images, encoding typical color images in this way results in lowly-entangled states, which are well approximated by tensor-network states. We numerically demonstrate this in App.~\ref{app:other_image_encodings}, where we consider the MCRQI and two other color image encodings, and show that for all of them the bipartite entanglement entropy does not grow with increasing image resolution. We choose to work with the MCRQI in the following, since it has the lowest entanglement entropy out of the three encodings. Due to their low entanglement, preparing color image states \emph{approximately} can be done with a circuit whose number of gates scales only linearly with the number of qubits. We will demonstrate this numerically in the next section.

\textbf{Interpolating to the Rotation Encoding.}
As mentioned in the introduction, the rotation encoding not only enjoys popularity in QML due to its efficient state preparation circuit, but also because from the point of view of kernel-based machine learning it implements a nonlinear function leading to a potentially more powerful feature map~\cite{schuld2021supervisedquantummachinelearning, schuld2022machine}. As such, it has also been successfully used in tensor-network-based machine learning~\cite{Stoudenmire2017}. To try to keep some of the benefits of the rotation encoding while keeping the number of qubits manageable, we can interpolate between the FRQI- or MCRQI-based encoding and the rotation encoding by splitting the original $2^n$-pixel image into $N_p$ patches and encoding each patch separately~\cite{Dilip2022}---this is shown for four patches in Fig.~\ref{fig:patching}. The number of required qubits is then $n = N_p \cdot \big(\log_2(2^n/N_p) + n_c\big)$, with either $n_c=1$, a single color qubit, for the FRQI or $n_c=3$ color qubits for the MCRQI. Using this \emph{patched encoding}, a single patch, $N_p=1$, corresponds to encoding the full image, while using as many patches as pixels, $N_p=2^n$, corresponds to encoding each pixel individually. For the FRQI, this recovers the rotation encoding exactly, with each qubit being rotated as $e^{-i\frac{\pi}{2}\sigma^y_j x_j}\ket{0} = \cos({\textstyle\frac{\pi}{2}} {x}_j) \ket{0} + \sin({\textstyle\frac{\pi}{2}} {x}_j) \ket{1}$. For the MCRQI this leads to a product state of three-qubit states encoding the RGB values of each pixel separately. By varying the number of patches $N_p$, we can gradually interpolate between the two limiting cases of encoding the entire image into a single entangled state or representing it as a product state of individual pixels.

\textbf{Multiple Copies of the Encoding.}
Another popular technique in QML to make models more powerful is to start from a product state of multiple copies of the state encoding the input data. This can be seen as a particular instance of the more general data reuploading framework, based on which certain universality results for the outputs of quantum classifiers can be proved~\cite{P_rez_Salinas_2020, Goto_2021, Schuld_2021}. Another way of seeing the possible advantage of this approach is that this allows to interpret the output state of a unitary operation that entangles the different copies as a nonlinear transformation of the matrix elements of a single copy of the input density matrix~\cite{Holmes2023}. Thus, the \emph{multi-copy encoding} can give the quantum model access to otherwise possibly missing nonlinearities required for solving many tasks.
\section{Optimizing Quantum Circuits for Efficient State Preparation}
\label{sec:efficient_encoding}

In the preceding section, we introduced encodings of classical image data that, at least for typical images, result in states with low entanglement that are effectively represented by tensor-network states. In this section, we are concerned with finding efficient circuits that (approximately) prepare those states on a quantum computer.

The efficient state preparation is possible because images have an internal structure that can be exploited for further compression. For instance, typical images exhibit a concentrated Fourier spectrum as the low-frequency components often already faithfully capture the subject and high-frequency components are perceived as noise. Truncating this Fourier spectrum leads to a representation as an MPS, whose bond dimension does not need to grow with increasing image resolution to keep the approximation error below some threshold~\cite{jobst2023efficientmpsrepresentationsquantum}. This MPS representation of the state then allows for preparation on a quantum computer with a number of gates that scale only linearly with the number of qubits. While an MPS can be directly mapped to a linear-depth quantum circuit employing multi-qubit gates~\cite{Schoen2005, Schoen2007, Lubasch2020, Smith2022, Lin2021, Barratt2021}, we can further reduce the required number of gates by choosing an MPS-inspired circuit architecture consisting of two-qubit gates only~\cite{Dilip2022, iaconis2023tensornetworkbasedefficient, jobst2023efficientmpsrepresentationsquantum, shen2024classification,maxwell_2024}. The parameters in the circuit are subsequently optimized to maximize the fidelity with the target state. We will introduce the specific circuit architecture and our optimization algorithm in the following.

\subsection{Circuit Architecture}
\label{sec:circuit_architecture}

\begin{figure}[t]
    \centering
    \includegraphics[width=\linewidth]{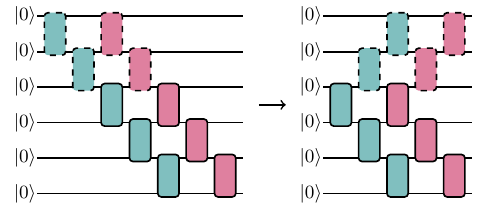}
    
    \caption{\textbf{Illustration of quantum circuits inspired by MPSs.} The left side shows a circuit with a staircase pattern with two layers (represented in turquoise and pink), where two-qubit gates are applied sequentially, corresponding to a right-canonical MPS. The right side shows the proposed circuit architecture corresponding to an MPS in mixed canonical form. By effectively shifting the gates with the dashed outlines to the right, the gates are applied sequentially outward starting from the center. This reduces the circuit depth while maintaining its expressivity.
    }
    \label{fig:architecture}
\end{figure}

A widely used quantum circuit architecture consists of multiple layers of two-qubit gates sequentially arranged in a staircase pattern~\cite{Lin2021, Barratt2021, Dilip2022, iaconis2023tensornetworkbasedefficient, jobst2023efficientmpsrepresentationsquantum, shen2024classification, PhysRevA.101.032310, Rudolph2023_2, Ben-Dov2024}, shown graphically in Fig.~\ref{fig:architecture} on the left. This pattern is inspired by the right-canonical form of MPSs~\cite{SCHOLLWOCK201196, Or_s_2014}. Indeed, an MPS with a bond dimension of two corresponds exactly to a single layer of such a sequential circuit, since in right-canonical form all MPS tensors are isometric and can be implemented by two-qubit gates---for larger bond dimensions, multi-qubit gates are needed for this correspondence~\cite{Schoen2005, Schoen2007, Lubasch2020, Smith2022, Lin2021, Barratt2021}. To avoid having to decompose multi-qubit gates, the more popular approach is to simply stack several layers of the circuit corresponding to an MPS with a bond dimension of two.

Instead of the sequential circuit obtained from a right-canonical MPS, we can also consider an MPS in mixed canonical form, where the canonical center is a central site of the MPS. For an MPS with a bond dimension of two, this now corresponds to a sequential circuit where two staircases of two-qubit gates begin in the center of the system and move outwards~\cite{Wei2022, Bohun2024}. One layer of such a circuit is shown in Fig.~\ref{fig:architecture} on the right in turquoise, and we refer to it as a \emph{center-sequential} circuit. Compared to the original quantum circuit, we now obtain a circuit with reduced depth but the same expressivity. To increase the expressivity of the circuit without requiring multi-qubit gates, just as for the original sequential circuit, we can simply stack more layers of the circuit (see Fig.~\ref{fig:architecture} on the right, turquoise and pink layers).

The two-qubit gates appearing in the circuit can be parametrized in various ways, which can differ in the number of CNOTs required to implement them on quantum hardware~\cite{PhysRevA.70.012310}. They thus offer a trade-off in expressivity versus CNOT gate count efficiency, which might be desirable since, on many hardware platforms, CNOT gates require longer pulse sequences and exhibit higher error rates than single-qubit gates. Specifically, we consider three different two-qubit primitives: (i)~arbitrary unitary matrices in $SU(4)$ requiring three CNOTs to be implemented~\cite{PhysRevA.69.062321}, (ii)~special orthogonal matrices in $SO(4)$ requiring only two CNOTs~\cite{Wei:2012zgm}, and (iii)~hardware-efficient gates parametrized by a single CNOT enclosed by single-qubit gates, which we will refer to as \emph{sparse} gates~\cite{shen2024classification}. We give further details on the parametrization in the description of the algorithm below.

\subsection{Optimization Algorithm}
\label{sec:algorithm}

After fixing the structure of the circuit to the center-sequential layout, the goal is to optimize the gates in the circuit to maximize its fidelity with the target state. We assume the target state is given as an MPS, since---as discussed in the preceding section---the target states we consider are lowly entangled and their MPS representation can be computed with acceptable cost. The optimization of the circuit then consists of four steps:
\begin{enumerate}[itemsep=-1ex,partopsep=1ex,parsep=1ex]
    \item[(1)] We initialize the gates in the circuit via an approximate \emph{analytical decomposition} of the target state~\cite{PhysRevA.101.032310}.
    \item[(2)] By \emph{sweeping} through the circuit, we iteratively replace individual two-qubit gates with the ones maximizing the fidelity~\cite{Evenbly2009, Lin2021}.
    \item[(3)] We then \emph{decompose the two-qubit gates} into CNOTs and single-qubit gates~\cite{PhysRevA.69.062321, PhysRevA.70.012310, Wei:2012zgm}.
    \item[(4)] Finally, we further optimize the parametrized single-qubit gates using the \emph{Broyden–\allowbreak Fletcher–\allowbreak Goldfarb–\allowbreak Shanno (BFGS) algorithm}~\cite{NoceWrig06}.
\end{enumerate}
This combines previous findings that the analytic initialization of the gates followed by the sweeping algorithm yields high-fidelity quantum circuits~\cite{Rudolph2023_2, Ben-Dov2024}, and that switching from an optimization algorithm with local updates like the sweeping algorithm to a global, second-order quasi-Newton method like BFGS in the latter stages of the optimization can speed up convergence and further improve the quality of the optimized state significantly~\cite{Hauru_2021}. In the following, we describe each algorithm step in more detail.

\begin{figure}[!bt]
    \centering
    \includegraphics[width=\linewidth]{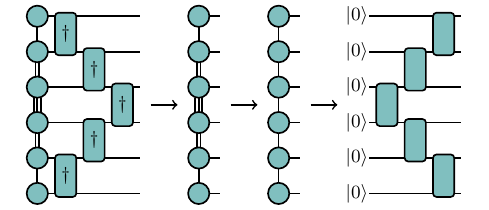}
    
    \caption{\textbf{Approximate analytical decomposition of the target state to initialize circuit layers.} In the first step, if we already have layers of the circuit, we contract their conjugate transpose with the target state, partially disentangling it. Next, we have the partially disentangled target state and truncate it to an MPS with a bond dimension of two. This truncated MPS corresponds exactly to one layer of a center-sequential circuit, which we can use as a new first layer of the multi-layer circuit.}
    \label{fig:algorithm_init}
\end{figure}

\textbf{Analytical Decomposition.}
When optimizing a quantum circuit, it has been shown that it is important to initialize the parameters close to an optimum to aid in convergence~\cite{Dborin2022, Zhang2022, Kashif2023, Rudolph2023, Park2024, Wang2024, Rudolph2023_2, Puig2025, Mhiri2025}. In our case, since the target state is given as an MPS, we can make use of the correspondence between one layer of the circuit and an MPS with bond dimension two to iteratively disentangle the target state and construct new layers of the circuit~\cite{PhysRevA.101.032310}. For a given target state, we can truncate its MPS representation to a bond dimension of two, which yields an MPS that we can exactly write as a single layer of our circuit ansatz. This highly-truncated state should already capture some relevant features of the target state. We can initialize more layers of the circuit by contracting the conjugate transpose of the already obtained circuit layers with the target state, which gradually disentangles the target state towards the product state $\ket{\vec{0}}\equiv\ket{0}^{\otimes (n+n_c)}$. The resulting disentangled state can then again be truncated to an MPS with bond dimension two and mapped to one layer of the center-sequential circuit, which is then used as a new first layer in the multi-layer circuit---these steps are graphically shown in Fig.~\ref{fig:algorithm_init}. We can thus add an additional layer to a given pre-existing circuit whose layers do not stem directly from this initialization technique, e.g., one obtained from a previous optimization with one fewer layer.

\begin{figure*}[!t]
    \centering
    \includegraphics[width=\linewidth]{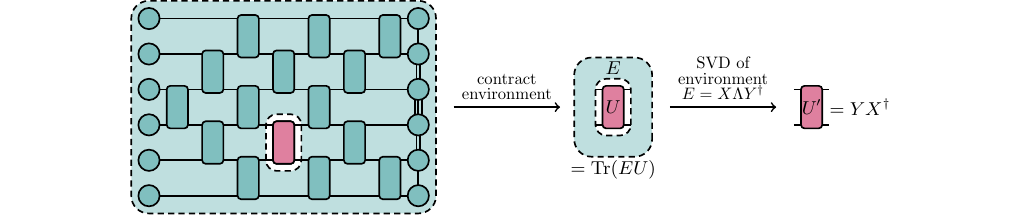}
    
    \caption{\textbf{Sweeping optimization algorithm.} The sweeping algorithm maximizes the fidelity of the state generated by the quantum circuit and the target state by one-by-one replacing each unitary gate in the circuit with the one maximizing the fidelity. For each unitary $U$, we can compute its environment tensor $E$, such that the fidelity is given by $\Tr(E U)$. If we consider general two-qubit gates $U \in SU(4)$, the updated gate can be obtained from the singular value decomposition (SVD) of the environment $E = X \Lambda Y^{\dagger}$ as $U \to U' = YX^{\dagger}$~\cite{Evenbly2009, Lin2021, Higham1989}. Similar update rules exist for the other gate sets we consider. The algorithm is efficient because we can cache partial contractions of the environment tensor as an MPS since both the target state and the quantum circuit are lowly entangled states.}
    \label{fig:algorithm_sweep}
\end{figure*}

Note that the initialization technique above works for both the general unitary and special orthogonal two-qubit gates we consider. For general gates in $SU(4)$ we simply drop an irrelevant overall complex phase from the unitary matrices of the truncated, mixed-canonical MPS. For the special orthogonal gates in $SO(4)$, note that the target state is always real so that the resulting matrices from the truncated MPS will always be orthogonal matrices either in $SO(4)$ (as needed) or in $O(4) \backslash SO(4)$ (i.e., real gates with determinant $-1$ instead of $+1$). In the latter case, since one of the input qubits of the gate will always act on the state $\ket{0}$, we can multiply the gate by a CNOT gate controlled by the qubit in state $\ket{0}$, such that the action of the gate remains unchanged but its determinant changes from $-1$ to $+1$ as required.

For the sparse gates, we follow the same idea of disentangling the target state, but slightly adapt the procedure. The circuit is made of gates of the form $(V_1 \otimes V_2) \mathrm{CNOT}$, with general single-qubit gates $V_1, V_2 \in SU(2)$, arranged in the center-sequential layout. At the beginning of the circuit, there is an additional initial layer of general single-qubit gates; like this, every CNOT gate in the circuit is effectively surrounded by four single-qubit gates. To add a new layer of the circuit, we initialize the new sparse gates of the layer in the inverse order that they act in the circuit. For each sparse gate, we obtain its two single-qubit-gate components by truncating the MPS of the target state to a product state, and selecting the two single-qubit gates that locally rotate the state $\ket{0}$ into the single-qubit state of the truncated MPS. The new sparse gate is obtained from combining these single-qubit gates with a CNOT gate. Before we move to the next gate, we update the target state by contracting the conjugate transpose of the new sparse gate with it. Once all sparse gates in the new layer have been initialized this way, we truncate the updated target state one final time to a product state, and set the new initial layer of single-qubit gates to be the ones that prepare this product state from the state $\ket{\vec{0}}$.

\textbf{Sweeping Optimization Algorithm.}
The sweeping algorithm iteratively updates the circuit by sweeping through its gates and replacing one two-qubit gate at a time by the one maximizing the fidelity of the state generated by the quantum circuit and the target state. Each iteration starts at the first layer of the circuit and moves on layer by layer to the final one. Within each layer, we first start from the central gate (the leftmost gate in the $<$-shape of the layer in Fig.~\ref{fig:algorithm_sweep}) and update gates by sweeping along the right-canonical arm of the center-sequential circuit, followed by updating the remaining gates in the layer starting from the center sweeping along the left-canonical arm.

To update a given gate $U$, we compute its environment $E$ by contracting all surrounding tensors such that the fidelity is given by $\Tr(E U)$---see Fig.~\ref{fig:algorithm_sweep}. We can then select the gate $U$---from one of the three gate sets we consider---that maximizes this expression. The contraction of the environment is efficient because we can cache and reuse partial contractions of the tensor network; since the target state, as well as the circuit that is optimized to approximate the target state, are lowly entangled, these partial contractions themselves can always be efficiently stored as an MPS. When we consider general unitary two-qubit gates ${U \in SU(4)}$, we can obtain the updated gate from the singular value decomposition (SVD) of the environment $E = X \Lambda Y^{\dagger}$ as $U \to U' = Y X^{\dagger}$. This is the unitary matrix guaranteed to maximize $\Tr(E U)$, i.e., the fidelity for a given environment $E$~\cite{Higham1989, Evenbly2009}.

If we restrict the two-qubit gates to be special orthogonal $U \in SO(4)$, we must ensure that the updated gate is real and has a unit determinant. The first point is always guaranteed since the target state and all gates in the circuit are real. To ensure the second point, after computing the SVD of the environment as $E=X \Lambda Y^{\dagger}$, we flip the sign of one of the singular vectors associated with the smallest singular value in $\Lambda$, if the new gate does not have the required determinant. This means the update rule for the special orthogonal case---assuming the singular values in $\Lambda$ are sorted in descending order---is given by $U \rightarrow U' = Y C X^{\dagger}$ with the diagonal matrix $C = \text{diag}\!\left(\!\begin{array}{cccc}1&1&1&\det(YX^{\dagger})\end{array}\!\right)$, where $\det(YX^{\dagger}) = \pm1$ because $X$ and $Y$ are orthogonal matrices. The resulting matrix is the special orthogonal matrix that is guaranteed to maximize the fidelity~\cite{myronenko2009closedformsolutionrotationmatrix}.

If we further restrict to sparse two-qubit gates that are parametrized by a single CNOT gate and some single-qubit gates in $SU(2)$, we can update these single-qubit gates in essentially the same way as the general two-qubit gates in $SU(4)$ before. The only difference is that the matrix dimensions of the contracted environment tensors are smaller.

In practice, before moving on to the next steps of the algorithm, we iterate the two steps above to grow the number of layers to the desired number~\cite{Rudolph2023_2}. In each iteration, we initialize an additional layer according to the analytical decomposition and optimize all layers for some iterations with the sweeping algorithm.

\textbf{Two-Qubit Gate Decomposition.}
In the preceding steps, we performed the optimization using (i)~general two-qubit gates in $SU(4)$, (ii)~special orthogonal two-qubit gates in $SO(4)$ and (iii)~sparse two-qubit gates using only a single CNOT gate. Until now, though, we have stored them as $4\times4$ matrices, so we need to decompose them into hardware-compatible CNOTs and parametrized single-qubit gates.

General two-qubit gates in $SU(4)$ can be decomposed into three CNOTs and several single-qubit gates~\cite{PhysRevA.69.062321}. The first and last gates in this decomposition are single-qubit gates, so after decomposing every two-qubit gate in the circuit, there are consecutively appearing single-qubit gates that can be merged into a single one. After this merging of single-qubit gates, we have one layer of single-qubit gates acting on the initial state $\ket{\vec{0}}$, and the remaining two-qubit gates without the redundant single-qubit gates are then parametrized as
\begin{equation}
    \vcenter{\hbox{\includegraphics[height=1cm]{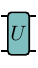}}} = \vcenter{\hbox{
        \includegraphics[height=1cm]{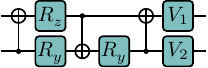}}},
\end{equation}
where $R_{\alpha}$ denote single-qubit rotations about the axis $\alpha\in\{x,y,z\}$ and $V_i = R_z(\phi_i)R_y(\theta_i)R_z(\lambda_i)$ are general single-qubit gates in $SU(2)$. Thus, there are nine free parameters per two-qubit gate corresponding to the rotation angles. Together with the parameters of the initial layer of single-qubit gates, this gives a total of $9(L-1)d+2L$ parameters for a center-sequential circuit with $L$ qubits and $d$ layers.

Two-qubit gates from the special orthogonal group $SO(4)$ can be decomposed using two CNOTs and six single-qubit $y$-rotation gates~\cite{Wei:2012zgm}. After merging consecutive single-qubit gates, we again have an initial layer of single-qubit gates acting on the initial state $\ket{\vec{0}}$, and two-qubit gates parametrized as
\begin{equation}
    \vcenter{\hbox{\includegraphics[height=1cm]{figures_visualizations/u.pdf}}}
     = \vcenter{\hbox{
        \includegraphics[height=1cm]{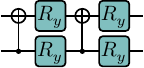}}},
\end{equation}
which now only require four $R_y$ gates in addition to the two CNOT gates. In total, a center-sequential circuit with $L$ qubits and $d$ layers using special orthogonal two-qubit gates thus has $4(L-1)d+L$ free parameters.

Finally, the sparse gates, by definition, consist of only a single CNOT gate and single-qubit gates. We again have an initial layer of single-qubit gates acting on the initial state $\ket{\vec{0}}$ and each two-qubit gate is then parametrized as
\begin{equation}
    \vcenter{\hbox{\includegraphics[height=1cm]{figures_visualizations/u.pdf}}} = \vcenter{\hbox{
        \includegraphics[height=1cm]{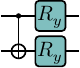}}}.
\end{equation}
Note that in our case, we can always choose the general single-qubit gates to be $R_y$ gates, since the single-qubit gates obtained from the sweeping algorithm are real and we can modify them to ensure a positive determinant. For this, we focus on one qubit at a time, starting from the uppermost qubit (cf. Fig.~\ref{fig:architecture}) and then moving down one by one, and for each qubit separately we consider the set of single-qubit gates acting on it. We start with the last single-qubit gate acting on the given qubit, update the gate if necessary, and afterwards successively update the preceding single-qubit gate if necessary. To check whether we need to update a single-qubit gate, we can check its determinant: If it is positive, the single-qubit gate is an $R_y$ gate and we can move on to the next gate. If its determinant is negative, we need to modify the gates in the circuit. We decompose the gate as $R_y\,X$ and move the Pauli-$X$ gate past the preceding CNOT gate. If it needs to move past the control of the CNOT gate, we use the identity $(X \otimes \Id) \text{CNOT} = \text{CNOT} (X \otimes X)$ which moves the Pauli-$X$ gate past the CNOT gate and creates another Pauli-$X$ gate on the lower neighboring qubit. The single-qubit gates in the circuit can be updated by absorbing these Pauli-$X$ gates, and then we move on to the next single-qubit gate (i.e., the upper one of the two that were just updated). In the other case, where the Pauli-$X$ gate needs to move past the target of the CNOT gate, the gates commute and we can simply update the preceding single-qubit gate by absorbing the Pauli-$X$ gate; then, we move on to the next single-qubit gate (i.e., the one that was just updated). Once we arrive at the first single-qubit gate acting on the qubit, i.e., the one that acts on the state $\ket{0}$, we can eliminate a potentially needed Pauli-$X$ gate by making use of the relation $X\ket{0} = \ket{1} = R_y(\pi)\ket{0}$ and merging the $y$-rotation gates. Once we reach the last (bottommost in Fig.~\ref{fig:architecture}) qubit, note that only the target of CNOT gates acts on this qubit, so any Pauli-$X$ gates can be commuted past the CNOT gates and no new Pauli-$X$ gates are created by this commutation. Thus, the algorithm terminates after these final substitutions and afterwards the circuit consists of only $R_y$ and CNOT gates. The center-sequential circuit using sparse gates with $L$ qubits and $d$ layers therefore has $2(L-1)d+L$ free parameters in total.

\textbf{Broyden-Fletcher-Goldfarb-Shanno Optimization.}\\
Far from the optimal circuit, the sweeping algorithm substantially improves the fidelity with the target state with each update. However, close to convergence, the sweeping algorithm performs only small update steps in the direction of a maximum, and thus acts similarly to a simple gradient ascent optimizer~\cite{Hauru_2021}. Gradient ascent only makes use of information about the gradient of the function (making it a first-order method) leading to a linear convergence rate. Moreover, we only update a single tensor at a time, which can be seen as a variant of coordinate descent. This leads to very slow convergence close to a maximum of the fidelity. To speed up convergence, we therefore switch to the Broyden-Fletcher-Goldfarb-Shanno (BFGS) algorithm~\cite{NoceWrig06} to optimize the parameters in the circuit after the gate decomposition. The BFGS algorithm is also an iterative method, similar to gradient ascent, except that the update rule additionally incorporates information about the Hessian of the function, making it a second-order method. It belongs to the class of quasi-Newton methods, since it approximates Newton’s method by estimating the inverse Hessian based on gradient evaluations. Theoretically, Newton's method promises a faster, quadratic convergence. Moreover, at each step, we update all parameters at once allowing to traverse the cost landscape more efficiently than the coordinate descent of the sweeping algorithm. Combined, this leads to a faster convergence and higher fidelity after the optimization compared to only using the sweeping algorithm.

Note that, in principle, we can use the same methods as for the sweeping algorithm to compute the gradient of each gate in the circuit by contracting its environment and caching partial contractions of the circuit as MPSs. This approach formally scales only polynomially with the system size and the number of parameters. However, current state-of-the-art implementations of statevector simulations are so highly optimized that---even though they formally scale exponentially with the number of qubits---for the system sizes we consider here (up to $19$ qubits) they still outperform our Python implementation of the tensor-network-based gradient computation. Thus, for the results presented in the following section, we use statevector simulators and automatic differentiation for the gradient computation for the BFGS optimization.

\subsection{Efficient Circuit Representations of Typical Machine Learning Datasets}
\label{sec:typical_ml_datasets}

We apply the algorithm described in Sec.~\ref{sec:algorithm} to find circuits which encode four standard machine learning datasets: the MNIST~\cite{Lecun1998, deng2012mnist}, Fashion-MNIST~\cite{FashionMNIST}, CIFAR-10~\cite{CIFAR10}, and Imagenette~\cite{imagenette} datasets. We will use the same datasets to show the results of quantum classifiers in Sec.~\ref{sec:classifiers}. For the implementation of the algorithm, we use Ray~\cite{moritz_ray2018} to parallelize the execution of the circuit optimization tasks. For the gradient-based optimization, we use PennyLane~\cite{bergholm2022pennylaneautomaticdifferentiationhybrid} in combination with the just-in-time compilation and vectorization capabilities of JAX~\cite{jax}, and SciPy's BFGS implementation~\cite{scipy2020}.

\begin{figure*}[t]
    \centering
    \includegraphics[
        width=\linewidth,
        trim={0.6cm 0.7cm 0.5cm 0.5cm},
        clip
    ]{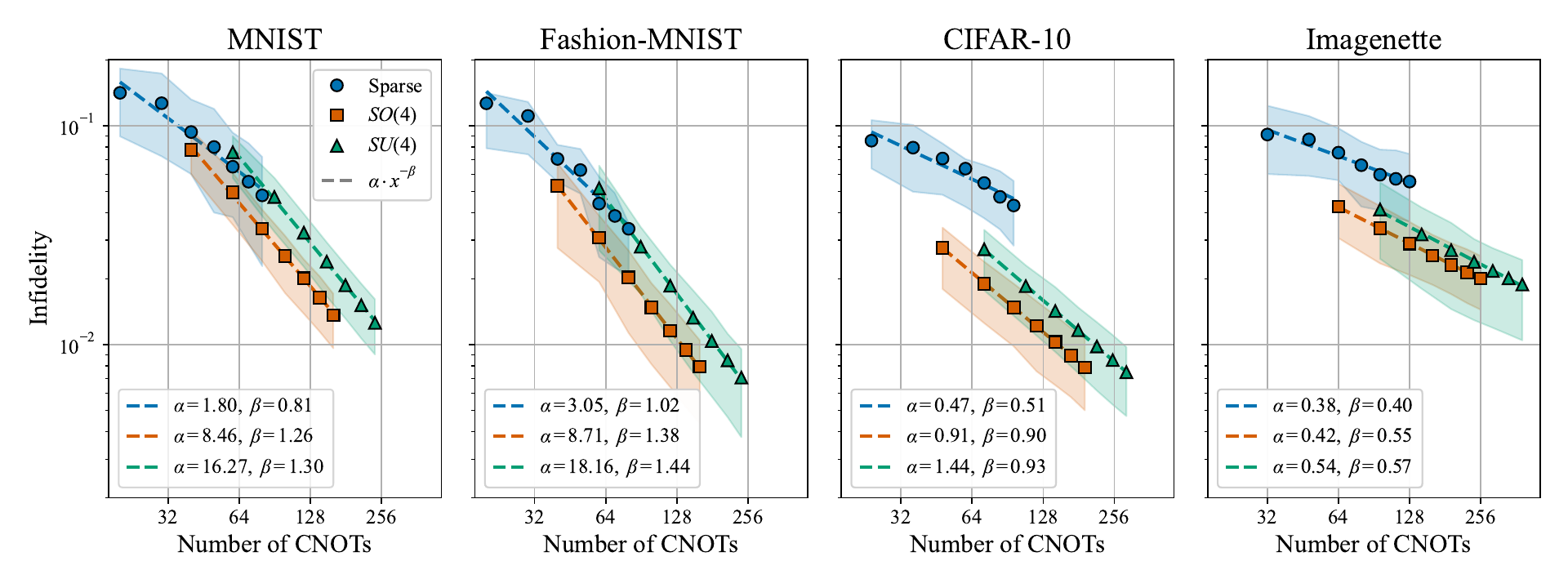}
    
    \caption{\textbf{Scaling of the infidelity with an increasing number of CNOTs.} We encode $100$ randomly selected images ($10$ per class) from the MNIST, Fashion-MNIST, CIFAR-10 and Imagenette datasets as discussed in Sec.~\ref{sec:encoding}. The markers show the average infidelity between the exactly encoded states and their quantum circuit approximation (defined as $1 - \left|\braket{\psi_{\text{exact}}}{\psi_{\text{circ.}}}\right|^2$) plotted against the number of CNOTs in the circuit using sparse (blue circles), special orthogonal (orange squares) and general unitary (green triangles) two-qubit gates. The shaded areas show the $25$th--$75$th percentiles, the dashed lines show the results of fitting an algebraic decay $f(x)=\alpha \, x^{-\beta}$ to the data. The fitted values, including their uncertainties, are given in Table~\ref{tab:encoding_cnot_infid} in the appendix. We observe that the infidelity of the circuits using special orthogonal or unitary gates decays in the same way, with the difference that the circuits utilizing special orthogonal gates save on CNOT gates due to their cheaper decomposition.}
    \label{fig:infidelity_vs_cnot}
\end{figure*}

The MNIST dataset~\cite{Lecun1998, deng2012mnist} is a simple and widely used dataset for training machine learning models. It contains grayscale images of handwritten digits between `$0$' and `$9$', and associated labels indicating the correct digit. The original images have $28\times28$ pixels. Here, we use bilinear interpolation to resize them to $32\times32$ pixels making them suitable for processing on a quantum computer. The class distribution over the $\num{70000}$ images is approximately uniform, with each class representing between $9\%$ and $11\%$ of the dataset. Current state-of-the-art classical convolutional neural networks achieve a validation accuracy of above $99.7\%$ on this dataset~\cite{Ciregan2012, Wan2013, Hasanpour2016, Rajasegaran2019}.

The Fashion-MNIST dataset~\cite{FashionMNIST} was introduced as a more challenging alternative to MNIST, after it became apparent that MNIST was too easily solved and no longer posed a significant challenge for more sophisticated classification models. The dataset also features $\num{70000}$ grayscale images with an original resolution of $28\times28$ pixels, which we again resize to $32\times32$ pixels using bilinear interpolation. Instead of handwritten digits, the images feature different clothing articles and the associated labels denote the ten different classes of fashion articles, e.g., T-shirts, trousers or sneakers. The images are equally distributed over the ten classes. While Fashion-MNIST is a more complex dataset than MNIST, current classical state-of-the-art neural networks nonetheless achieve around $94\%$--$98\%$ validation accuracy~\cite{Rajasegaran2019, Meshkini2020, Kayed2020}, making it still a relatively simple toy dataset.

The CIFAR-10 dataset~\cite{CIFAR10} is a more realistic dataset, and also widely used as a benchmark in machine learning and computer vision. It consists of $\num{60000}$ $32\times32$-pixel color images. The images are photographs of objects belonging to ten different classes, such as airplanes, automobiles or birds, with an equal number of images per class. Being a more challenging dataset, classical state-of-the-art models `only' achieve a validation accuracy of $92\%$--$95\%$~\cite{Rajasegaran2019, Hasanpour2016, Huang2017}.

Finally, we consider the Imagenette dataset~\cite{imagenette}, which is a subset of the more well-known and larger ImageNet dataset~\cite{ImageNet}. It includes color images of ten more easily differentiated classes from the ImageNet dataset, e.g., cassette players, garbage trucks and parachutes, and is intended as a smaller and simpler stand-in for the huge ImageNet dataset. It comprises only $\num{13394}$ images. The resolutions and aspect ratios vary between images, so to work with a uniform resolution we take the largest square section of each image and resize them to $128\times128$ pixels using bilinear interpolation. Classical neural networks achieve around $94\%$ validation accuracy on this dataset~\cite{imagenette}.

We test the optimization algorithm described in the previous subsection on images from these datasets. First, we want to investigate how the approximation quality of the optimized circuit improves when increasing the number of layers in the circuit. For this, we consider the infidelity of the circuit with its target state, defined as $1 - \left|\braket{\psi_{\text{exact}}}{\psi_{\text{circ.}}}\right|^2$. To compare the circuits using different two-qubit-gate primitives, i.e., general unitary gates, special orthogonal gates or sparse gates, we compare them by the total number of CNOTs used in the circuit. This is because CNOT gates typically incur larger errors than single-qubit gates in many hardware realizations, so for a faithful implementation on a quantum computer we want to minimize the total number of CNOT gates. We focus on CNOT gates as a metric because we intend to offer circuits that can be executed on today's noisy hardware, as has been demonstrated for similar circuit architectures in Refs.~\cite{iaconis2023tensornetworkbasedefficient, shen2024classification}. A complementary view would be to see this state preparation as a subroutine on a future fault-tolerant quantum computer, where (when working with a Clifford+$T$ gate set) implementing Clifford gates is free but the implementation of $T$ gates becomes costly. We show a similar comparison with the number of $T$ gates in App.~\ref{app:t_count}.

As a representative example for this comparison, we take $100$ images from each of the four datasets, $10$ random images from each class, and optimize circuits with $2$--$8$ layers for each of the three different two-qubit gates. The results are shown in Fig.~\ref{fig:infidelity_vs_cnot}, where blue circles show the average infidelity for circuits using sparse two-qubit gates, orange squares the average infidelity for special orthogonal two-qubit gates and green triangles the average infidelity for general unitary two-qubit gates. The shaded areas correspond to the $25$th--$75$th percentiles. The dashed lines show algebraic fits to the data, the obtained values including uncertainties are summarized in Table~\ref{tab:encoding_cnot_infid} in the appendix.

We can see for all datasets that the infidelity decays algebraically with the number of CNOTs, which is proportional to the number of layers of the circuit. This is not unexpected, as these circuits are similar to MPSs with the number of layers playing a role akin to the bond dimension, and for MPSs such an algebraically decaying infidelity can be derived~\cite{jobst2023efficientmpsrepresentationsquantum}. For the circuits using the sparse gates, we see that while they use the fewest CNOTs per layer, they also make use of them in the least efficient way, displaying the largest infidelity for a fixed number of CNOT gates out of all three gate types. The infidelities of the circuits using general unitary or special orthogonal two-qubit gates decay at the same rate. Since the target states are real, it seems reasonable to expect that a circuit using real gates can approximate the target states just as well as a circuit using complex-valued gates with the same number of layers. From this intuition, one would expect the ratio of prefactors of the algebraic decay to roughly match the ratio $2/3\approx0.67$ of CNOTs required for implementing a special orthogonal or general unitary gate. For the low-resolution datasets with $32\times32$ pixels, the empirical ratios are even smaller with $8.46/16.27 \approx 0.52$ for MNIST, $8.71/18.16 \approx 0.48$ for Fashion-MNIST, and $0.91/1.44 \approx 0.63$ for CIFAR-10, indicating that the circuits with special orthogonal gates are even more efficient than expected on these datasets. However, on the higher resolution dataset Imagenette with $128\times128$ pixels, the ratio of prefactors is $0.42/0.54 \approx 0.78$ which is larger than the expected ratio. Still, across all four datasets, the special orthogonal gates clearly perform best out of the three considered gate variants, so in the following we will focus on this gate set.

\begin{figure}[t]
    \centering
    \includegraphics[
        width=\linewidth,
        trim={0.35cm 0.05cm 1.55cm 0.95cm},
        clip
    ]{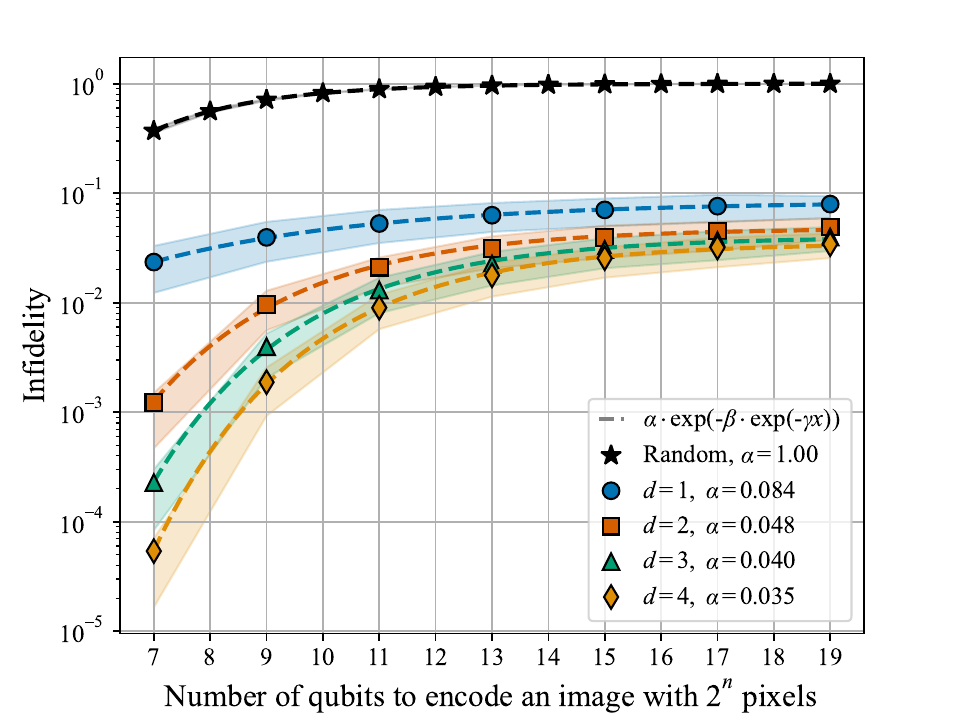}

    \caption{\textbf{Scaling of the infidelity with increasing image resolution.} We consider $100$ randomly selected color images (with an original resolution of at least $256\times256$) from the Imagenette dataset and rescale them using bilinear interpolation to different resolutions ranging from $4\times4$ to $256\times256$ pixels. The plot shows the average infidelity between the exactly encoded states and their quantum circuit approximation (using special orthogonal two-qubit gates) as a function of the number of qubits, which is directly related to the image resolution. Different marker shapes and colors denote circuits with an increasing number of layers $d$, the shaded areas show the $25$th--$75$th percentiles. The black stars show results for approximating real Haar-random states with quantum circuits using special orthogonal two-qubit gates with $d=4$ layers. While for the random states the infidelity quickly saturates at one, indicating that they cannot be effectively approximated by a low-depth circuit, the infidelity of the image states seems to saturate for a fixed number of layers as the image resolution increases, suggesting that the circuit approximation works even for high-resolution images. The dashed lines act as a guide to the eye, they correspond to a fit of a Gompertz function [see Eq.~\eqref{eq:Gompertz}] to the data, where $\alpha$ captures the saturation value of the infidelity. The exact values of all fit parameters are given in Table~\ref{tab:infidelity_vs_resolution} in the appendix.}
    \label{fig:infidelity_vs_resolution}
\end{figure}

Next, we want to examine the scaling of the infidelity with the image resolution for a fixed number of layers. For this, we consider $100$ randomly selected images from the Imagenette dataset which originally have a resolution higher than $256\times256$ pixels, so that we can reasonably tune the image resolution. This is in contrast to the previously used images from the MNIST, Fashion-MNIST and CIFAR-10 datasets, which at $32\times32$ pixels already have a rather low resolution. Fig.~\ref{fig:infidelity_vs_resolution} shows the resulting scaling of the average infidelity when changing the $2^{n/2}\times2^{n/2}$-pixel image resolution through $\frac{n}{2}\in\{2,3,\ldots,8\}$. Different colors and marker shapes show results for different numbers of layers, the shaded regions show the $25$th--$75$th percentiles. The black stars show the results for optimizing circuits with four layers to approximate states with amplitudes sampled from a normal distribution before normalization, i.e., real Haar-random states. While such random states cannot be effectively described by low-depth circuits, the states encoding image data are well-captured with far smaller infidelities. Generally, we see the trend that more circuit layers help capture the state more accurately (as already seen in Fig.~\ref{fig:infidelity_vs_cnot}), while a higher resolution introduces a greater complexity into the image making the approximation more difficult.

As we change the image resolution, we expect two distinct behaviors. For a high image resolution, a further increase of the resolution does not introduce a lot of new features to the image, but only sharpens the already visible details. As such, the state should not be much harder to represent than at a lower resolution. From the MPS representation, where a truncated Fourier series of the image can be represented exactly, we expect that for a high resolution there are only finite-size effects that are exponentially small in the number of qubits~\cite{jobst2023efficientmpsrepresentationsquantum}. For very low image resolutions, changing the resolution can still drastically change the image and introduce a lot of new information which has to be captured by the approximation. Thus, at low image resolutions we expect the generic behavior where the infidelity of two distinct quantum states grows exponentially with the number of qubits. However, the actual value of the infidelity must be very small in this regime, as a circuit with sufficiently many layers can represent the target state exactly for sufficiently few qubits (e.g., for two qubits a single-layer circuit can prepare the state exactly). At some intermediate value of the infidelity, there has to be a crossover between the two regimes. A function that captures such a crossover is the Gompertz function, defined as
\begin{equation}
	f(x) = \alpha \, \exp(-\beta \, \exp(-\gamma \, x)).
    \label{eq:Gompertz}
\end{equation}
For small inputs $\gamma\, x \ll 1$, this becomes an exponential increase: $\alpha \, \exp(-\beta \, (1 -\gamma \, x)) \propto \exp(\beta\gamma\,x)$. For large inputs $\gamma\, x \gg 1$ this converges to the value of the prefactor $\alpha$ up to exponentially small corrections, as $\alpha \, (1 - \beta \, \exp(-\gamma \, x)) = \alpha + \bigO{\exp(-\gamma\,x)}$. The dashed lines in Fig.~\ref{fig:infidelity_vs_resolution} show a fit of the Gompertz function to the data. All values of the fit parameters and their uncertainties are given in Table~\ref{tab:infidelity_vs_resolution} in the appendix. We see that it phenomenologically captures the behavior in all cases. Importantly, it indicates that the average infidelity of the states seems to saturate to some value smaller than one as the image resolution is increased. This suggests that the number of layers in the circuit does not need to grow asymptotically with the system size to keep the infidelity below a certain error threshold, and thus means that the state preparation of MCRQI-encoded images with fixed error only scales linearly with the number of qubits.

Making use of the success that the circuits employing special orthogonal gates have shown in these two settings, we have optimized circuits using them for all images in the MNIST~\cite{Lecun1998, deng2012mnist}, Fashion-MNIST~\cite{FashionMNIST}, CIFAR-10~\cite{CIFAR10}, and Imagenette~\cite{imagenette} datasets. We provide the optimized circuits as PennyLane datasets at \pennylanedataset.
\section{Benchmarking Quantum Classifiers}
\label{sec:classifiers}

After bringing the classical data from the MNIST~\cite{Lecun1998, deng2012mnist}, Fashion-MNIST~\cite{FashionMNIST}, CIFAR-10~\cite{CIFAR10} and Imagenette~\cite{imagenette} datasets into a form in which we can efficiently load them on a quantum computer, we use them to evaluate the performance of different classifiers. While finding a quantum classifier that is competitive with classical methods is beyond the scope of this work, these benchmarks demonstrate the utility of the datasets while showing some shortcomings of current quantum classifiers and highlighting potentially fruitful directions to explore. The tested classifiers include variational quantum circuits (VQCs)---both standard and a nonlinear variant---as well as support vector machines (SVMs) with kernels induced by the data encoding. When we use the patched or multi-copy encoding, the number of qubits grows, and therefore the simulation cost increases, so we switch from circuit-based to tensor-network-based classifiers. We also compare against a classical convolutional neural network (CNN) representing a state-of-the-art classical method. To ensure a robust evaluation, we split the data into training and validation sets using five-fold cross-validation. After evaluating the models on the uncompressed datasets, we assess the performance of the classifiers on the compressed datasets.

For the training of the models in the following, we use Ray~\cite{moritz_ray2018} and GNU Parallel~\cite{tange_2024_13826092} to parallelize multiple model instances. For state vector simulations, we use PennyLane~\cite{bergholm2022pennylaneautomaticdifferentiationhybrid} in combination with the just-in-time compilation and vectorization capabilities of JAX~\cite{jax}. The tensor-network simulations are likewise implemented directly in JAX, leveraging the same performance-optimizing features.

\subsection{Quantum Classifiers}

In the following, we introduce the quantum classifiers considered in this work. First, we discuss the \emph{explicit} approach to learning classification tasks on a quantum computer, where a VQC is trained as a quantum circuit classifier to construct a decision boundary for the classification problem.
For this setup, we also propose a way to introduce \emph{nonlinearities} into each layer of the VQC processing the input data, similar to how classical neural networks process data nonlinearly at each step.
Then, we consider kernel-based classifiers, which can be seen as an \emph{implicit} implementation of the linear quantum circuit classifier.

\textbf{Quantum Circuit Classifiers.}
For a supervised classification problem like we consider here, the goal is to learn a decision function $f_{\ell}(\vec{x};\vec{\theta})$ parametrized by some variational parameters $\vec{\theta}$ that, given an input $\vec{x}$, outputs a number for each possible class label $\ell$. The largest number of these outputs should correspond to the correct class. The model is trained by tuning the parameters $\vec{\theta}$ such that they minimize some loss function, which quantifies the error the model is making on some training set, usually via gradient descent methods. The typical approach in variational quantum computing is to express this learnable decision function $f_{\ell}(\vec{x};\vec{\theta})$ as the expectation value of a quantum circuit: starting from a state that encodes a data point $\ket{\psi(\vec{x})}$, we apply some parametrized quantum circuit $V(\vec{\theta})$ and, finally, measure an observable $O_{\ell}$ for each class~\cite{PhysRevA.98.032309, farhi2018classificationquantumneuralnetworks, Schuld_2020}. This way, the decision function is
\begin{equation}
\begin{aligned}
    f_{\ell}(\vec{x})
    &= \bra{\psi(\vec{x})}
    V(\vec{\theta})^{\dagger} O_{\ell} V(\vec{\theta})
    \ket{\psi(\vec{x})}\\
    &= \Tr[O_{\ell}(\vec{\theta}) \, \rho(\vec{x})],
\end{aligned}
\end{equation}
where we have collected the input state into a density matrix $\rho(\vec{x}) \equiv \ketbra{\psi(\vec{x})}{\psi(\vec{x})}$ and reinterpreted the variational circuit as a basis transformation of the measured observable $O_{\ell}(\vec{\theta}) \equiv V(\vec{\theta})^{\dagger} O_{\ell} V(\vec{\theta})$. From this rewriting, we can directly see that the quantum model will be a linear function of the input density matrix. Each observable $O_{\ell}(\vec{\theta})$ draws a hyperplane in the space of Hermitian matrices separating data points corresponding to label $\ell$ from data points corresponding to other classes~\cite{Schuld_2019b, Havl_ek_2019, schuld2021supervisedquantummachinelearning}. So, for the classifier to capture sufficiently complex relations it is expected that a more nonlinear mapping from the input data $\vec{x}$ to the quantum state $\rho(\vec{x})$, like the patched or multi-copy encodings discussed in Sec.~\ref{sec:encoding}, is needed.

\begin{figure}[t]
    \centering
    \includegraphics[width=\linewidth]{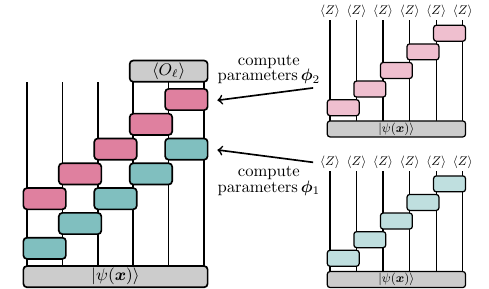}
    
    \caption{\textbf{Depiction of the nonlinear VQC.} The left circuit depicts the quantum model that is used to infer a prediction for the class label of a given input state $\ket{\psi(\vec{x})}$. If the gates in the circuit are simply parametrized by variational parameters, we have a standard VQC that is a linear classifier with respect to the density matrix of the input state. For the nonlinear VQC, we instead compute the parameters of the circuit, $\vec{\phi}_1$ for the first layer of the circuit and $\vec{\phi}_2$ for the second layer, from the expectation values measured with the two parameter-retrieval circuits shown on the right. As the parameters $\vec{\phi}_1$ and $\vec{\phi}_2$ now implicitly depend on the input state $\ket{\psi(\vec{x})}$, the classifier becomes a nonlinear function of the input state (or input density matrix) at the cost of evaluating multiple circuits.}
    \label{fig:nonlinear}
\end{figure}

In our case, the input states are given by the states discussed in the preceding sections. The circuit architecture of the VQC is a sequential circuit (cf. Fig.~\ref{fig:nonlinear}, on the left without the additional circuits on the right) with general two-qubit gates in $SU(4)$~\cite{Dilip2022, shen2024classification}. We parametrize the gate by explicitly expressing the generator as a linear combination of the fifteen non-identity Pauli strings. When the system has $L$ qubits, there are thus $15(L-1)$ trainable parameters per layer. Note that, if the input states are also given as sequential circuits, this circuit architecture avoids barren plateaus for a number of layers up to logarithmic in the number of qubits~\cite{Liu2022, Zhang2024, Barthel2025}. For the measured observables, we take a weighted sum of the probabilities of all sixteen bitstrings in the computational basis on the last four qubits. Specifically, the observable is given by
\begin{equation}
    O_{\ell} = \sum_{j=0}^{15} A_{\ell j} \ketbra{j}{j} + b_{\ell}\,\Id,
    \label{eq:observable}
\end{equation}
with the $10\times16$-matrix of weights $A$ and the $10$-dimensional bias vector $\vec{b}$, that are also optimized during training~\cite{shen2024classification}. 

For training, we feed the output of the model into a softmax function with a temperature of $T=1/128$ before using the categorical cross-entropy loss. (This combination is sometimes also referred to as the softmax loss function.) We train for $100$ epochs using the Adam optimizer~\cite{kingma2015adam} with a learning rate of $8\cdot10^{-4}$ and a batch size of $100$. Both the variational parameters of the quantum circuit and the parameters of the observable are initialized uniformly at random in the range $[0, 1]$.

\textbf{Nonlinear Quantum Circuit Classifiers.}
To address the limitations of linear models, the framework of data reuploading has been proposed~\cite{P_rez_Salinas_2020, Goto_2021, Schuld_2021}. Repeating data-dependent gates several times in the circuit can make the resulting state depend very nonlinearly on the input data. The data-dependent operations can be repeated in space, for example like the multi-copy encoding discussed in Sec.~\ref{sec:encoding} (which we will consider in the following subsection), or repeated in time, when the same gate appears several times at different depths of the circuit. The latter strategy allows to show that even a single qubit can approximate any continuous function~\cite{P_rez_Salinas_2020}, rendering it a universal classifier in principle, analogous to the universal approximation theorem for neural networks~\cite{hornik_1991}. However, this specific strategy is not applicable to the encodings considered in this work, since our circuits do not implement a general basis transformation. Instead, they only realize an isometry that maps the initial state~$\ket{\vec{0}}$ to the quantum state approximately representing the image.

Here, we explore an alternative route to add nonlinearity to the classifier, and propose a quantum circuit model that nonlinearly processes the input density matrix by making the circuit's gates explicitly dependent on the input state. A schematic of the model is depicted in Fig.~\ref{fig:nonlinear}. The nonlinear quantum circuit classifier comprises two main components: a parametrized quantum circuit that acts as the classifier and parameter-retrieval circuits that are used to compute the parameters of the gates in the classifier circuit.

For the classifier quantum circuit, we use the same setup as for the usual, linear VQC classifier. That is, we use a sequential circuit (cf. Fig.~\ref{fig:nonlinear}, left) with several layers, and each gate a general two-qubit gate in $SU(4)$. For a system with $L$ qubits, this means $15(L-1)$ parameters per layer. The observable is also the same trainable weighted sum of probabilities of the computational-basis bitstrings of the final four qubits as for the linear model---see Eq.~\eqref{eq:observable}.

The distinctive feature of this model is that the gates in the classifier quantum circuit do not have fixed parameter values, but rather the parameters are calculated for each new input state $\ket{\psi(\vec{x})}$ via the parameter-retrieval circuits. For each layer $m$ of the main classifier circuit, there is a separate parameter-retrieval circuit $V_m(\vec{\theta}_m)$ (with variational parameters $\vec{\theta}_m$). To compute the parameters for the $m$th layer of the main classifier circuit, we first compute the vector of single-qubit Pauli-$Z$ expectation values $\vec{\sigma}_m$ with respect to the state obtained by applying the parameter-retrieval circuit to the input state, i.e.,
\begin{equation}
    \sigma_{m,j} = \bra{\psi(\vec{x})}V_m(\vec{\theta}_m)^{\dagger} \, Z_j \, V_m(\vec{\theta}_m) \ket{\psi(\vec{x})}.
\end{equation}
For a system with $L$ qubits, this gives only $L$ numbers, while we need $15(L-1)$ parameters for the classifier circuit. To match the dimensions, we multiply the vector of expectation values by a $15(L-1) \times L$-dimensional weight matrix $W_m$. The final parameters for the classifier circuit are then obtained as
\begin{equation}
    \vec{\phi}_m = \pi \tanh(W_m \vec{\sigma}_{m}),
\end{equation}
where the factor of $\pi$ and the hyperbolic tangent are applied element-wise to rescale the gate parameters between $-\pi$ and $\pi$. We keep the weight matrix $W_m$ as additional variational parameters of the model. The parameters $\vec{\phi}_m$, which implicitly depend on the input state $\ket{\psi(\vec{x})}$, are then used as the gate parameters of the $m$th layer of the main classifier circuit. Importantly, the dependence of the parameters $\vec{\phi}_m$ on the input state makes the resulting classifier circuit a nonlinear function of the input density matrix---this advantage, however, comes at the cost of having to evaluate an increased number of circuits on the quantum computer.

For the optimization of the model, we use the same setup and hyperparameters as for the training of the linear quantum circuit classifier.

\begin{figure*}[t]
    \centering
    \includegraphics[
        width=\linewidth,
        trim={0.55cm 0.6cm 0.55cm 0.55cm},
        clip
    ]{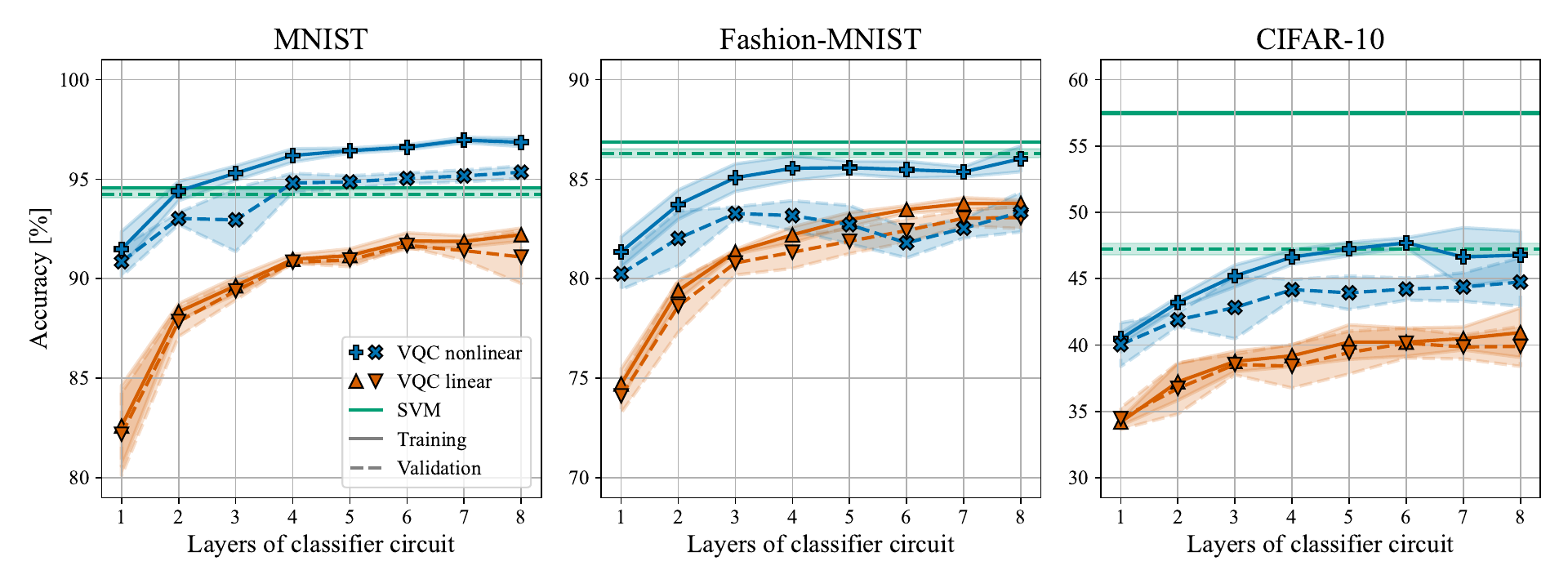}
    
    \caption{\textbf{Accuracy against the number of layers in the circuit for the linear and nonlinear variants of the VQC.} The circuit ansatz is shown in Fig.~\ref{fig:nonlinear} (the linear variant uses the same ansatz without the parameter-retrieval circuits), the gates in the circuits are general $SU(4)$ gates. The three subplots show, in order from left to right, results of the VQCs and an SVM classifier on the MNIST, Fashion-MNIST and CIFAR-10 datasets. Training accuracies are shown as solid lines and validation accuracies as dashed lines. Results on the Imagenette dataset are omitted as training the circuits on the full-scale dataset was not feasible. Since the datasets contain ten classes, the baseline accuracy for random guessing is $10\%$. The error bands represent the standard deviation from five-fold cross-validation. The nonlinear VQC outperforms the linear VQC on both the training and validation sets, and performs similarly as the SVM classifier.}
    \label{fig:acc(non)linear}
\end{figure*}

\textbf{Quantum Kernel Methods.}
Instead of constructing a quantum circuit that explicitly computes the label of an encoded data point, we can also implicitly use a quantum computer to infer the label, by first using a quantum computer to find inner products of a new data point with a reference set of labeled data points and then classically obtaining the label from weighting the inner products~\cite{Schuld_2019b, Havl_ek_2019, Jerbi2023}. This is also the idea of kernel methods~\cite{Schoelkopf2001}, the inner product leading to the specific kernel function $\kappa(\vec{x}, \vec{y}) = \Tr[\rho(\vec{x}) \rho(\vec{y})]$~\cite{Schuld_2019b, Havl_ek_2019, schuld2021supervisedquantummachinelearning}. The classifier is then built as a linear combination of kernel functions, where one of the inputs is fixed to elements $\vec{x}_n$ of the training set, as
\begin{equation}
    f_{\ell}(\vec{x};\vec{\alpha}) = \sum_n \alpha_{\ell n} \, \kappa(\vec{x}_n, \vec{x}).
\end{equation}
If the training set is unrestricted and comprises all possible input data, this recovers the same set of linear classifiers as the linear quantum circuits discussed before~\cite{schuld2021supervisedquantummachinelearning, Jerbi2023}. However, even for a restricted training set, the representer theorem guarantees that a model of the form above achieves minimal loss when evaluated on the same training set~\cite{Schoelkopf2001}, and there are known methods for finding the optimal parameters of such a model. Still, while the training loss may be optimal, no such statement can be made about the loss on unseen samples like the validation set, and kernel methods tend to overfit quite strongly in practice~\cite{Jerbi2023}.

With the data encodings we use, the resulting quantum states remain lowly entangled and can be represented as MPSs. Therefore, the inner products can be computed efficiently classically and this computation does not require a quantum computer. Nonetheless, the kernel-based classifier is closely related to the explicit quantum circuit classifiers, and serves as an interesting model for comparison. As a specific kernel-based model, we consider a support vector machine (SVM). We use the \texttt{svm.SVC} implementation of scikit-learn~\cite{scikit-learn}, which implements a polynomial kernel
\begin{equation}
    \kappa(\vec{x}, \vec{y}) = (\gamma \cdot \braket{\psi(\vec{x})}{\psi(\vec{y})}  + C)^d
\end{equation}
where we set the kernel prefactor $\gamma=1$, the regularization parameter $C=0$ and the degree $d=2$ to reproduce the quantum kernel. (Other parameters of the scikit-learn implementation are kept at the default values.) Multi-class classification is handled via a one-vs-rest strategy. 

\textbf{Performance of the Quantum Classifiers.}
Fig.~\ref{fig:acc(non)linear} compares the classification accuracies of the linear VQC (orange), the nonlinear VQC (blue) and the SVM with quantum kernel (turquoise) on the three benchmark datasets MNIST, Fashion-MNIST and CIFAR-10. Training the circuit classifiers on the complete Imagenette dataset is not feasible because of the larger number of qubits needed. The solid lines indicate the accuracy on the training set, while dashed lines show the accuracy on the validation set, and shaded areas depict one standard deviation from five-fold cross-validation.

In addition to training the quantum models, we also trained a linear neural network (i.e., a matrix transformation) to identify how easily the datasets can be classified linearly. We use the same resolution of the images as for the quantum datasets and standardize them to zero mean and unit variance. As for the quantum classifiers, to compute the mean and the standard deviation we employ five-fold cross-validation. We train the model for $100$ epochs using the Adam optimizer with default parameters, and, in the following discussion, report the training and validation accuracy of the epoch with the highest validation accuracy.

On the MNIST dataset (left panel), all classifiers achieve a high accuracy above $90\%$ for sufficiently many layers. The nonlinear VQC reaches the best accuracy, which saturates around $95\%$ validation accuracy for moderately deep circuits ($4$--$8$ layers). With this, it outperforms both linear classifiers, the linear VQC, which saturates a few percentage points lower at $92\%$, and also narrowly the SVM, which achieves an accuracy of $94.2\pm0.2\%$. The linear neural network achieves an accuracy of $93.5\pm0.2\%$ on the training and $92.3\pm0.2\%$ on the validation set, which is similar to accuracy of the linear quantum models.

The increased complexity of the Fashion-MNIST dataset (central panel) leads to all models achieving lower accuracies than on the MNIST dataset, but overall the performance is still promising. When using only a few layers, the nonlinear VQC still outperforms its linear variant, however, the validation accuracies of both classifiers saturate at a similar value, and only the training accuracy of the nonlinear VQC remains about $3\%$ higher at larger depths. Both classifiers are beaten by the SVM, which achieves the highest accuracy of $86.3\pm0.3\%$. With $86.7\pm0.5\%$ on the training and $85.6\pm0.4\%$ on the validation set, the linear neural network closely matches the performance of the linear quantum variants.

The performance of all classifiers decreases significantly on the more realistic CIFAR-10 dataset (right panel), with validation accuracies below $48\%$ for all classifiers. While this is still notably better than the baseline of $10\%$ for random guessing, it is also much worse than the performance on the previous two datasets. We again observe the trend that the nonlinear VQC outperforms the linear VQC, saturating at almost $45\%$ validation accuracy compared to around $40\%$ for the linear model. The nonlinear VQC, however, does not beat the SVM classifier, which achieves a slightly higher validation accuracy of $47.2 \pm 0.5\%$; the SVM classifier, though, also overfits strongly with a testing accuracy of $57.5\pm0.2\%$. The linear neural network baseline attains $43.76\pm0.5\%$ on the training and $40.4\pm0.5\%$ on the validation set, comparable to the linear quantum models.

Overall, we observe that the nonlinear VQC tends to outperform the linear VQC on all three datasets, suggesting that the added expressivity of allowing the circuit parameters to depend on the input state leads to a higher classification accuracy. Note that while the nonlinear VQC also has significantly more variational parameters for the same number of layers, the difference in performance does indeed seem to stem from the added nonlinearity. The accuracy of the linear VQC saturates for a large number of layers, indicating that adding more layers until the number of parameters in the two models matches will not improve performance. This saturation value, however, is distinctly smaller than the accuracy of the nonlinear VQC.
While the improved accuracy of the nonlinear VQC over the linear VQC makes it a promising model, the added nonlinearity still does not make it reliably beat the SVM classifier (even though their performance is very close). Smarter parameter initialization schemes or more structured methods for computing the circuit parameters from the expectation values of the parameter-retrieval circuits could potentially push the model beyond the accuracy of the SVM classifier.

From these results, we also see that it is important to test the quantum classifiers on a range of different datasets. Had we considered only the MNIST dataset, we might have concluded that the nonlinear VQC already outperforms the SVM classifier, while the other two datasets show that this is not reliably the case. On the other hand, had we only tested on the Fashion-MNIST dataset we might have assumed that the nonlinear VQC shows no improvement over the linear VQC, since both converge to the same validation accuracy on this dataset with an increased number of layers.
More generally, we see that the classifiers perform well on the simple datasets, especially the MNIST and to some extent also on the Fashion-MNIST dataset, but this does not translate to realistic datasets like CIFAR-10, where the performance drops off significantly.

\subsection{Tensor-Network Classifiers}

The FRQI- and MCRQI-based encodings of the data offer a resource-efficient mapping from the data into the Hilbert space, using only slightly more qubits than are needed for the Hilbert space dimension to match the data dimension. However, this mapping might be too efficient, in the sense that it offers too little freedom to subsequently allow linear quantum classifiers to draw a hyperplane that is able to separate the different classes. By increasing the number of qubits used in the encoding, the data is embedded in a higher-dimensional space and it becomes easier to find a hyperplane that adequately separates the data. In the following, we consider the effect of the patched and multi-copy encoding introduced in Sec.~\ref{sec:encoding}, which are precisely such mappings that lead to more qubits and a larger Hilbert space dimension. As the number of qubits increases, so does the simulation cost, and we transition from direct simulation of the circuits to using tensor-network-based classifiers for these problems.

\begin{figure}[t]
    \centering
    \begin{subfigure}[t]{\linewidth}
        \centering
        \includegraphics[width=\linewidth]{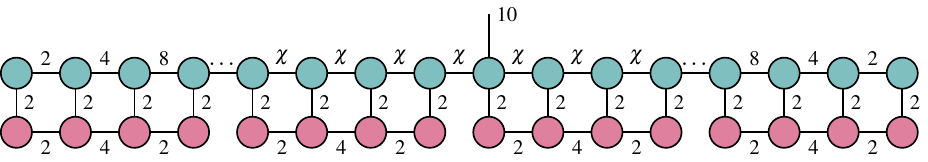}
        \caption{MPS classifier}
        \label{fig:mpsarchitecture-a}
    \end{subfigure}%
    \vspace{2mm}
    \begin{subfigure}[t]{\linewidth}
        \centering
        \includegraphics[width=\linewidth]{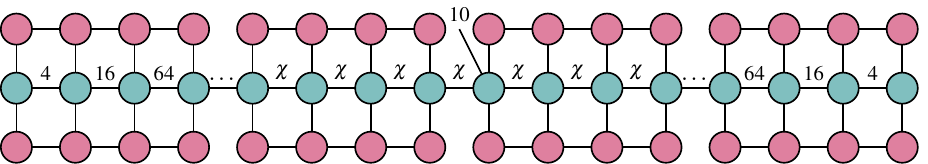}
        \caption{MPO classifier}
        \label{fig:mpsarchitecture-b}
    \end{subfigure}
    
    \caption{\textbf{Architecture of the tensor-network classifiers.} The example shows a case where the input image is either split into four patches or loaded with four copies. The bond dimension of the classifier is constrained by a maximum bond dimension $\chi$, for the results presented in the following we choose $\chi=32$. The additional protruding leg in the center of the tensor network has dimension ten, and corresponds to the output for the ten classes of the considered datasets.}
    \label{fig:mpsarchitecture}
\end{figure}

\textbf{Matrix-Product State Classifier.}
The \emph{MPS classifier}~\cite{Stoudenmire2017, Novikov2018, Huggins2019, Efthymiou2019, Dilip2022}, illustrated in Fig.~\ref{fig:mpsarchitecture-a}, provides a quantum-inspired classical model for supervised learning tasks. The input state for the classifier is given by the FRQI- and MCRQI-encoded states discussed in Sec.~\ref{sec:encoding}; it can be either a single MPS representing the entire state, or a product state of several MPSs when there are multiple patches or copies of the input image. The classifier is also an MPS, with an additional physical leg on one of the central tensors, whose dimension corresponds to the number of distinct labels in the dataset. Contracting an encoded image with the MPS classifier thus yields a vector with the same dimension as the number of labels, which is the decision function of the model---similarly to the quantum classifiers we already discussed. As before, the predicted class is the label associated with the largest entry in the output vector.

% Both random init and pretraining
When initializing the entries of the MPS classifier, we consider a random initialization and a pretraining step for the initialization.
For the random initialization, the elements of the MPS tensors are sampled from a Gaussian distribution with zero mean and standard deviation $\sigma = 1/\sqrt{\chi}$, where the relevant bond dimension $\chi$ is the one pointing away from the classification leg and we set a maximal bond dimension $\chi=32$. The central tensor including the classification leg is initialized with standard deviation $\sigma = 1/\sqrt{\chi_l\chi_r}$, where $\chi_l$ and $\chi_r$ are the bond dimensions of the left and right bonds, respectively. Like this, we ensure that the overlap between the classifier and the input states has unit variance~\cite{Barratt2022}. We give a proof for this normalization in App.~\ref{app:init}.
Apart from the random initialization, we also consider a pretraining algorithm as a warm start for training~\cite{Lin:2023put}, where a subset $\num{1000}$ images in the training set are summed up as MPSs~\cite{SCHOLLWOCK201196} and then compressed to the target bond dimension. To keep the intermediate bond dimension tractable without too strong truncation effects, we sum the states in the training set in batches, and compress them first with a larger bond dimension of $\num{100}$, before summing up the different batches and compressing to the final bond dimension of $\chi=32$ for the resulting classifier. To avoid a fluctuating norm of the MPS classifier that depends on the training samples used for pretraining, we normalize it to have norm $\sqrt{2^L\,10}$, the same norm as for the random initialization.

For training, we feed the output of the MPS classifier into a softmax function before using the categorical cross-entropy loss. We train for $100$ epochs using the Adam optimizer~\cite{kingma2015adam} with a learning rate of $10^{-4}$ and a batch size of $100$.

\textbf{Matrix-Product Operator Classifier.}
The second tensor-network classifier we consider, very similar to the MPS classifier, is the \emph{matrix-product operator (MPO) classifier}---see Fig.~\ref{fig:mpsarchitecture-b}. It extends the MPS classifier by introducing a second physical leg per MPS tensor and another copy of the input state for the contraction with a data point. This makes the classifier more similar to the quantum classifiers we discussed before: note that the MPS classifier was linear in the input quantum state, while the quantum classifiers were linear in the input density matrix. The MPO classifier is now also properly linear in the input density matrix, and thus corresponds to a quantum classifier. (We show that this classifier indeed corresponds to the expectation value of a Hermitian operator, even without restricting the MPO to be Hermitian, in App.~\ref{app:MPO_is_quantum}.) Apart from the larger tensor dimensions, which means more variational parameters for the same bond dimension, this is a possible advantage of the MPO classifier over the MPS classifier because the former can construct linear combinations of quadratic functions of the input state compared to only linear functions of the input state for the latter.

For the MPO classifier, we consider the same initialization techniques as for the MPS classifier, i.e., random initialization and pretraining. For the random initialization, the entries of the MPO tensors are drawn from the same normal distribution with zero mean and standard deviation $\sigma=1/\sqrt{\chi}$ as the MPS tensors. For the initialization via pretraining, the only change is that we now sum the outer products of the input states instead of the states directly, so the output is an MPO instead of an MPS, and we afterwards normalize the MPO to have Frobenius norm $\sqrt{2^{2L}\,10}$. Also for the MPO classifier we set the maximal bond dimension to $\chi=32$.

For the training of the MPO classifier, we use the same setup and hyperparameters as for the MPS classifier.

\begin{figure*}[p]
    \centering
    \includegraphics[
        width=\linewidth,
        trim={0.55cm 0.7cm 0.55cm 0.55cm},
        clip
    ]{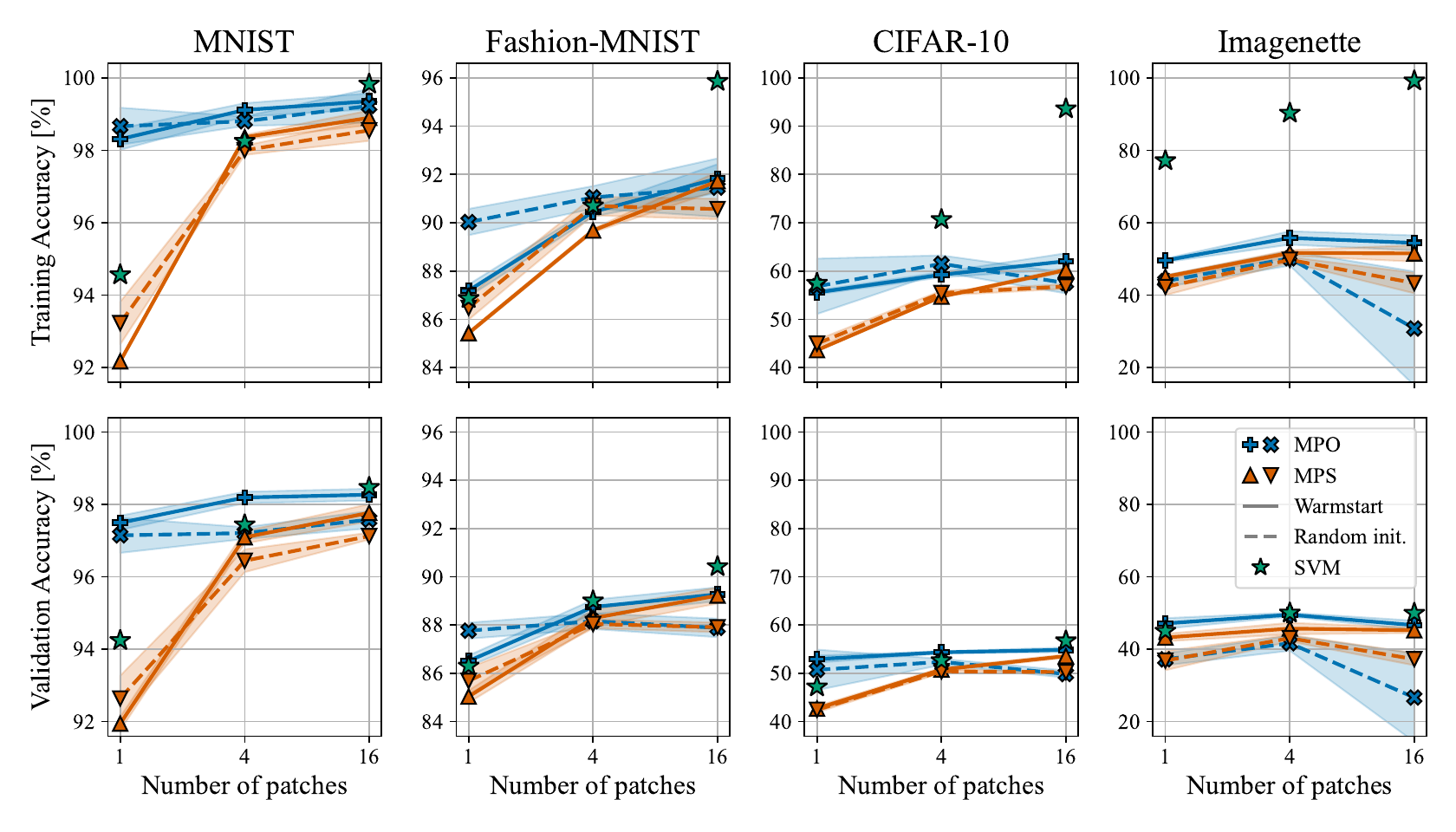}
    
    \caption{\textbf{Accuracy of the tensor-network and SVM classifiers versus the number of patches used in the encoding.} Both the training (top row) and validation accuracy (bottom row) are shown, left to right, on the four datasets MNIST, Fashion-MNIST, CIFAR-10 and Imagenette. Results for the MPS and MPO classifiers are shown in orange and blue, for both the warm start initialization as solid lines and random initialization as dashed lines. The results for the SVM are shown as green stars. Since the datasets contain ten classes, the baseline accuracy for random guessing is $10\%$. The error bands represent the standard deviation from five-fold cross-validation.}
    \label{fig:tn_patching}
    \vspace{4mm}%
    \includegraphics[
        width=\linewidth,
        trim={0.55cm 0.6cm 0.55cm 0.55cm},
        clip
    ]{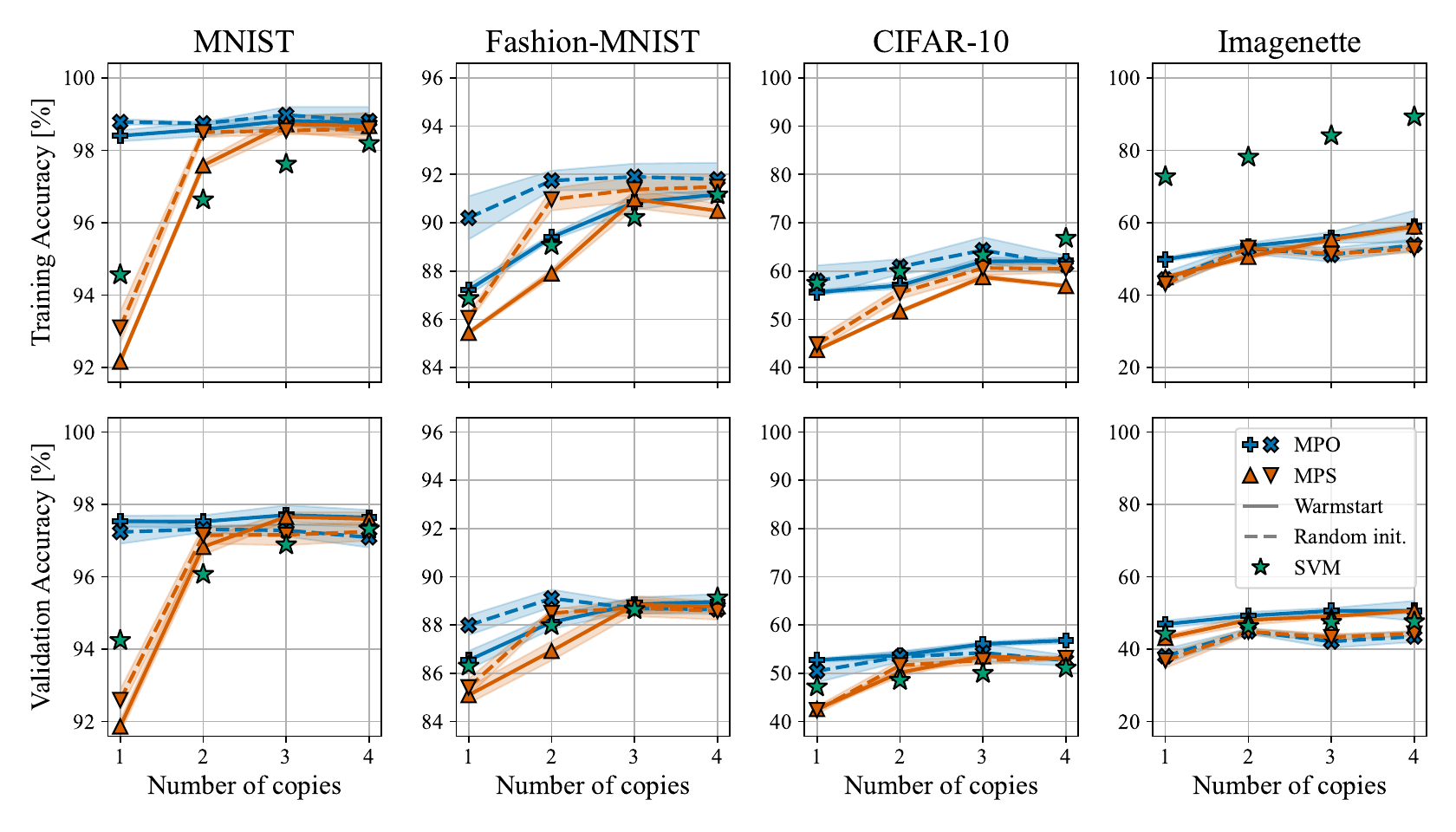}
    
    \caption{\textbf{Accuracy of the tensor-network and SVM classifiers versus the number of copies used in the encoding.} Both the training (top row) and validation accuracy (bottom row) are shown on the four datasets MNIST, Fashion-MNIST, CIFAR-10 and Imagenette. Results are shown for the MPS (orange) and MPO (blue) classifiers, for both the warm start initialization (solid lines) and random initialization (dashed lines). The results for the SVM are shown as green stars. The baseline accuracy for random guessing is $10\%$. The error bands represent the standard deviation from five-fold cross-validation.}
    \label{fig:tn_multicopy}
\end{figure*}

\textbf{Performance of the Tensor-Network Classifiers.}
In Fig.~\ref{fig:tn_patching} we compare the performance of the MPS and MPO classifiers and an SVM classifier on the four datasets, when the input image is encoded using multiple patches (see Sec.~\ref{sec:encoding}). Both the training (top row) and validation (bottom row) accuracies are shown. We show results for the tensor-network classifiers initialized with two different strategies before training: random initialization (dashed lines, crosses and down-facing triangles) and a warm start through pretraining (solid lines, pluses and up-facing triangles). The results for the MPO classifier are shown in blue, for the MPS classifier in orange, and for the SVM classifier as green stars.

On the MNIST dataset, all classifiers achieve a high accuracy: for a single patch it is already above $90\%$, and grows above $96\%$ when using multiple patches. For all classifiers both training and validation accuracies increase when splitting the image into increasingly more patches. The MPO classifier outperforms the other models when encoding the full image as a single patch, however, for more patches they perform similarly well. For more patches, we can also see that the warm-start initialized tensor-network classifiers tend to perform slightly better than the randomly initialized ones. The overall similarity between the achieved training and validation accuracies suggests that none of the classifiers are overfitting strongly.

The slightly more complex Fashion-MNIST dataset leads to a decreased accuracy of all classifiers, but their overall performance is still good with a validation accuracy of around $89\%$ on $16$ patches. Compared with the results on the MNIST dataset, the difference in performance between the MPS and MPO classifiers is less pronounced. Still, we observe the general trends that more patches tend to improve performance, and that for many patches the warm-start initialization outperforms random initialization. On a single patch the MPO classifier performs best, but on more patches the SVM achieves the highest validation accuracy, reaching $90.4\pm0.2\%$ on sixteen patches---considering its training accuracy, however, it also begins to overfit significantly.

Moving to the more realistic CIFAR-10 dataset, the first of the datasets to consist of color images, we see the accuracy of all classifiers drop considerably compared to the two previous datasets. Nonetheless, we observe some similar trends as before: The MPO classifier outperforms the other models for a small number of patches, but this difference becomes less significant when using more patches. For sixteen patches the SVM classifier eventually beats the MPO classifier with an accuracy of $56.7\pm0.1\%$; however, it also drastically overfits when compared with its training accuracy of $95.6\pm0.1\%$. Unlike before, increasing the number of patches does not improve performance for all models, for the randomly initialized tensor-network classifiers the validation accuracy stagnates or even drops when changing from four patches to sixteen.

Finally, we consider the Imagenette dataset, which includes color images with a higher resolution than the other datasets. On this dataset, the performance of all classifiers is quite poor. Previously, the MPO classifiers using either initialization technique outperformed the MPS classifiers (at least for a small number of patches), but now both warm-started tensor-network classifiers outperform the randomly initialized ones. Also, while the accuracy improves when going from one to four patches, it does not further improve when using sixteen patches: it stagnates for the SVM and warm-started MPS classifiers, and even decreases for the randomly initialized MPS and both MPO classifiers. The best accuracy is achieved by both the SVM classifier with $50.0\pm0.3\%$ and the warm-started MPO classifier for four patches with $49.5\pm0.6\%$; however, the SVM again overfits drastically more than the MPO classifier.

Fig.~\ref{fig:tn_multicopy} shows analogous results when a product state of multiple copies of the image, rather than multiple patches, is provided as an input to the tensor network (cf. Sec.~\ref{sec:encoding}). The top row shows the training accuracy and the bottom row the validation accuracy, as the number of copies of the image is increased from one to four. (Note that the results for a single copy are the same as for a single patch in Fig.~\ref{fig:tn_patching}.) The results for the MPO classifier are again shown in blue, the results for the MPS classifier again in orange. Warm-started and randomly initialized tensor-network classifiers are distinguished by solid and dashed lines, respectively, and different marker symbols. The results for the SVM classifier are shown as green stars.

As in the case of the patched encoding, all classifiers achieve a high accuracy on the MNIST dataset, although the best-performing models with just under $98\%$ accuracy remain below the accuracy of the best-performing models with the patched encoding that achieved just over $98\%$. We observe that introducing more copies increases the performance for all models. The MPO classifiers outperform the MPS classifiers with the same initialization, and we see that for more copies the warm-started tensor-network classifiers outperform their randomly initialized counterparts. In contrast to what we saw for the patched encoding, here the SVM classifier performs worse than the tensor-network classifiers, even though for more copies this difference diminishes.

On the Fashion-MNIST dataset, the performance of all classifiers slightly drops. The resulting accuracies are comparable to those obtained using the patched encoding, but the best-performing models remain below the accuracy of the best-performing models on the patched dataset. We still see the trend that the accuracy of the classifiers increases as the number of copies is increased, however for three or more copies this saturates and all models perform similarly well.

Compared to the first two datasets, the accuracy on the CIFAR-10 dataset is significantly lower, but remains comparable to the accuracy on the patch-encoded CIFAR-10 dataset. Similar to before, we see the trend that an increasing number of copies tends to improve the performance of the classifier. The MPO classifier outperforms the MPS classifier; for the MPO classifier the warm-starting works better than random initialization, while for the MPS classifier both work similarly well. In contrast to the patched encoding, the SVM is surprisingly the worst performing classifier out of the three models.

On the higher-resolution Imagenette dataset, the accuracy drops even further and remains below $50\%$, as it similarly did for the patched encoding. Increasing the number of copies of the input image only moderately increases the performance and quickly saturates. For this dataset wee see that the warm-started tensor-network classifiers outperform the randomly initialized ones, and the SVM classifier lies somewhere in between. This is also the only dataset for the multi-copy encoding where the SVM classifier very strongly overfits, reaching testing accuracies twice as large as the validation accuracy---for the patched encoding strong overfitting of the SVM classifier was also visible on the Fashion-MNIST and CIFAR-10 datasets.

Overall, we see that it is important to test these classifiers on a range of datasets, including on realistic datasets. Results can change when moving away from toy datasets, exemplified here by the drastic decrease of performance when moving from the rather simple MNIST and Fashion-MNIST datasets to the more realistic CIFAR-10 and Imagenette datasets.

Some general observations seem to hold on all datasets: Adding nonlinearities to the classifiers, either by increasing the number of patches or the number of copies, tends to increase performance---but there are diminishing returns for more patches or copies. The benefit of the nonlinearity is also exemplified by both the SVM and MPO classifiers outperforming the MPS classifiers on the patch-encoded datasets, as the latter can only construct linear functions of the input state amplitudes while the former can construct nonlinear quadratic functions. This performance difference is most pronounced for a small number of patches, as with an increasing number of patches also the MPS classifier becomes increasingly nonlinear with respect to the original pixel values. On the multi-copy encoded datasets the SVM only outperforms the MPS classifier on a single patch, where the same argument holds. For more than one copy, also the MPS classifier has access to quadratic functions of the input state, and higher order polynomials seem to be not much more helpful as the accuracy of all classifiers does not increase much beyond two copies of the input state.
Viewing the classifier as a polynomial of the input state with its degree set by the number of copies is equivalent to the more common perspective of the output of a classifier as a Fourier series with respect to the input data,%
\footnote{%
    The output of the classifier is a polynomial of the state amplitudes, which consist of products of $\cos(\pi x_j/2)^{2k}$ or $\sin(\pi x_j/2)^{2k}$, where $k$ can be up to the number of copies of the input state. Equivalently, when viewed as a function of the data components $x_j$, these terms are part of a Fourier series including terms up to $e^{\pm i 2k \, \frac{\pi}{2} x_j}$ with frequency $2k$.
}
where the range of accessible Fourier frequencies grows with the number of copies~\cite{Schuld_2021}. So, naively, one would expect the expressivity to directly grow with the number of copies of the state. However, this is not necessarily the case as, depending on the structure of the circuit ansatz, the coefficients of the resulting Fourier series can be quite constrained~\cite{Nemkov2023, Okumura2023, Barthe2024}.

On the larger models that require many qubits---either because there are many patches or copies of the image, or because the images are in color and have a higher resolution---we observe that the warm-start initialized tensor-network classifiers tend to outperform the randomly initialized ones. This might be because for models with many qubits the output of the classifier is extremely sensitive to the norm of the tensors forming the tensor network: the overlap with an input state either vanishes or explodes exponentially if they are too small or too large. While our specific choice of random initialization fixes the standard deviations so that the output is not exponentially small or large, after a few steps of the optimization algorithm we can no longer make such a statement. The pretraining initializes the classifier in a more controlled way, which seems to place it in a more well-behaved region of the cost landscape.

A final observation is that, while all classifiers perform better on the training than on the validation set, only the SVM overfits very strongly. This is less pronounced on the multi-copy encoded datasets than on the patch-encoded datasets, but visible for both approaches and most significant as the number of qubits becomes large. As discussed in the section on quantum kernel methods, this is somewhat expected from the way this classifier is constructed. In particular, as the number of qubits grows, distinct states become increasingly more orthogonal and so the quantum kernel function approaches a Kronecker delta, which would lead to the SVM being able to classify the training set perfectly but being unable to predict anything in the validation set. The explicit approaches to the classifier do not necessarily have this problem.

\subsection{Performance on the Compressed Datasets}

% Motivation 
For the performance evaluation on the compressed datasets, we use the model configurations of the VQCs, the SVM, and tensor-network classifiers that achieved high accuracies on average in earlier experiments. We focus on the multi-copy encoding over the patched encoding since approximating only small patches should be easier than approximating the full images, and we are precisely interested in the error arising from the imperfect approximation. Specifically, we use an SVM with three copies of the input state, an MPO classifier with three copies and warm-start initialization, and a nonlinear VQC with four layers trained on a single copy. All models are trained using the same hyperparameters as in the previous experiments.
In addition, we also train a classical CNN on the compressed datasets to see whether any loss of accuracy is due to information loss during the compression, or because the quantum classifiers are sensitive to the compression-induced distortion of the input state.

% CNN
\textbf{Convolutional Neural Networks.}
For a classical baseline, we compare the classification performance of the quantum models with the performance of a convolutional neural network (CNN)~\cite{NIPS1989_53c3bce6}. CNNs are a well-established model for processing image data, leveraging local connectivity and parameter sharing to extract spatial hierarchies of features. While recent models like vision transformers~\cite{visiontransormer2023} have gained popularity, CNNs still demonstrate superior performance on small datasets when transfer learning is not employed.

To restore spatial locality, we reshape the encoded states and invert the hierarchical ordering. For the grayscale images of the MNIST and Fashion-MNIST datasets this results in inputs of shape $(32,32,2)$, and shapes $(32,32,8)$ and $(128,128,8)$ for the color images of the CIFAR-10 and Imagenette datasets. The last dimension corresponds to the amplitudes of the quantum state encoding the color values of each pixel---two amplitudes from the single color qubit used for the grayscale encoding and eight amplitudes from the three color qubits used for the color encoding. Since no preprocessing is used for training the quantum classifiers, no transformations are applied to the input; all values are real and relatively small as a result of the full quantum state normalization.

The CNN architecture resembles the one of SimpleNet~\cite{Hasanpour2016} and we only adjust the number of input channels from the standard three (RGB images) to the number of basis states of the color qubits (i.e., two for grayscale and eight for color images). For all configurations this results in a parameter count of about $5.5$ million. The overall architecture comprises multiple convolutional blocks, each consisting of a convolutional layer, batch normalization, and ReLU activation. Additionally, intermediate max pooling and dropout layers are applied. Like the original study, we use Adadelta~\cite{zeiler2012adadeltaadaptivelearningrate} as the optimization algorithm and Xavier uniform initialization~\cite{pmlr-v9-glorot10a} for the convolutional layers. We use a multi-step learning rate scheduler, where the learning rate is decreased by a factor of 0.1 at predefined milestone epochs. Given that inputs are represented as quantum states rather than RGB color images, we performed a hyperparameter optimization. Like in the original publication, we performed the optimization only for CIFAR-10 (on uncompressed states) and used the same hyperparameter configuration for all four datasets. We optimized the learning rate, dropout probability, and weight decay. The best-performing configuration used a learning rate of $0.17$, a dropout probability of $0.3$, and a weight decay of $4 \cdot 10^{-4}$. For this purpose, we sampled 200 hyperparameter configurations and employed the Asynchronous Successive Halving Algorithm (ASHA) scheduler~\cite{MLSYS2020_a06f20b3} in connection with the hyperband optimizer~\cite{pmlr-v28-bergstra13}. These tools incorporate early stopping and Bayesian-based hyperparameter optimization.

\begin{figure*}[!t]
    \centering
    \includegraphics[
        width=\linewidth,
        trim={0.55cm 0.6cm 0.55cm 0.55cm},
        clip
    ]{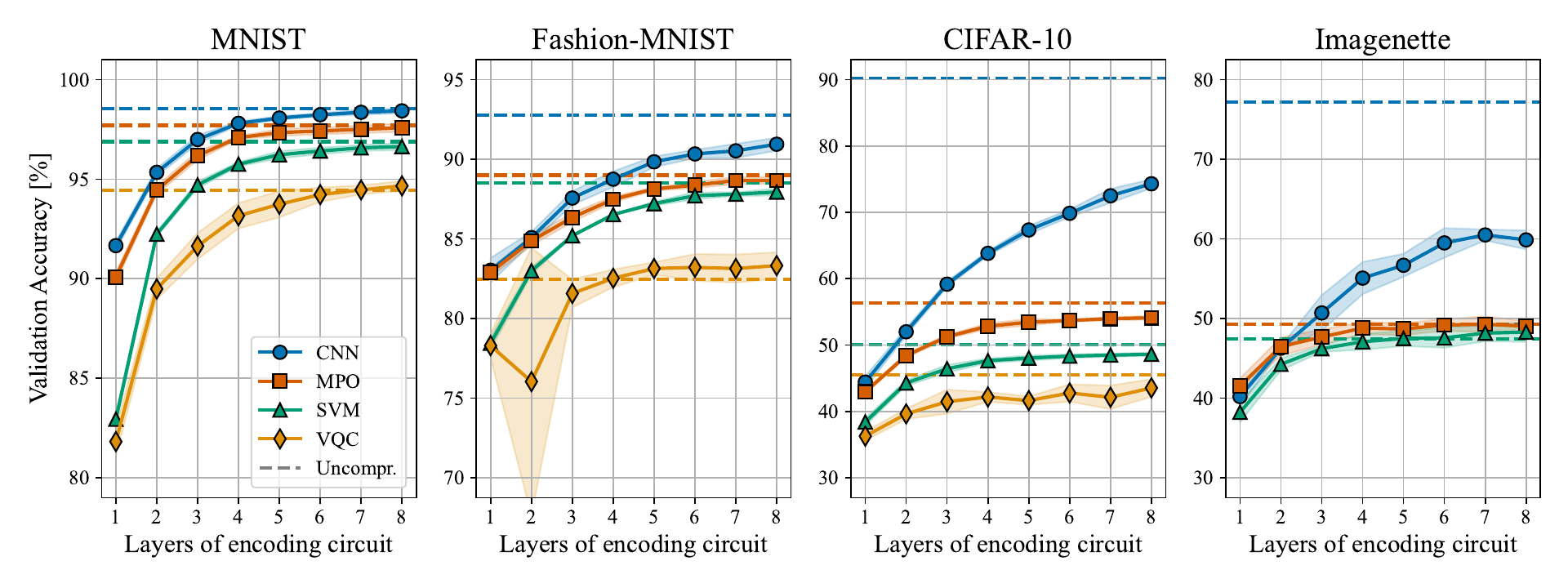}
    
    \caption{\textbf{Validation accuracy on the compressed datasets as a function of encoding circuit depth.} The results are shown for the CNN (blue circles), MPO classifier (orange squares), SVM classifier (green triangles), and nonlinear VQC (yellow diamonds). The dashed horizontal lines indicate the corresponding validation accuracy on the uncompressed datasets. From left to right, the four panels show results on MNIST, Fashion-MNIST, CIFAR-10, and Imagenette datasets. Training of the nonlinear VQC on the full Imagenette dataset was not feasible due to the increased number of qubits, so these data points are omitted. The error bands denote the standard deviation from five-fold cross-validation. Across all datasets, more layers for the encoding tend to improve accuracy, with the CNN consistently outperforming quantum models.}
    \label{fig:compressed}
\end{figure*}

% Discussion of the plot
\textbf{Performance on the Compressed Datasets.}
Fig.~\ref{fig:compressed} shows the validation accuracy of four classifiers---the CNN (blue circles), MPO classifier (orange squares), SVM  classifier (green triangles), and nonlinear VQC (yellow diamonds)---on the compressed versions of the MNIST, Fashion-MNIST, CIFAR-10, and Imagenette datasets as a function of the number of layers of the image-encoding circuit. Shaded areas denote one standard deviation from five-fold cross validation. The dashed lines show the performance on the uncompressed dataset as a reference value.

On the compressed MNIST dataset, as the number of layers of the encoding circuit increases, all models quickly reach a similar validation accuracy as they did on the uncompressed dataset. When using around eight layers for the encoding circuit, all models achieve the same accuracy as on the uncompressed dataset.

Also on the Fashion-MNIST dataset all models achieve a similar accuracy on the compressed and uncompressed datasets when a moderate number of layers is used for the image encoding circuit. For the MPO and SVM classifiers, the same accuracy as on the uncompressed dataset is reached at around seven to eight layers; for the nonlinear VQC the compression acts as a regularization and when the encoding circuit uses five or more layers it even helps to outperform the nonlinear VQC trained on the uncompressed dataset. Only the CNN does not reach the same accuracy as on the uncompressed dataset, but even this worse performance achieves significantly higher accuracies than all other models (both those trained on the compressed or uncompressed datasets).

The CIFAR-10 dataset is the first dataset where the quantum classifiers start to struggle on the uncompressed dataset. We also see this in the performance on the compressed dataset, which with an increasing number of layers of the encoding circuit quickly reaches a similar level as on the uncompressed dataset but never exactly the same accuracy. The CNN does not struggle with the uncompressed CIFAR-10 dataset and reaches $90\%$ accuracy. However, it is more affected by the compression of the dataset than the quantum classifiers, and it does not reach similar accuracies on the compressed version. Still, the accuracies reached on the compressed dataset are much larger than any of the other models, even when those are trained on the uncompressed dataset. The dip in performance of all models on the compressed CIFAR-10 dataset may be partially due to these images having quite a lot of detail for their comparably low resolution. Preserving all of these minute details is hard during the compression, but may be necessary for correct classification.

On the Imagenette dataset the MPO and SVM classifiers reach the same (or larger) accuracy as on the uncompressed dataset when the dataset is compressed with more than four layers of the encoding circuit. Results for the nonlinear VQC are omitted as training on the full-scale Imagenette dataset was not feasible due to the increased number of qubits. The CNN does perform worse on the compressed dataset than it does on the uncompressed one, but even the accuracies achieved on the compressed dataset are significantly better than any of the other models' accuracies on either the compressed or uncompressed datasets.

Overall, we see that the CNN outperforms all other models---whether they are trained on the compressed or uncompressed datasets. With the CNN we do not quite reach the state-of-the-art levels reported in Sec.~\ref{sec:typical_ml_datasets}; this is not unexpected, as we feed the normalized quantum states into the CNN and not the images directly. For the Imagenette dataset, we further crop and resize the images which may additionally reduce the accuracy.

The purpose of the CNN comparison is not to achieve state-of-the-art performance or to directly evaluate whether CNNs outperform the quantum classifiers; this outcome is expected, given that CNN architectures and training procedures have been refined over decades. Furthermore, the CNN models used here have the highest parameter count among all models considered. Rather, the objective is to assess how much information is preserved during compression: specifically, whether the essential features required for classification remain present in the quantum states after encoding the image data into circuit form.

While the CNN is affected the most by the compression out of all classifiers, this is most likely because we directly optimize the fidelity (related to the usual two-norm) when compressing the states. The fidelity is directly related to the kernel functions that are evaluated for the SVM classifiers, and to the expectation values measured with either the quantum circuit classifiers or the tensor-network classifiers. Thus, optimizing the fidelity leads to directly minimizing deviations in the outputs of these classifiers. Conversely, the CNN is not directly related to the two-norm as it is a highly nonlinear function. While optimizing the fidelity should result in similar enough compressed images that the classifier can still recognize them, there are no guarantees that the deviation of the CNN output is minimally affected like for the quantum models.

Nonetheless, we see that the CNN trained on the compressed datasets clearly outperforms all other models, even those trained on the uncompressed datasets. Thus, we see that the quantum models in general are not limited by the amount of information present in the input state, but rather by the way they process that information. This highlights the need for new ideas and architectures for quantum models, which can better extract the available information from the quantum state. As we have seen in the previous subsections, exploring the integration of nonlinearities into these models---either during the processing of the quantum circuit or before that in the input state---may be a fruitful direction.

\section{Conclusion}
\label{sec:conclusion}

Loading classical data on a quantum computer is a major bottleneck for many QML algorithms. Here, we have investigated a method that allows to efficiently load classical image data, and other data with a similar structure. We considered two different encodings, the FRQI for greyscale images and the MCRQI for color images. These encodings lead to lowly-entangled quantum states, which are well-captured by MPS. This, in turn, means that low-depth MPS-inspired quantum circuits can approximate the encoded image data with a high fidelity. We have also presented an optimization algorithm capable of finding these low-depth quantum circuits.

We have applied this algorithm to four standard machine learning datasets, the MNIST, Fashion-MNIST, CIFAR-10 and Imagenette datasets, to test the effectiveness of our approach. We found that these low-depth circuits indeed capture the encoded images with a high fidelity. The fidelity can be systematically increased by growing the number of layers in the circuit, without signs of the fidelity saturating. Further, even as the image resolution is increased, we found that a fixed number of layers of the circuit was able to capture the state, suggesting that this method also scales to higher-resolution image data. The optimized quantum circuits for loading the four image datasets are available via PennyLane at \pennylanedataset. We invite researchers to use them to run their own benchmarks.

% classifiers
Additionally, we presented benchmark results from training quantum classifiers on these datasets, first on the uncompressed and then on the compressed version. These classifiers include quantum kernel methods, variational quantum circuits, and tensor-network-based classifiers. On the uncompressed dataset, we observed that the classifiers perform well on the simple MNIST and Fashion-MNIST datasets, but become less effective on the more complex CIFAR-10 and Imagenette datasets. Shifting from the linear to the nonlinear VQC, or employing the patched or multi-copy encoding to introduce nonlinearities into the tensor-network classifiers, showed promising results for enhancing the models' performance, including on the more complex datasets. Ultimately, however, it is not enough to bring them in line with state-of-the-art performance.

On the compressed dataset the quantum classifiers reach roughly the same validation accuracy, even if the circuits encoding the images only consist of a few layers---in some cases the accuracy is a bit worse, but sometimes, surprisingly, the accuracy even increases on the compressed dataset. We also trained a classical CNN to ensure that the relevant information needed for classifying the images survived the compression step. The CNN achieves a great performance on the uncompressed data, and even on the compressed complex datasets it performs significantly better than the quantum models. This suggests that the quantum models are not limited by the amount of information available in the input state---that information is preserved by the encoding and compression steps. Rather, the quantum models have insufficient capabilities to separate the complex datasets. Indeed, a consistent observation across our benchmarks was that the models perform better as they get access to nonlinearities. Thus, a promising future research direction might be to focus on designing quantum architectures that can effectively incorporate nonlinearities.

Also feature maps can give rise to nonlinearities. Here we have focused on the FRQI and MCRQI for mapping the image data to quantum states, and used the patched or multi-copy encoding to achieve more complex feature maps. It would be interesting to see if there are other mappings from the classical data to quantum states, that realize the necessary nonlinearities but remain lowly entangled such that they can be efficiently loaded in the same way.

Finally, the compressed datasets are readily available via PennyLane. This enables researchers to benchmark novel ideas on the same collection of datasets. We hope that this facilitates comparison among different QML models, and allows to identify promising methods that also work as the complexity of the dataset is increased. Instead of benchmarking QML models, these datasets could also be used to benchmark quantum hardware by running established QML algorithms on them.

\section*{Code Availability}
The four datasets are available via PennyLane at \pennylanedataset. The code for circuit optimization and benchmarking of the classifiers is available at \gitrepo.

\begin{acknowledgments}
    The authors thank Diego Guala and Ben Lau from Xanadu for their support regarding the distribution of the data, and Lukas Lechner for generously granting permission to use their photograph in Fig.~\ref{fig:patching}. B.J., F.P. and C.A.R. thank Kevin Shen and Elvira Shishenina for their collaboration on related previous projects. B.J. thanks Sheng-Hsuan Lin for interesting discussions about the optimization of quantum circuits. F.P. acknowledges the support of the Deutsche Forschungsgemeinschaft (DFG, German Research Foundation) under Germany's Excellence Strategy EXC-2111-390814868, the European Research Council (ERC) under the European Union's Horizon 2020 research and innovation program (grant agreement No. 771537), as well as the Munich Quantum Valley, which the Bavarian state government supports with funds from the Hightech Agenda Bayern Plus. Additionally, the work received support from NERSC (ERCAP0032477/ERCAP0029512: ``Characterization and Middleware for Hybrid Quantum-HPC Applications'') and the Bavarian State Ministry of Economic Affairs (BenchQC project, Grant DIK0425/03).
\end{acknowledgments}

\appendix
\section{Other Color Image Encodings}
\label{app:other_image_encodings}

In the main text we solely focused on the MCRQI to encode color images, but, of course, it is not the only way to encode the color information of an image into the amplitudes of a quantum state. Here, we consider and compare other similar encoding strategies. We also consider different ways to index the pixels in the image, as in the main text we claimed that the Z-order we used leads to the smallest entanglement entropies. We find that all encodings with all indexing variants follow an area law for the entanglement entropy, i.e., the bipartite entanglement entropy of these states does not grow with the image resolution (which sets the number of qubits). The MCRQI with hierarchical indexing on average has the smallest entanglement entropy in our numerical results, suggesting that they can be represented by tensor networks most effectively, which is why we focus on this encoding in the main text.

All states we discuss are of the same form as in the main text, cf. Eq.~\eqref{eq:target_state}, that is they can be written as
\begin{equation}
    \ket{\psi(\vec{x})} = \frac{1}{\sqrt{2^n}} \sum_{j=0}^{2^{n}-1} \ket{c(\vec{x}_j)} \otimes \ket{j}.
\end{equation}
For the different encodings, only the state of the color qubits $\ket{c(\vec{x}_j)}$ will be different. We will also consider three different indexing variants, which change how the pixels are labeled by the index $j$. The simplest variant is when the pixels are counted left to right within each row, and row by row from top to bottom~\cite{Le2011, Le2011_2, jobst2023efficientmpsrepresentationsquantum}; in the hierarchical variant that we use in the main text the pixels are indexed by the Z-order (see Fig.~\ref{fig:hierarchical_order})~\cite{latorre2005imagecompressionentanglement, Le2011, Le2011_2, jobst2023efficientmpsrepresentationsquantum}; and, finally, we consider a snake ordering where rows are ordered top to bottom as in the row-by-row indexing, but every other row is traversed in the opposite direction~\cite{Dilip2022, jobst2023efficientmpsrepresentationsquantum}.

Taking the FRQI for encoding grayscale images as motivation, as a first example we could just consider a state with three color qubits that are simply a tensor product of the single-color-qubit states used in the FRQI, each encoding one of the three different RGB channels. Then, the state of the color qubits is
\begin{equation}
\begin{aligned}
    \ket{c(\bm{x}_j)} =
    &\left(\cos({\textstyle\frac{\pi}{2}} x^R_j)|0\rangle
         + \sin({\textstyle\frac{\pi}{2}} x^R_j)|1\rangle\right)\\
    \otimes
    &\left(\cos({\textstyle\frac{\pi}{2}} x^G_j)|0\rangle
         + \sin({\textstyle\frac{\pi}{2}} x^G_j)|1\rangle\right)\\
    \otimes
    &\left(\cos({\textstyle\frac{\pi}{2}} x^B_j)|0\rangle
         + \sin({\textstyle\frac{\pi}{2}} x^B_j)|1\rangle\right),
\end{aligned}
\end{equation}
and we call this encoding \emph{$\otimes$-multi FRQI}.

The state above uses all eight different basis states, and once it is entangled with the address register the bond dimension across that bond will generically be maximal. By taking a direct sum of the different FRQI components, instead of the tensor product, we get only six contributing basis states, limiting the bond dimension of this bond. This sets the state of the color qubit to be
\begin{equation}
\begin{aligned}
    \hspace{-0.2em}\ket{c(\bm{x}_j)} = \frac{1}{\sqrt{3}}\Big(\!
     &\cos({\textstyle\frac{\pi}{2}} x^R_j)|000\rangle
    + \sin({\textstyle\frac{\pi}{2}} x^R_j)|100\rangle\\
    +&\cos({\textstyle\frac{\pi}{2}} x^G_j)|001\rangle
    + \sin({\textstyle\frac{\pi}{2}} x^G_j)|101\rangle\\
    +&\cos({\textstyle\frac{\pi}{2}} x^B_j)|010\rangle
    + \sin({\textstyle\frac{\pi}{2}} x^B_j)|110\rangle\Big),\hspace{-0.7em}
\end{aligned}
\end{equation}
and we call this encoding \emph{$\oplus$-multi FRQI}.

And finally, for completeness, the MCRQI amounts to adding another $\alpha$ channel to the previous encoding, but fixing all $x^{\alpha}_j=0$. The state of the color qubits is then
\begin{equation}
\begin{aligned}
    \ket{c(\bm{x}_j)} = \frac{1}{2}\Big(\!
     &\cos({\textstyle\frac{\pi}{2}} x^R_j)|000\rangle
    + \sin({\textstyle\frac{\pi}{2}} x^R_j)|100\rangle\\
    +&\cos({\textstyle\frac{\pi}{2}} x^G_j)|001\rangle
    + \sin({\textstyle\frac{\pi}{2}} x^G_j)|101\rangle\\
    +&\cos({\textstyle\frac{\pi}{2}} x^B_j)|010\rangle
    + \sin({\textstyle\frac{\pi}{2}} x^B_j)|110\rangle\\
    +&\cos({\textstyle\frac{\pi}{2}} x^{\alpha}_j)|011\rangle
    + \sin({\textstyle\frac{\pi}{2}} x^{\alpha}_j)|111\rangle\Big),
\end{aligned}
\end{equation}
as in the main text---cf. Eq.~\eqref{eq:RGB_color_state}.

\begin{figure}[t]
    \centering
    \includegraphics[width=\linewidth]{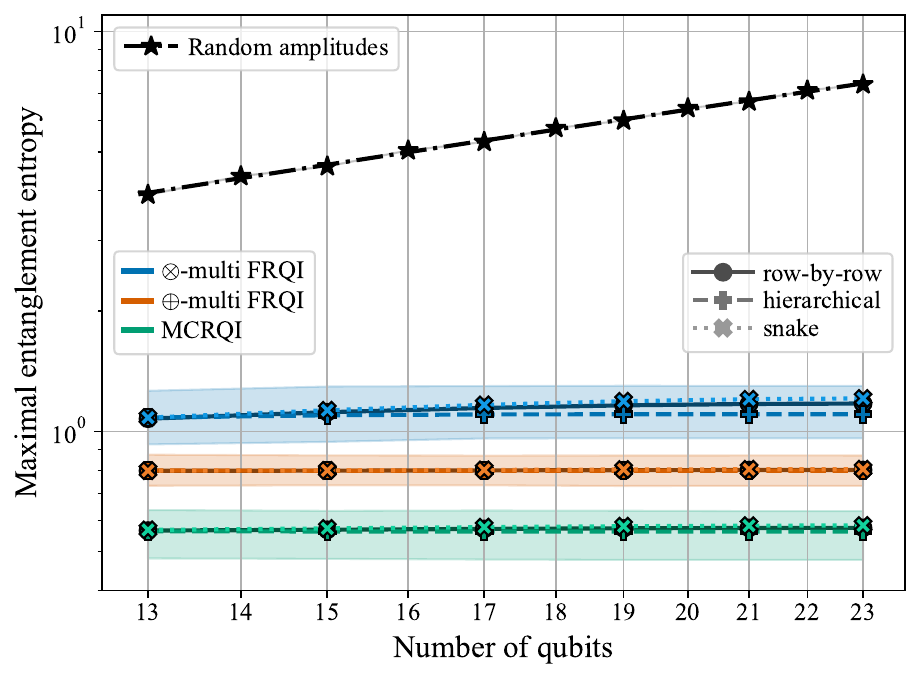}
    
    \caption{\textbf{Scaling of the entanglement entropy for different color image encodings.} The plot shows the scaling of the maximal entanglement entropy over any bipartition of the state into two contiguous halves with the number of qubits in the state. The quantum states are obtained from encoding the $174$ highest-resolution images of the Imagenette dataset~\cite{imagenette} with the three different image encodings---\mbox{$\otimes$-multi} FRQI (blue), \mbox{$\oplus$-multi} FRQI (orange) and the MCRQI (green)---and the three different indexing variants---row-by-row (circles with solid lines), hierarchical (pluses with dashed lines) and snake (crosses with dotted lines)---discussed in the text. The shaded areas show the $25$th--$75$th percentiles for the different encodings using the hierarchical indexing. For reference, we also show the average half-chain entanglement entropy for $200$ states whose amplitudes were drawn from a normal distribution before normalization as black stars. The black shaded area shows the $25$th--$75$th percentiles and the dash-dotted line shows a linear fit to the data.}
    \label{fig:entanglement_scaling}
\end{figure}

The three different encodings are compared in Fig.~\ref{fig:entanglement_scaling}. As a representative dataset, we consider the $174$ images in the Imagenette dataset~\cite{imagenette} that originally have a higher resolution than $1024\times1024$ pixels. The figure shows the maximal entanglement entropy over all bipartitions of the state into two contiguous halves, averaged over all images in the dataset, plotted against the number of qubits needed to represent the images at increasingly higher resolutions. The different colors denote the different encodings, i.e., blue shows the data for the $\otimes$-multi FRQI, orange the data for the $\oplus$-multi FRQI and green the data for the MCRQI. The marker shapes and line styles denote the different ways the pixels are indexed; circles connected by a solid line show the data for row-by-row indexing, pluses connected by a dashed line the data for hierarchical indexing and crosses connected by a dotted line the data for snake indexing. The shaded areas show the $25$th--$75$th percentiles for the different image encodings using the hierarchical indexing; we omit the shaded areas for the data points corresponding to the other indexing variants for visual clarity, the results would look similar. As a reference value, we also plot the half-chain entanglement entropy of states with amplitudes that are sampled randomly from a normal distribution with zero mean and a variance of one before normalization. The average of $200$ realizations is marked by black stars, the $25$th--$75$th percentiles are shaded in black. The black dash-dotted line shows a linear fit to the data, illustrating the generic case where the entanglement entropy of quantum states grows linearly with the subsystem size~\cite{Page1993}.

From the numerical data, we can see that the states encoding image data behave quite distinctly from random or typical quantum states. In particular, we can see that for all encodings and indexing variants the bipartite entanglement entropy seems to saturate to a constant value as the number of qubits is increased, in contrast to the linear growth for random states. We can also see that, on average, the MCRQI states are the least entangled out of the three encodings, and the $\oplus$-multi FRQI states are less entangled than the $\otimes$-multi FRQI states. The choice of indexing makes less of a difference, but we can still see that for high resolutions there is a tendency that the hierarchical encoding leads to more lowly-entangled states than the row-by-row indexing, which leads to more lowly-entangled states than the snake indexing. Overall, the combination which on average leads to the smallest entanglement entropies, and thus should be the most compressible, is the MCRQI encoding with hierarchical indexing---this is what we use in the main text.

\begin{figure*}[t]
    \centering
    \includegraphics[
        width=\linewidth,
        trim={0.6cm 0.7cm 0.5cm 0.5cm},
        clip
    ]{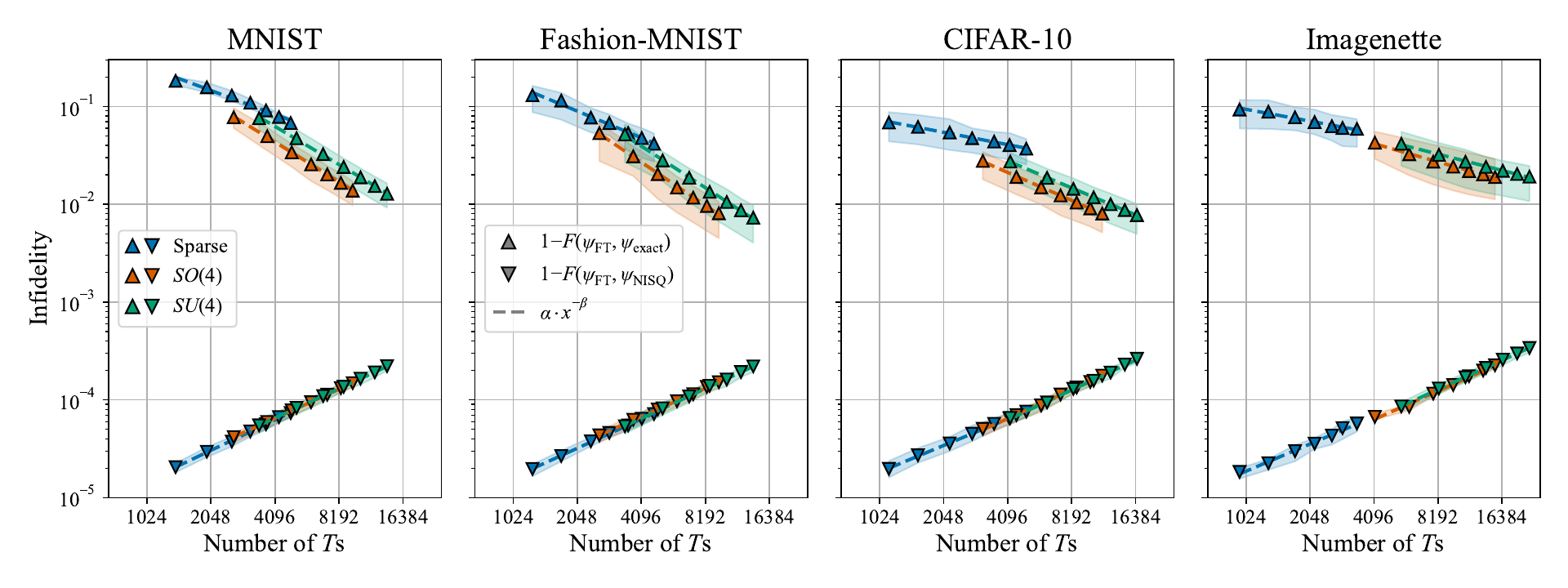}

    \caption{\textbf{Scaling of the infidelity with an increasing number of $\bm{T}$ gates.} We encode $100$ randomly selected images ($10$ per class) from the MNIST, Fashion-MNIST, CIFAR-10 and Imagenette datasets as discussed in Sec.~\ref{sec:encoding}. The upwards-facing triangles show the average infidelity between the exactly encoded states and their quantum circuit approximations compiled into the Clifford+$T$ gate set, the downwards-facing triangles show the average infidelity between the compiled quantum circuits and the same quantum circuits pre-compilation. Both are plotted against the number of $T$ gates in the circuit after compilation. Blue markers show the data for circuits using sparse two-qubit gates, orange markers for those using special orthogonal two-qubit gates and green markers for those using general unitary two-qubit gates. The shaded areas show the $25$th--$75$th percentiles of the infidelities, the variance of the $T$-gate count between different images is negligible. The dashed lines show the results of fitting an algebraic decay or growth $f(x)=\alpha \, x^{\mp\beta}$ to the data. The fitted values, including their uncertainties, are given in Tables~\ref{tab:encoding_t_infid_target} and~\ref{tab:encoding_t_infid_compressed}.}
    \label{fig:infidelity_vs_t}
\end{figure*}

\section{Compiling the Circuits to the Clifford+$T$ Gate Set}
\label{app:t_count}

Apart from the scaling of the infidelity with an increasing number of CNOTs, as shown in Fig.~\ref{fig:infidelity_vs_cnot}, one might also be interested in the infidelity as a function of the number of $T$ gates when compiling the circuits into the Clifford+$T$ gate set suitable for a fault-tolerant quantum computer. There, Clifford operations are effectively free, and the resource overhead is dominated by the $T$ gates, which have to be implemented, e.g., via a costly magic–state distillation~\cite{Bravyi_2005}. Here we consider this fault-tolerant setup, and show the resulting $T$-gate counts in Fig.~\ref{fig:infidelity_vs_t}.

To compile the circuits into the Clifford+$T$ gate set, we note that we already have the circuits in the form of CNOT gates, which are Clifford gates, and single-qubit gates. For the circuits using the sparse or special orthogonal two-qubit gates, the single-qubit gates are simply $y$-rotations. For the circuits using general two-qubit gates also general single-qubit gates appear, which can further be decomposed into single-qubit rotation gates. Up to a basis transformation composed of Clifford operations, these single-qubit rotations are equivalent to a $z$-rotation. Thus, we only have to consider the decomposition of these single-qubit $z$-rotations, for which there exist very efficient decomposition algorithms~\cite{Ross_2014}.

We use the Python implementation~\cite{pygridsynth} of the Gridsynth algorithm~\cite{Ross_2014} to transform the $R_z$ gates into a sequence of single-qubit Clifford and $T$ gates. The typical number of $T$ gates returned by this algorithm to reach a precision $\varepsilon$ is $3\log_{2}(1/\varepsilon) + O\bigl(\log\log(1/\varepsilon)\bigr)$. For our numerical results we use a precision of $\varepsilon=10^{-3}$ to ensure that the combined approximation errors arising from the gate decompositions play a subleading role compared to the approximation error from the circuit optimization.

Fig.~\ref{fig:infidelity_vs_t} shows the infidelities of the circuits compiled with the Clifford+$T$ gate set against the mean number of $T$ gates used. The data points in the upper half of the plot shown as upwards-facing triangles correspond to the infidelity of the compiled circuit with the uncompressed quantum state of the image, and the data points in the lower half shown as downwards-facing triangles correspond to the infidelity of the compiled circuit with the quantum circuit pre-compilation containing arbitrary-angle rotation gates. As in the main text, blue markers show the data for the circuits optimized using the sparse gates, orange markers for circuits optimized with special orthogonal gates and green markers for circuits optimized with general unitary gates. The shaded areas show the $25$th--$75$th percentiles of the infidelity. In comparison, the variance in the required number of $T$ gates for the different images is negligible and thus we omit plotting it. The dashed lines show the result of an algebraic fit to the data, the obtained fitted parameters and their uncertainties are summarized in Tables~\ref{tab:encoding_t_infid_target} and~\ref{tab:encoding_t_infid_compressed}.

Overall, the results for the infidelities with respect to the uncompressed states as a function of the $T$-gate count looks similar to those for the CNOT count (cf. Fig.~\ref{fig:infidelity_vs_cnot}). Even the exponents of the power laws obtained obtained from fitting to the data are very close when plotting against the CNOT count or $T$-gate count (cf. Tables~\ref{tab:encoding_cnot_infid} and~\ref{tab:encoding_t_infid_target}). Ignoring the initial layer of single-qubit gates, sparse and special orthogonal gates require two single-qubit rotations per CNOT, and general unitary gates require three per gate. Assuming the same mean number of $T$ gates for the three different gate types, we expect the same relative behavior between sparse and special orthogonal gates as when plotting against the CNOT count. Only the results for the general unitary gates would be expected to appear slightly shifted towards larger values. At a closer look, however, the mean number of $T$ gates is not the same for all gate types, as for the sparse gates we observe that the total number of $T$ gates actually decreases with image resolution (i.e., going from MNIST to Imagenette), while it grows for the other two gate types. This artifact arises because, in circuits using sparse gates, many single-qubit gates acting on the few least-significant address qubits in the MCRQI are nearly identity operations and are therefore compiled as identity gates. We could remove this artifact by setting the compilation precision to a lower value of $\varepsilon=10^{-5}$, but already at a precision of $\varepsilon=10^{-3}$ the error arising from the compilation is orders of magnitudes smaller than the error that stems from the circuit approximation error. The low approximation error arising from the compilation to the Clifford+$T$ gate set can be seen from the data in the lower half of the plot: for a circuit with two layers, the infidelity of the compiled circuit with its pre-compiled counterpart is roughly three orders of magnitude lower than the infidelity of the compiled circuit with its target state, and for the eight-layer circuits this difference is still roughly two orders of magnitude.

We can speculate on why so many near-identity gates appear for the circuits using sparse gates, more so than for the other two gate types. Note that most of these near-identity gates appear when acting on the least significant bits of the address qubits, which control the finest length scales of the image (cf. Fig.~\ref{fig:hierarchical_order}). For a high-resolution image, a good approximation to the image might be a coarser version obtained by blocks of pixels with uniform color. This corresponds to completely disentangling the qubits corresponding to the few least-significant bits and preparing them in an equal-weight superposition. The finer details are then obtained by weakly entangling these qubits with the address qubits corresponding to more significant bits and the color qubits. We can imagine that after the first layer of the circuit is initialized during the optimization algorithm, such a state is prepared, where the qubits corresponding to the least-significant bits are close to a product state made of equal-weight superpositions and they are only weakly entangled with the rest of the system. The next few layers of the circuit initialized during the optimization should then only marginally increase the correlations of the least-significant bits with the rest of the system, keeping the overall weakly-entangled structure intact. However, since the sparse gates are just a single-CNOT with single-qubit gates, they are always a maximally entangling gate that generically increases the entanglement entropy significantly. Thus, it might turn out best to keep the single-qubit gates very close to an identity, so that many of the CNOTs act on the state $\ket{\vec{0}}$ and do not increase the entanglement entropy. Conversely, special orthogonal and general unitary gates can locally be tuned to the identity by choosing suitable single-qubit gates, which should allow for a finer control of how much entanglement and what kind of correlations are added to the least significant bits.

\section{Fit Parameters}
\label{app:parameters}

% Fitted parameters from main text
\begin{table}[!th]
\setlength{\tabcolsep}{1pt}
\centering
\begin{tabular*}{\linewidth}{@{\extracolsep{\fill}}l c c}
\hline
\rule[-1.5mm]{0pt}{5mm}%
\textbf{Setting} & $\alpha$ & $\beta$ \\
\hline
\rule{0pt}{4mm}%
\textbf{MNIST} & & \\
\quad Sparse     & $1.80 \pm 0.48$  & $0.812 \pm 0.069$ \\
\quad $SO(4)$    & $8.46 \pm 0.77$  & $1.263 \pm 0.020$ \\
\quad $SU(4)$     & $16.27 \pm 1.60$ & $1.304 \pm 0.020$ \\
\rule{0pt}{4mm}%
\textbf{Fashion-MNIST} & & \\
\quad Sparse     & $3.05 \pm 1.04$  & $1.02 \pm 0.088$ \\
\quad $SO(4)$    & $8.71 \pm 0.41$  & $1.382 \pm 0.010$ \\
\quad $SU(4)$     & $18.16 \pm 1.45$ & $1.436 \pm 0.016$ \\
\rule{0pt}{4mm}%
\textbf{CIFAR-10} & & \\
\quad Sparse     & $0.47 \pm 0.12$  & $0.506 \pm 0.063$ \\
\quad $SO(4)$    & $0.91 \pm 0.03$  & $0.902 \pm 0.006$ \\
\quad $SU(4)$     & $1.44 \pm 0.02$  & $0.929 \pm 0.003$ \\
\rule{0pt}{4mm}%
\textbf{Imagenette} & & \\
\quad Sparse     & $0.38 \pm 0.06$  & $0.400 \pm 0.035$ \\
\quad $SO(4)$    & $0.42 \pm 0.01$  & $0.548 \pm 0.005$ \\
\quad $SU(4)$     & $0.54 \pm 0.04$  & $0.566 \pm 0.013$%
\rule[-2mm]{0pt}{1mm} \\
\hline
\end{tabular*}
\caption{\textbf{Parameters obtained from fitting an algebraic decay $\bm{f(x) = \alpha \, x^{-\beta}}$ to the infidelities of the circuit approximations as a function of the number of CNOTs in the circuit.} We show results for states encoding images from the four datasets MNIST, Fashion-MNIST, CIFAR-10 and Imagenette, and for circuits using sparse, special orthogonal and general unitary gates. These parameters correspond to the algebraic decays shown as dashed lines in Fig.~\ref{fig:infidelity_vs_cnot}.}
\label{tab:encoding_cnot_infid}
\end{table}

Here we give the detailed fitted parameters for the scaling of the infidelities of the circuit approximations plotted in the main text in Sec.~\ref{sec:efficient_encoding}.
The parameters obtained from fitting an algebraic decay to the infidelities as a function of the CNOT count as shown in Fig.~\ref{fig:infidelity_vs_cnot} are listed in Table~\ref{tab:encoding_cnot_infid}.
The fitted parameters for the Gompertz functions that describe the scaling of the infidelities with image resolution in Fig.~\ref{fig:infidelity_vs_resolution} are summarized in Table~\ref{tab:infidelity_vs_resolution}.

% Fitted parameters from main text
\begin{table}[!th]
\setlength{\tabcolsep}{1pt}
\centering
\begin{tabular*}{\linewidth}{@{\extracolsep{\fill}}lccc}
\hline
\rule[-1.5mm]{0pt}{5mm}%
\textbf{Setting} & $\alpha$ & $\beta$ & $\gamma$ \\
\hline
\rule{0pt}{4mm}Random  & $1.002 \pm 0.000$ & $42.9 \pm 0.2$ & $0.538 \pm 0.001$ \\
\rule{0pt}{4mm}$d=1$  & $0.084 \pm 0.001$ & $7.4 \pm 0.3$  & $0.252 \pm 0.007$ \\
\rule{0pt}{4mm}$d=2$  & $0.048 \pm 0.002$ & $54.3 \pm 7.9$ & $0.386 \pm 0.021$ \\
\rule{0pt}{4mm}$d=3$  & $0.040 \pm 0.002$ & $77.5 \pm 6.9$ & $0.387 \pm 0.013$ \\
\rule{0pt}{4mm}$d=4$  & $0.035 \pm 0.001$ & $98.8 \pm 5.6$ & $0.389 \pm 0.008$ \\
\hline
\end{tabular*}
\caption{\textbf{Parameters obtained from fitting a Gompertz function $\bm{f(x) = \alpha \, \text{exp}(-\beta \, \text{exp}(-\gamma \, x))}$ to the scaling of the infidelities of the circuit approximations with increasing image resolution.}
We show results for circuits using special orthogonal gates with $d \in \{1, 2, 3, 4\}$ layers approximating states encoding high-resolution images from the Imagenette dataset, and for a circuit using special orthonormal gates with $d=4$ layers approximating real-valued Haar random states. These parameters correspond to the Gompertz functions shown as dashed lines in Fig.~\ref{fig:infidelity_vs_resolution}.}
\label{tab:infidelity_vs_resolution}
\end{table}

We also show the detailed obtained parameters from fitting the infidelities as a function of the $T$ count as considered in App.~\ref{app:t_count}. For the infidelities of the circuits compiled into Clifford and $T$ gates with respect to the uncompressed quantum states encoding the image data as shown in Fig.~\ref{fig:infidelity_vs_t} (upper half, upwards-facing triangles), the obtained parameters are listed in Table~\ref{tab:encoding_t_infid_target}. The fitted parameters for the infidelity between the compiled circuits and the circuits pre-compilation also shown in Fig.~\ref{fig:infidelity_vs_t} (lower half, downwards-facing triangles) are listed in Table~\ref{tab:encoding_t_infid_compressed}.

% Fitted parameters from appendix
\begin{table}[!t]
\setlength{\tabcolsep}{1pt}
\centering
\begin{tabular*}{\linewidth}{@{\extracolsep{\fill}}l c c}
\hline
\rule[-1.5mm]{0pt}{5mm}%
\textbf{Setting} & $\alpha$ & $\beta$ \\
\hline
\rule{0pt}{4mm}%
\textbf{MNIST} & & \\
\quad Sparse     & $69.74 \pm 33.21$   & $0.809 \pm 0.060$ \\
\quad $SO(4)$    & $3222.11 \pm 367.78$ & $1.349 \pm 0.013$ \\
\quad $U(4)$     & $2919.40 \pm 476.45$ & $1.293 \pm 0.018$ \\
\rule{0pt}{4mm}%
\textbf{Fashion-MNIST} & & \\
\quad Sparse     & $89.28 \pm 32.44$    & $0.905 \pm 0.046$ \\
\quad $SO(4)$    & $5330.57 \pm 924.25$  & $1.466 \pm 0.020$ \\
\quad $U(4)$     & $4982.32 \pm 878.37$  & $1.413 \pm 0.020$ \\
\rule{0pt}{4mm}%
\textbf{CIFAR-10} & & \\
\quad Sparse     & $1.32 \pm 0.13$       & $0.418 \pm 0.012$ \\
\quad $SO(4)$    & $56.86 \pm 4.98$      & $0.949 \pm 0.010$ \\
\quad $U(4)$     & $56.01 \pm 3.27$      & $0.915 \pm 0.006$ \\
\rule{0pt}{4mm}%
\textbf{Imagenette} & & \\
\quad Sparse     & $1.51 \pm 0.32$       & $0.401 \pm 0.028$ \\
\quad $SO(4)$    & $7.15 \pm 1.31$       & $0.619 \pm 0.020$ \\
\quad $U(4)$     & $4.84 \pm 0.65$       & $0.555 \pm 0.014$%
\rule[-2mm]{0pt}{1mm} \\
\hline
\end{tabular*}
\caption{\textbf{Parameters obtained from fitting an algebraic decay $\bm{f(x) = \alpha \, x^{-\beta}}$ as a function of the $\bm{T}$-gate count to the infidelities between the circuits compiled into Clifford and $\bm{T}$ gates and the uncompressed states encoding the image data.} We show results for states encoding images from the four datasets MNIST, Fashion-MNIST, CIFAR-10 and Imagenette, and for circuits using sparse, special orthogonal and general unitary gates pre-compilation. These parameters correspond to the algebraic decays shown as dashed lines in the upper half of Fig.~\ref{fig:infidelity_vs_t}.}
\label{tab:encoding_t_infid_target}
\end{table}

\begin{table}[!t]
\setlength{\tabcolsep}{1pt}
\centering
\begin{tabular*}{\linewidth}{@{\extracolsep{\fill}}l c c}
\hline
\rule[-1.5mm]{0pt}{5mm}%
\textbf{Setting} & $\alpha \cdot 10^8$ & $\beta$ \\
\hline
\rule{0pt}{4mm}%
\textbf{MNIST} & & \\
\quad Sparse     & $1.16 \pm 0.16$   & $1.03 \pm 0.017$ \\
\quad $SO(4)$    & $1.75 \pm 0.08$   & $0.99 \pm 0.005$ \\
\quad $SU(4)$    & $1.60 \pm 0.07$   & $1.00 \pm 0.005$ \\
\rule{0pt}{4mm}%
\textbf{Fashion-MNIST} & & \\
\quad Sparse     & $1.68 \pm 0.17$   & $0.99 \pm 0.013$ \\
\quad $SO(4)$    & $2.22 \pm 0.19$   & $0.96 \pm 0.010$ \\
\quad $SU(4)$    & $1.44 \pm 0.14$   & $1.01 \pm 0.011$ \\
\rule{0pt}{4mm}%
\textbf{CIFAR-10} & & \\
\quad Sparse     & $3.51 \pm 0.31$   & $0.90 \pm 0.011$ \\
\quad $SO(4)$    & $1.95 \pm 0.36$   & $0.97 \pm 0.021$ \\
\quad $SU(4)$    & $1.33 \pm 0.22$   & $1.02 \pm 0.018$ \\
\rule{0pt}{4mm}%
\textbf{Imagenette} & & \\
\quad Sparse     & $3.09 \pm 0.85$   & $0.92 \pm 0.036$ \\
\quad $SO(4)$    & $2.00 \pm 0.68$   & $0.97 \pm 0.038$ \\
\quad $SU(4)$    & $1.70 \pm 0.12$   & $0.99 \pm 0.007$%
\rule[-2mm]{0pt}{1mm} \\
\hline
\end{tabular*}
\caption{
\textbf{Parameters obtained from fitting an algebraic growth $\bm{f(x) = \alpha \, x^{\beta}}$ as a function of the $\bm{T}$-gate count to the infidelities between the circuits compiled into Clifford and $\bm{T}$ gates and the same circuits pre-compilation.} We show results for states encoding images from the four datasets MNIST, Fashion-MNIST, CIFAR-10 and Imagenette, and for circuits using sparse, special orthogonal and general unitary gates. These parameters correspond to the algebraic growths shown as dashed lines in the lower half of Fig.~\ref{fig:infidelity_vs_t}.}
\label{tab:encoding_t_infid_compressed}
\end{table}

\section{Details of the Tensor-Network Classifiers}

Here we give some more details on the tensor-network classifiers used in the main text. First, we show that the random initialization we use indeed leads to an output with unit variance. Then, we connect the MPO classifiers with the usual variational quantum classifiers that compute the expectation value of some Hermitian observable.

\subsection{Numerically Stable Random Initialization}
\label{app:init}

In the main text we presented a method to randomly initialize the entries of an MPS classifier, and claimed that this ensured that the variance of the output of MPS classifier had unit variance. Here we are going to prove this claim for both the MPS and MPO classifier.

We consider an MPS classifier as shown in Fig.~\ref{fig:mpsarchitecture} in the main text. The MPS classifier has $L$ sites, with the tensor on site $i$ having an additional leg for the label index $\ell$. We refer to the bond dimension on the $m$th bond as $\chi_m$ with $m=1,\ldots,L-1$ and consider arbitrary physical dimensions $d_m$ on each site $m$. We will denote the entries of the MPS tensors $M_{\alpha_m\beta_m}^{j_m}$ on sites $m=1,\ldots,i-1,i+1,\ldots,L$ and $M_{\alpha_i\beta_i}^{j_i\ell}$ on site $i$, where $\alpha_m,\beta_m$ are the virtual indices running over the local bond dimension, $j_m$ is the physical index labeling the basis of the local $d_m$-dimensional Hilbert space and $\ell$ is the label index. For the initialization of these tensors, we assume that all entries are drawn independently from a Gaussian distribution with zero mean and a site-dependent variance $\sigma^2_m$, $m=1,\ldots,L$. Further, we consider the case where the input state $\psi_{j_1,\dots,j_L}$ that is contracted with the MPS classifier is real and normalized as $\sum_{j_1,\ldots,j_L} \psi_{j_1,\ldots,j_L}^2 = 1$. For the MPS classifier, the input state is the normalized pure quantum state $\psi_{j_1,\dots,j_L} = \braket{j_1,\ldots,j_L}{\psi(\vec{x})}$ encoding the data. Note that we can treat the MPO classifier in the same framework, by combining the physical indices $k_m$, $k_m'$ of the MPO tensor into one index $j_m = (k_m,k_m')$ with twice the size. Then, also the input state, which corresponds to the density matrix $\psi_{j_1,\dots,j_L} = \bra{k_1,\ldots,k_L}\rho(\vec{x})\ket{k'_1,\ldots,k'_L} = \braket{k_1,\ldots,k_L}{\psi(\vec{x})}\braket{\psi(\vec{x})}{k'_1,\ldots,k'_L}$ of the encoded data, is properly normalized since the density matrix corresponds to a pure state:
\begin{equation}
\begin{aligned}
    \sum_{j_1,\ldots,j_L}\! \psi_{j_1,\ldots,j_L}^2
    =\! &\sum_{k_1,\ldots,k_L}
    \braket{k_1,\ldots,k_L}{\psi(\vec{x})}^2\\
    \times\! &\sum_{k_1',\ldots,k_L'}\! 
    \braket{\psi(\vec{x})}{k'_1,\ldots,k'_L}^2 = 1.
\end{aligned}
\end{equation}
Thus, in the following we can focus on the MPS classifier but all results also follow for the MPO classifier, which is essentially an MPS classifier with a doubled dimension of the physical Hilbert space.

Before computing the expectation value and variance of the MPS classifier contracted with an input state, we recall some simple identities for the expectation value and variance of sums and products of random variables. First, the expectation value is a linear function, so for a linear combination of multiple random variables $X_n$ we have
\begin{equation}
    \E\Big[\sum_n a_n X_n\Big] = \sum_n a_n \E[X_n].
    \label{eq:E_sum}
\end{equation}
For independent random variables, the expectation value of a product becomes
\begin{equation}
    \E\Big[\prod_n X_n\Big] = \prod_n \E[X_n].
    \label{eq:E_prod}
\end{equation}
The variance of a sum of independent random variables behaves as
\begin{equation}
    \Var\Big[\sum_n a_n X_n\Big] = \sum_n a_n^2 \Var[X_n],
    \label{eq:Var_sum}
\end{equation}
and for the product of independent random variables which also have zero mean the variance is
\begin{equation}
    \Var\Big[\prod_n X_n\Big] = \prod_n \Var[X_n].
    \label{eq:Var_prod}
\end{equation}
In our case, the random variables are the entries of the MPS tensors $M_{\alpha_m\beta_m}^{j_m}$ and $M_{\alpha_i\beta_i}^{j_i\ell}$, which are indeed sampled independently and have zero mean, so we can make use of the relations above.

We denote the result of the contraction of the MPS classifier with an input state as $f_{\ell}(\vec{x})$---the decision function of the model. Then, for the expectation value we have
\begin{widetext}
\begin{equation}
\begin{aligned}
    \E[f_{\ell}(\vec{x})]
    &= \E\Bigg[
    \sum_{j_1,\ldots,j_L}\!
    \sum_{\substack{\alpha_2,\ldots,\alpha_{L}\\\beta_1,\ldots,\beta_{L-1}}}
    \hspace{-0.6em}\psi_{j_1,\ldots, j_L}
    \left(\prod_{k=1}^{L-1}\delta_{\beta_k,\alpha_{k+1}}\right)
    \left(\prod_{m=1}^{i-1}M_{\alpha_m,\beta_m}^{j_m}\right)
    M_{\alpha_i\beta_i}^{j_i\ell}
    \left(\prod_{n=i+1}^{L}M_{\alpha_n,\beta_n}^{j_n}\right)
    \Bigg]\\
    &= \sum_{j_1,\ldots,j_L}\!
    \sum_{\substack{\alpha_2,\ldots,\alpha_{L}\\\beta_1,\ldots,\beta_{L-1}}}
    \hspace{-0.6em}\psi_{j_1,\ldots, j_L}
    \left(\prod_{k=1}^{L-1}\delta_{\beta_k,\alpha_{k+1}}\right)
    \E\Bigg[
    \left(\prod_{m=1}^{i-1}M_{\alpha_m,\beta_m}^{j_m}\right)
    M_{\alpha_i\beta_i}^{j_i\ell}
    \left(\prod_{n=i+1}^{L}M_{\alpha_n,\beta_n}^{j_n}\right)
    \Bigg]\\
    &= \sum_{j_1,\ldots,j_L}\!
    \sum_{\substack{\alpha_2,\ldots,\alpha_{L}\\\beta_1,\ldots,\beta_{L-1}}}
    \hspace{-0.6em}\psi_{j_1,\ldots, j_L}
    \left(\prod_{k=1}^{L-1}\delta_{\beta_k,\alpha_{k+1}}\right)
    \left(\prod_{m=1}^{i-1} \E\Big[M_{\alpha_m,\beta_m}^{j_m}\Big]\right)
    \E\Big[M_{\alpha_i\beta_i}^{j_i\ell}\Big]
    \left(\prod_{n=i+1}^{L} \E\Big[M_{\alpha_n,\beta_n}^{j_n}\Big]\right)\\
    &= 0.
\end{aligned}
\end{equation}
The first line simply writes out the expression for contracting the MPS classifier with its input state. To get to the second line we used the linearity of the expectation value; to get to the third line we used the independence of the random entries of the MPS tensors; and, for the final result, we used that the random variables have zero mean.

For the variance of the decision function $f_{\ell}(\vec{x})$ we have
\begin{equation}
\begin{aligned}
    \Var[f_{\ell}(\vec{x})]
    &= \Var\Bigg[
    \sum_{j_1,\ldots,j_L}\!
    \sum_{\substack{\alpha_2,\ldots,\alpha_{L}\\\beta_1,\ldots,\beta_{L-1}}}
    \hspace{-0.6em}\psi_{j_1,\ldots, j_L}
    \left(\prod_{k=1}^{L-1}\delta_{\beta_k,\alpha_{k+1}}\right)
    \left(\prod_{m=1}^{i-1}M_{\alpha_m,\beta_m}^{j_m}\right)
    M_{\alpha_i\beta_i}^{j_i\ell}
    \left(\prod_{n=i+1}^{L}M_{\alpha_n,\beta_n}^{j_n}\right)
    \Bigg]\\
    &= \sum_{j_1,\ldots,j_L}\!
    \sum_{\substack{\alpha_2,\ldots,\alpha_{L}\\\beta_1,\ldots,\beta_{L-1}}}
    \hspace{-0.6em}\psi_{j_1,\ldots, j_L}^2
    \left(\prod_{k=1}^{L-1}\delta_{\beta_k,\alpha_{k+1}}\right)
    \Var\Bigg[
    \left(\prod_{m=1}^{i-1}M_{\alpha_m,\beta_m}^{j_m}\right)
    M_{\alpha_i\beta_i}^{j_i\ell}
    \left(\prod_{n=i+1}^{L}M_{\alpha_n,\beta_n}^{j_n}\right)
    \Bigg]\\
    &= \sum_{j_1,\ldots,j_L}\!
    \sum_{\substack{\alpha_2,\ldots,\alpha_{L}\\\beta_1,\ldots,\beta_{L-1}}}
    \hspace{-0.7em}\psi_{j_1,\ldots, j_L}^2\!
    \left(\prod_{k=1}^{L-1}\delta_{\beta_k,\alpha_{k+1}}\!\right)\!
    \left(\prod_{m=1}^{i-1}\!\Var\Big[M_{\alpha_m,\beta_m}^{j_m}\Big]\!\right)\!
    \Var\Big[M_{\alpha_i\beta_i}^{j_i\ell}\Big]\!
    \left(\prod_{n=i+1}^{L}\hspace{-0.5em}\Var\Big[M_{\alpha_n,\beta_n}^{j_n}\Big]\!\right)\\
    &= \sum_{j_1,\ldots,j_L} \psi_{j_1,\ldots, j_L}^2
    \sum_{\substack{\alpha_2,\ldots,\alpha_{L}\\\beta_1,\ldots,\beta_{L-1}}}
    \left(\prod_{k=1}^{L-1}\delta_{\beta_k,\alpha_{k+1}}\right)
    \left(\prod_{m=1}^{L}\sigma^2_m\right),
\end{aligned}
\end{equation}
\end{widetext}
\vspace*{-3em}
where the first line again just writes out the expression for the contraction of the MPS; to get to the second line we made use of the sum-rule for variances in Eq.~\eqref{eq:Var_sum} (note that $\delta_{\alpha_k,\beta_{k+1}}^2=\delta_{\alpha_k,\beta_{k+1}}$); to get to the third line we used the product-rule for independent random variables with zero mean in Eq.~\eqref{eq:Var_prod}; and in the last line we plugged in the variances of the random entries of the MPS tensors. To simplify the result further, we can use that the input state $\psi_{j_1,\ldots,j_L}$ is normalized and we can get rid of the sums over $\beta_k$ by making use of the Kronecker deltas. This yields
\begin{equation}
\begin{aligned}
    \Var[f_{\ell}(\vec{x})]
    &= \sum_{\alpha_2,\ldots,\alpha_{L}}
    \left(\prod_{m=1}^{L}\sigma^2_m\right)\\
    &= \left(\prod_{n=1}^{L-1} \chi_n\right)\left(\prod_{m=1}^{L}\sigma^2_m\right),
\end{aligned}
\end{equation}
where $\chi_n$ is the bond dimension on the $n$th bond.

To ensure that the final output has unit variance, we can rescale the standard deviations of the Gaussian distributions used for initializing the MPS tensors as described in the main text. For MPS tensors left of the one containing the label index $\ell$, we choose the standard deviation as $\sigma_m=1/\sqrt{\chi_{m-1}}$ where $\chi_{m-1}$ is the bond dimension of the bond left of the tensor and we set $\chi_0=1$. For MPS tensors right of the label index $\ell$, we choose the standard deviations as $\sigma_m=1/\sqrt{\chi_{m}}$, where $\chi_{m}$ is the bond dimension of the bond right of the tensor and we set $\chi_L=1$. The tensor containing the label index $\ell$ is initialized with standard deviation $\sigma_i = 1/\sqrt{\chi_{i-1}\chi_i}$. Plugging these choices for the standard deviations into the result for the variance above, we obtain
\begin{equation}
\begin{aligned}
    &\hspace{-0.1em}\Var[f_{\ell}(\vec{x})] =\\
    &\hspace{0.6em}= \left(\prod_{k=1}^{L-1} \chi_k\!\right)\!\!
    \left(\prod_{m=1}^{i-1} \frac{1}{\chi_{m-1}}\!\right)\!
    \frac{1}{\chi_{i-1}\chi_i}\!
    \left(\prod_{n=i+1}^{L} \frac{1}{\chi_{n}}\!\right)\!\\
    &\hspace{0.6em}= 1.
\end{aligned}
\end{equation}
Hence, the output has unit variance as claimed in the main text.

\subsection{MPO Classifiers are a Subset of General Quantum Classifiers}
\label{app:MPO_is_quantum}

In the main text we have introduced the MPO classifier because its decision function is a linear function with respect to the input density matrix, just like a quantum circuit classifier. However, the expectation values of the quantum circuit classifiers are usually taken with respect to a Hermitian observable, and the operator implemented by the MPO is not constrained to be Hermitian. Here we show that only the Hermitian part of the MPO contributes to the output, so the decision function of the MPO classifier can be viewed as the expectation value of a Hermitian observable---making the MPO classifier a proper subset of the most general quantum circuit classifiers.

Let us denote the operator implemented by the MPO classifier as $O_{\ell}$, where the index $\ell$ is the label index and corresponds to the one additional leg of the MPO that is not contracted with the input state. The operator $O_{\ell}$ is purely real, as all MPO tensors are real, and we can split it into a symmetric part $O_{\ell}^+ = \frac{1}{2}(O_{\ell} + O_{\ell}^\transpose)$ (which is also Hermitian) and an anti-symmetric part $O_{\ell}^- = \frac{1}{2}(O_{\ell} - O_{\ell}^\transpose)$. We recover the original operator as $O_{\ell} = O_{\ell}^+ + O_{\ell}^-$. For the expectation value of the anti-symmetric part with respect to a purely real-valued input state $\ket{\psi(\vec{x})}$ we have
\begin{equation}
\begin{aligned}
    &\bra{\psi(\vec{x})}O_{\ell}^-\ket{\psi(\vec{x})}
    = (\bra{\psi(\vec{x})}O_{\ell}^-\ket{\psi(\vec{x})})^\transpose\\
    &\hspace{0.5em}= \bra{\psi(\vec{x})} (O_{\ell}^-)^\transpose \ket{\psi(\vec{x})}
    = - \bra{\psi(\vec{x})} O_{\ell}^- \ket{\psi(\vec{x})}
\end{aligned}
\end{equation}
and thus $\bra{\psi(\vec{x})}O_{\ell}^-\ket{\psi(\vec{x})} = 0$. Therefore, the expectation value of the full operator $O_{\ell}$ is
\begin{equation}
\begin{aligned}
    &\bra{\psi(\vec{x})}O_{\ell}\ket{\psi(\vec{x})} =\\
    &\qquad= \bra{\psi(\vec{x})} O_{\ell}^+ \ket{\psi(\vec{x})}
    + \bra{\psi(\vec{x})} O_{\ell}^- \ket{\psi(\vec{x})}\\
    &\qquad= \bra{\psi(\vec{x})} O_{\ell}^+ \ket{\psi(\vec{x})}).
\end{aligned}
\end{equation}
Since $O_{\ell}^+ = (O_{\ell}^+)^{\dagger}$ is a Hermitian operator, the decision function of the MPO classifier can be written as the expectation value of a Hermitian observable as claimed.

\bibliographystyle{unsrtnat}
\bibliography{references}

\begin{thebibliography}{137}
\providecommand{\natexlab}[1]{#1}
\providecommand{\url}[1]{\texttt{#1}}
\expandafter\ifx\csname urlstyle\endcsname\relax
  \providecommand{\doi}[1]{doi: #1}\else
  \providecommand{\doi}{doi: \begingroup \urlstyle{rm}\Url}\fi

\bibitem[Schuld and Petruccione(2022)]{schuld2022machine}
Maria Schuld and Francesco Petruccione.
\newblock \emph{Machine Learning with Quantum Computers}.
\newblock Springer International Publishing, Cham, 2nd edition, October 2022.
\newblock ISBN 978-3-030-83097-7.
\newblock \doi{10.1007/978-3-030-83098-4}.

\bibitem[Biamonte et~al.(2017)Biamonte, Wittek, Pancotti, Rebentrost, Wiebe,
  and Lloyd]{Biamonte_2017}
Jacob Biamonte, Peter Wittek, Nicola Pancotti, Patrick Rebentrost, Nathan
  Wiebe, and Seth Lloyd.
\newblock Quantum machine learning.
\newblock \emph{Nature}, 549\penalty0 (7671):\penalty0 195--202, September
  2017.
\newblock ISSN 1476-4687.
\newblock \doi{10.1038/nature23474}.

\bibitem[Liu et~al.(2021)Liu, Arunachalam, and Temme]{Liu2021}
Yunchao Liu, Srinivasan Arunachalam, and Kristan Temme.
\newblock A rigorous and robust quantum speed-up in supervised machine
  learning.
\newblock \emph{Nature Physics}, 17\penalty0 (9):\penalty0 1013--1017, July
  2021.
\newblock \doi{10.1038/s41567-021-01287-z}.

\bibitem[Sweke et~al.(2021)Sweke, Seifert, Hangleiter, and
  Eisert]{Sweke2021quantumversus}
Ryan Sweke, Jean-Pierre Seifert, Dominik Hangleiter, and Jens Eisert.
\newblock On the {Q}uantum versus {C}lassical {L}earnability of {D}iscrete
  {D}istributions.
\newblock \emph{{Quantum}}, 5:\penalty0 417, March 2021.
\newblock ISSN 2521-327X.
\newblock \doi{10.22331/q-2021-03-23-417}.

\bibitem[Huang et~al.(2021)Huang, Broughton, Mohseni, Babbush, Boixo, Neven,
  and McClean]{Huang_2021}
Hsin-Yuan Huang, Michael Broughton, Masoud Mohseni, Ryan Babbush, Sergio Boixo,
  Hartmut Neven, and Jarrod~R. McClean.
\newblock Power of data in quantum machine learning.
\newblock \emph{Nature Communications}, 12\penalty0 (1), May 2021.
\newblock ISSN 2041-1723.
\newblock \doi{10.1038/s41467-021-22539-9}.

\bibitem[Pirnay et~al.(2023)Pirnay, Sweke, Eisert, and
  Seifert]{PhysRevA.107.042416}
Niklas Pirnay, Ryan Sweke, Jens Eisert, and Jean-Pierre Seifert.
\newblock Superpolynomial quantum-classical separation for density modeling.
\newblock \emph{Physical Review A}, 107:\penalty0 042416, Apr 2023.
\newblock \doi{10.1103/PhysRevA.107.042416}.

\bibitem[Gyurik and Dunjko(2023)]{Gyurik2023}
Casper Gyurik and Vedran Dunjko.
\newblock Exponential separations between classical and quantum learners.
\newblock \emph{arXiv:2306.16028}, June 2023.
\newblock \doi{10.48550/arXiv.2306.16028}.

\bibitem[Gil-Fuster et~al.(2024)Gil-Fuster, Gyurik, Pérez-Salinas, and
  Dunjko]{Gil-Fuster2024}
Elies Gil-Fuster, Casper Gyurik, Adrián Pérez-Salinas, and Vedran Dunjko.
\newblock On the relation between trainability and dequantization of
  variational quantum learning models.
\newblock \emph{arXiv:2406.07072}, June 2024.
\newblock \doi{10.48550/arXiv.2406.07072}.

\bibitem[Zimborás et~al.(2025)Zimborás, Koczor, Holmes, Borrelli, Gilyén,
  Huang, Cai, Acín, Aolita, Banchi, Brandão, Cavalcanti, Cubitt, Filippov,
  García-Pérez, Goold, Kálmán, Kyoseva, Rossi, Sokolov, Tavernelli, and
  Maniscalco]{zimboras2025mythsquantumcomputationfault}
Zoltán Zimborás, Bálint Koczor, Zoë Holmes, Elsi-Mari Borrelli, András
  Gilyén, Hsin-Yuan Huang, Zhenyu Cai, Antonio Acín, Leandro Aolita, Leonardo
  Banchi, Fernando G. S.~L. Brandão, Daniel Cavalcanti, Toby Cubitt, Sergey~N.
  Filippov, Guillermo García-Pérez, John Goold, Orsolya Kálmán, Elica
  Kyoseva, Matteo A.~C. Rossi, Boris Sokolov, Ivano Tavernelli, and Sabrina
  Maniscalco.
\newblock Myths around quantum computation before full fault tolerance: What
  no-go theorems rule out and what they don't.
\newblock \emph{arXiv:2501.05694}, January 2025.
\newblock \doi{10.48550/arXiv.2501.05694}.

\bibitem[Stoudenmire and Schwab(2016)]{Stoudenmire2017}
E.~Miles Stoudenmire and David~J. Schwab.
\newblock Supervised learning with quantum-inspired tensor networks.
\newblock In D.~Lee, M.~Sugiyama, U.~Luxburg, I.~Guyon, and R.~Garnett,
  editors, \emph{Advances in Neural Information Processing Systems}, volume~29,
  pages 4799--4807. Curran Associates, Inc., December 2016.
\newblock URL
  \url{https://proceedings.neurips.cc/paper_files/paper/2016/file/5314b9674c86e3f9d1ba25ef9bb32895-Paper.pdf}.

\bibitem[Pérez-Salinas et~al.(2020)Pérez-Salinas, Cervera-Lierta, Gil-Fuster,
  and Latorre]{P_rez_Salinas_2020}
Adrián Pérez-Salinas, Alba Cervera-Lierta, Elies Gil-Fuster, and José~I.
  Latorre.
\newblock Data re-uploading for a universal quantum classifier.
\newblock \emph{Quantum}, 4:\penalty0 226, February 2020.
\newblock ISSN 2521-327X.
\newblock \doi{10.22331/q-2020-02-06-226}.

\bibitem[Goto et~al.(2021)Goto, Tran, and Nakajima]{Goto_2021}
Takahiro Goto, Quoc~Hoan Tran, and Kohei Nakajima.
\newblock Universal approximation property of quantum machine learning models
  in quantum-enhanced feature spaces.
\newblock \emph{Physical Review Letters}, 127\penalty0 (9), August 2021.
\newblock ISSN 1079-7114.
\newblock \doi{10.1103/physrevlett.127.090506}.

\bibitem[Schuld et~al.(2021)Schuld, Sweke, and Meyer]{Schuld_2021}
Maria Schuld, Ryan Sweke, and Johannes~Jakob Meyer.
\newblock Effect of data encoding on the expressive power of variational
  quantum-machine-learning models.
\newblock \emph{Physical Review A}, 103\penalty0 (3), March 2021.
\newblock ISSN 2469-9934.
\newblock \doi{10.1103/physreva.103.032430}.

\bibitem[Latorre(2005)]{latorre2005imagecompressionentanglement}
Jose~I. Latorre.
\newblock Image compression and entanglement.
\newblock \emph{arXiv:quant-ph/0510031}, October 2005.
\newblock \doi{10.48550/arXiv.quant-ph/0510031}.

\bibitem[Plesch and Brukner(2011)]{Plesch2011}
Martin Plesch and {\v{C}}aslav Brukner.
\newblock Quantum-state preparation with universal gate decompositions.
\newblock \emph{Physical Review A}, 83:\penalty0 032302, March 2011.
\newblock \doi{10.1103/PhysRevA.83.032302}.

\bibitem[Iten et~al.(2016)Iten, Colbeck, Kukuljan, Home, and
  Christandl]{Iten2016}
Raban Iten, Roger Colbeck, Ivan Kukuljan, Jonathan Home, and Matthias
  Christandl.
\newblock Quantum circuits for isometries.
\newblock \emph{Physical Review A}, 93:\penalty0 032318, March 2016.
\newblock \doi{10.1103/PhysRevA.93.032318}.

\bibitem[Grant et~al.(2018)Grant, Benedetti, Cao, Hallam, Lockhart, Stojevic,
  Green, and Severini]{Grant2018}
Edward Grant, Marcello Benedetti, Shuxiang Cao, Andrew Hallam, Joshua Lockhart,
  Vid Stojevic, Andrew~G. Green, and Simone Severini.
\newblock Hierarchical quantum classifiers.
\newblock \emph{npj Quantum Information}, 4\penalty0 (1), December 2018.
\newblock \doi{10.1038/s41534-018-0116-9}.

\bibitem[Havlíček et~al.(2019)Havlíček, Córcoles, Temme, Harrow, Kandala,
  Chow, and Gambetta]{Havl_ek_2019}
Vojtěch Havlíček, Antonio~D. Córcoles, Kristan Temme, Aram~W. Harrow,
  Abhinav Kandala, Jerry~M. Chow, and Jay~M. Gambetta.
\newblock Supervised learning with quantum-enhanced feature spaces.
\newblock \emph{Nature}, 567\penalty0 (7747):\penalty0 209--212, March 2019.
\newblock ISSN 1476-4687.
\newblock \doi{10.1038/s41586-019-0980-2}.

\bibitem[Schuld and Killoran(2019)]{Schuld_2019b}
Maria Schuld and Nathan Killoran.
\newblock Quantum machine learning in feature {Hilbert} spaces.
\newblock \emph{Physical Review Letters}, 122\penalty0 (4), February 2019.
\newblock ISSN 1079-7114.
\newblock \doi{10.1103/physrevlett.122.040504}.

\bibitem[Schuld et~al.(2020)Schuld, Bocharov, Svore, and Wiebe]{Schuld_2020}
Maria Schuld, Alex Bocharov, Krysta~M. Svore, and Nathan Wiebe.
\newblock Circuit-centric quantum classifiers.
\newblock \emph{Physical Review A}, 101\penalty0 (3), March 2020.
\newblock ISSN 2469-9934.
\newblock \doi{10.1103/physreva.101.032308}.

\bibitem[Bartkiewicz et~al.(2020)Bartkiewicz, Gneiting, Černoch, Jiráková,
  Lemr, and Nori]{Bartkiewicz2020}
Karol Bartkiewicz, Clemens Gneiting, Antonín Černoch, Kateřina Jiráková,
  Karel Lemr, and Franco Nori.
\newblock Experimental kernel-based quantum machine learning in finite feature
  space.
\newblock \emph{Scientific Reports}, 10\penalty0 (1):\penalty0 12356, July
  2020.
\newblock ISSN 2045-2322.
\newblock \doi{10.1038/s41598-020-68911-5}.

\bibitem[Kerenidis and Luongo(2020)]{Kerenidis2020}
Iordanis Kerenidis and Alessandro Luongo.
\newblock Quantum classification of the {MNIST} dataset with {Slow} {Feature}
  {Analysis}.
\newblock \emph{Physical Review A}, 101\penalty0 (6):\penalty0 062327, June
  2020.
\newblock ISSN 2469-9926, 2469-9934.
\newblock \doi{10.1103/PhysRevA.101.062327}.

\bibitem[Chalumuri et~al.(2021)Chalumuri, Kune, and Manoj]{Chalumuri2021}
Avinash Chalumuri, Raghavendra Kune, and B.~S. Manoj.
\newblock A hybrid classical-quantum approach for multi-class classification.
\newblock \emph{Quantum Information Processing}, 20:\penalty0 119, March 2021.
\newblock \doi{10.1007/s11128-021-03029-9}.

\bibitem[Johri et~al.(2021)Johri, Debnath, Mocherla, Singk, Prakash, Kim, and
  Kerenidis]{Johri2021}
Sonika Johri, Shantanu Debnath, Avinash Mocherla, Alexandros Singk, Anupam
  Prakash, Jungsang Kim, and Iordanis Kerenidis.
\newblock Nearest centroid classification on a trapped ion quantum computer.
\newblock \emph{npj Quantum Information}, 7\penalty0 (1):\penalty0 1--11,
  August 2021.
\newblock ISSN 2056-6387.
\newblock \doi{10.1038/s41534-021-00456-5}.

\bibitem[Peters et~al.(2021)Peters, Caldeira, Ho, Leichenauer, Mohseni, Neven,
  Spentzouris, Strain, and Perdue]{Peters2021}
Evan Peters, João Caldeira, Alan Ho, Stefan Leichenauer, Masoud Mohseni,
  Hartmut Neven, Panagiotis Spentzouris, Doug Strain, and Gabriel~N. Perdue.
\newblock Machine learning of high dimensional data on a noisy quantum
  processor.
\newblock \emph{npj Quantum Information}, 7\penalty0 (1):\penalty0 1--5,
  November 2021.
\newblock ISSN 2056-6387.
\newblock \doi{10.1038/s41534-021-00498-9}.

\bibitem[Bokhan et~al.(2022)Bokhan, Mastiukova, Boev, Trubnikov, and
  Fedorov]{Bokhan2022}
Denis Bokhan, Alena~S. Mastiukova, Aleksey~S. Boev, Dmitrii~N. Trubnikov, and
  Aleksey~K. Fedorov.
\newblock Multiclass classification using quantum convolutional neural networks
  with hybrid quantum-classical learning.
\newblock \emph{Frontiers in Physics}, 10:\penalty0 1069985, November 2022.
\newblock ISSN 2296-424X.
\newblock \doi{10.3389/fphy.2022.1069985}.

\bibitem[Hur et~al.(2022)Hur, Kim, and Park]{Hur2022}
Tak Hur, Leeseok Kim, and Daniel~K. Park.
\newblock Quantum convolutional neural network for classical data
  classification.
\newblock \emph{Quantum Machine Intelligence}, 4\penalty0 (1):\penalty0 3,
  February 2022.
\newblock ISSN 2524-4914.
\newblock \doi{10.1007/s42484-021-00061-x}.

\bibitem[Lu et~al.(2024)Lu, Jiao, Wolinski, Kornjača, Hu, Cantu, Liu, Yelin,
  and Wang]{Lu2025}
Jonathan~Z. Lu, Lucy Jiao, Kristina Wolinski, Milan Kornjača, Hong-Ye Hu,
  Sergio Cantu, Fangli Liu, Susanne~F Yelin, and Sheng-Tao Wang.
\newblock Digital–analog quantum learning on rydberg atom arrays.
\newblock \emph{Quantum Science and Technology}, 10\penalty0 (1):\penalty0
  015038, November 2024.
\newblock \doi{10.1088/2058-9565/ad9177}.

\bibitem[Bowles et~al.(2024)Bowles, Ahmed, and
  Schuld]{bowles2024betterclassicalsubtleart}
Joseph Bowles, Shahnawaz Ahmed, and Maria Schuld.
\newblock Better than classical? {T}he subtle art of benchmarking quantum
  machine learning models.
\newblock \emph{arXiv:2403.07059}, March 2024.
\newblock \doi{10.48550/arXiv.2403.07059}.

\bibitem[Zhou et~al.(2021)Zhou, Liu, and Du]{Zhou2021}
Nan-Run Zhou, Yu-Ling Liu, Xiu-Xun~Chen, and Ni-Suo Du.
\newblock Quantum k-nearest-neighbor image classification algorithm based on
  {K-L} transform.
\newblock \emph{International Journal of Theoretical Physics}, 60\penalty0
  (3):\penalty0 1209--1224, March 2021.
\newblock ISSN 1572-9575.
\newblock \doi{10.1007/s10773-021-04747-7}.

\bibitem[Riaz et~al.(2023)Riaz, Abdulla, Suzuki, Ganguly, Deo, and
  Hopkins]{Riaz2023}
Farina Riaz, Shahab Abdulla, Hajime Suzuki, Srinjoy Ganguly, Ravinesh~C. Deo,
  and Susan Hopkins.
\newblock Accurate image multi-class classification neural network model with
  quantum entanglement approach.
\newblock \emph{Sensors}, 23\penalty0 (5):\penalty0 2753, March 2023.
\newblock ISSN 1424-8220.
\newblock \doi{10.3390/s23052753}.

\bibitem[Baek et~al.(2023)Baek, Park, and Kim]{Baek2023}
Hankyul Baek, Soohyun Park, and Joongheon Kim.
\newblock Logarithmic dimension reduction for quantum neural networks.
\newblock In \emph{Proceedings of the 32nd ACM International Conference on
  Information and Knowledge Management}, CIKM '23, pages 3738--3742, New York,
  NY, USA, October 2023. Association for Computing Machinery.
\newblock ISBN 9798400701245.
\newblock \doi{10.1145/3583780.3615240}.

\bibitem[Khatun and Usman(2024)]{Khatun2024}
Amena Khatun and Muhammad Usman.
\newblock Quantum transfer learning with adversarial robustness for
  classification of high-resolution image datasets.
\newblock \emph{Advanced Quantum Technologies}, 8\penalty0 (1):\penalty0
  2400268, September 2024.
\newblock \doi{https://doi.org/10.1002/qute.202400268}.

\bibitem[Monbroussou et~al.(2024)Monbroussou, Landman, Wang, Grilo, and
  Kashefi]{Monbroussou2024}
Léo Monbroussou, Jonas Landman, Letao Wang, Alex~B. Grilo, and Elham Kashefi.
\newblock Subspace preserving quantum convolutional neural network
  architectures.
\newblock \emph{arXiv:2409.18918}, September 2024.
\newblock \doi{10.48550/arXiv.2409.18918}.

\bibitem[Russakovsky et~al.(2015)Russakovsky, Deng, Su, Krause, Satheesh, Ma,
  Huang, Karpathy, Khosla, Bernstein, Berg, and
  Fei-Fei]{russakovsky2015imagenetlargescalevisual}
Olga Russakovsky, Jia Deng, Hao Su, Jonathan Krause, Sanjeev Satheesh, Sean Ma,
  Zhiheng Huang, Andrej Karpathy, Aditya Khosla, Michael Bernstein,
  Alexander~C. Berg, and Li~Fei-Fei.
\newblock {ImageNet} large scale visual recognition challenge.
\newblock \emph{arXiv:1409.0575}, January 2015.
\newblock \doi{10.48550/arXiv.1409.0575}.

\bibitem[Krizhevsky et~al.(2012)Krizhevsky, Sutskever, and
  Hinton]{NIPS2012_c399862d}
Alex Krizhevsky, Ilya Sutskever, and Geoffrey~E Hinton.
\newblock {ImageNet} classification with deep convolutional neural networks.
\newblock In F.~Pereira, C.J. Burges, L.~Bottou, and K.Q. Weinberger, editors,
  \emph{Advances in Neural Information Processing Systems}, volume~25. Curran
  Associates, Inc., 2012.
\newblock URL
  \url{https://proceedings.neurips.cc/paper_files/paper/2012/file/c399862d3b9d6b76c8436e924a68c45b-Paper.pdf}.

\bibitem[Moussa et~al.(2022)Moussa, van Rijn, Bäck, and Dunjko]{Moussa2022}
Charles Moussa, Jan~N. van Rijn, Thomas Bäck, and Vedran Dunjko.
\newblock \emph{Hyperparameter Importance of Quantum Neural Networks Across
  Small Datasets}, pages 32--46.
\newblock Springer Nature Switzerland, November 2022.
\newblock ISBN 9783031188404.
\newblock \doi{10.1007/978-3-031-18840-4_3}.

\bibitem[Kashif et~al.(2023)Kashif, Rashid, Al-Kuwari, and
  Shafique]{Kashif2023}
Muhammad Kashif, Muhammad Rashid, Saif Al-Kuwari, and Muhammad Shafique.
\newblock Alleviating barren plateaus in parameterized quantum machine learning
  circuits: Investigating advanced parameter initialization strategies.
\newblock \emph{arXiv:2311.13218}, November 2023.
\newblock \doi{10.48550/arXiv.2311.13218}.

\bibitem[Fin\v{z}gar et~al.(2022)Fin\v{z}gar, Ross, Holscher, Klepsch, and
  Luckow]{Finzgar_2022}
Jernej~Rudi Fin\v{z}gar, Philipp Ross, Leonhard Holscher, Johannes Klepsch, and
  Andre Luckow.
\newblock {QUARK}: A framework for quantum computing application benchmarking.
\newblock In \emph{2022 {IEEE} International Conference on Quantum Computing
  and Engineering ({QCE})}. {IEEE}, 9 2022.
\newblock \doi{10.1109/qce53715.2022.00042}.

\bibitem[Kiwit et~al.(2023)Kiwit, Marso, Ross, Riofrío, Klepsch, and
  Luckow]{kiwit2023applicationoriented}
Florian~J. Kiwit, Marwa Marso, Philipp Ross, Carlos~A. Riofrío, Johannes
  Klepsch, and Andre Luckow.
\newblock Application-oriented benchmarking of quantum generative learning
  using {QUARK}.
\newblock In \emph{2023 IEEE International Conference on Quantum Computing and
  Engineering (QCE)}, volume~01, pages 475--484, 2023.
\newblock \doi{10.1109/QCE57702.2023.00061}.

\bibitem[Kiwit et~al.(2024)Kiwit, Wolf, Marso, Ross, Lorenz, Riofrío, and
  Luckow]{Kiwit_2024}
Florian~J. Kiwit, Maximilian~A. Wolf, Marwa Marso, Philipp Ross, Jeanette~M.
  Lorenz, Carlos~A. Riofrío, and Andre Luckow.
\newblock Benchmarking quantum generative learning: A study on scalability and
  noise resilience using {QUARK}.
\newblock \emph{KI - Künstliche Intelligenz}, August 2024.
\newblock ISSN 1610-1987.
\newblock \doi{10.1007/s13218-024-00864-7}.

\bibitem[Quetschlich et~al.(2023)Quetschlich, Burgholzer, and
  Wille]{Quetschlich_2023}
Nils Quetschlich, Lukas Burgholzer, and Robert Wille.
\newblock {MQT Bench}: Benchmarking software and design automation tools for
  quantum computing.
\newblock \emph{Quantum}, 7:\penalty0 1062, July 2023.
\newblock ISSN 2521-327X.
\newblock \doi{10.22331/q-2023-07-20-1062}.

\bibitem[Dilip et~al.(2022)Dilip, Liu, Smith, and Pollmann]{Dilip2022}
Rohit Dilip, Yu-Jie Liu, Adam Smith, and Frank Pollmann.
\newblock Data compression for quantum machine learning.
\newblock \emph{Physical Review Research}, 4:\penalty0 043007, October 2022.
\newblock \doi{10.1103/PhysRevResearch.4.043007}.

\bibitem[Jobst et~al.(2024)Jobst, Shen, Riofrío, Shishenina, and
  Pollmann]{jobst2023efficientmpsrepresentationsquantum}
Bernhard Jobst, Kevin Shen, Carlos~A. Riofrío, Elvira Shishenina, and Frank
  Pollmann.
\newblock Efficient {MPS} representations and quantum circuits from the
  {Fourier} modes of classical image data.
\newblock \emph{Quantum}, 8:\penalty0 1544, December 2024.
\newblock ISSN 2521-327X.
\newblock \doi{10.22331/q-2024-12-03-1544}.

\bibitem[Iaconis and Johri(2023)]{iaconis2023tensornetworkbasedefficient}
Jason Iaconis and Sonika Johri.
\newblock Tensor network based efficient quantum data loading of images.
\newblock \emph{arXiv:2310.05897}, October 2023.
\newblock \doi{10.48550/arXiv.2310.05897}.

\bibitem[Shen et~al.(2024)Shen, Jobst, Shishenina, and
  Pollmann]{shen2024classification}
Kevin Shen, Bernhard Jobst, Elvira Shishenina, and Frank Pollmann.
\newblock Classification of the {Fashion-MNIST} dataset on a quantum computer.
\newblock \emph{arXiv:2403.02405}, March 2024.
\newblock \doi{10.48550/arXiv.2403.02405}.

\bibitem[West et~al.(2024)West, Nakhl, Heredge, Creevey, Hollenberg, Sevior,
  and Usman]{maxwell_2024}
Maxwell~T. West, Azar~C. Nakhl, Jamie Heredge, Floyd~M. Creevey, Lloyd C.~L.
  Hollenberg, Martin Sevior, and Muhammad Usman.
\newblock Drastic circuit depth reductions with preserved adversarial
  robustness by approximate encoding for quantum machine learning.
\newblock \emph{Intelligent Computing}, 3:\penalty0 0100, 2024.
\newblock \doi{10.34133/icomputing.0100}.

\bibitem[Lecun et~al.(1998)Lecun, Bottou, Bengio, and Haffner]{Lecun1998}
Y.~Lecun, L.~Bottou, Y.~Bengio, and P.~Haffner.
\newblock Gradient-based learning applied to document recognition.
\newblock \emph{Proceedings of the IEEE}, 86\penalty0 (11):\penalty0
  2278--2324, November 1998.
\newblock \doi{10.1109/5.726791}.

\bibitem[Deng(2012)]{deng2012mnist}
Li~Deng.
\newblock The {MNIST} database of handwritten digit images for machine learning
  research.
\newblock \emph{IEEE Signal Processing Magazine}, 29\penalty0 (6):\penalty0
  141--142, November 2012.
\newblock \doi{10.1109/MSP.2012.2211477}.

\bibitem[Xiao et~al.(2017)Xiao, Rasul, and Vollgraf]{FashionMNIST}
Han Xiao, Kashif Rasul, and Roland Vollgraf.
\newblock Fashion-{MNIST}: a novel image dataset for benchmarking machine
  learning algorithms.
\newblock \emph{arXiv:1708.07747}, August 2017.
\newblock \doi{10.48550/arXiv.1708.07747}.
\newblock Dataset available at
  \url{https://github.com/zalandoresearch/fashion-mnist}.

\bibitem[Krizhevsky(2009)]{CIFAR10}
Alex Krizhevsky.
\newblock Learning multiple layers of features from tiny images, April 2009.
\newblock URL
  \url{https://www.cs.toronto.edu/~kriz/learning-features-2009-TR.pdf}.
\newblock Dataset available at
  \url{https://www.cs.toronto.edu/~kriz/cifar.html}.

\bibitem[Howard(2019)]{imagenette}
Jeremy Howard.
\newblock Imagenette, December 2019.
\newblock Dataset available at \url{https://github.com/fastai/imagenette}.

\bibitem[Bergholm et~al.(2022)Bergholm, Izaac, Schuld, Gogolin, Ahmed, Ajith,
  Alam, Alonso-Linaje, AkashNarayanan, Asadi, Arrazola, Azad, Banning, Blank,
  Bromley, Cordier, Ceroni, Delgado, Matteo, Dusko, Garg, Guala, Hayes, Hill,
  Ijaz, Isacsson, Ittah, Jahangiri, Jain, Jiang, Khandelwal, Kottmann, Lang,
  Lee, Loke, Lowe, McKiernan, Meyer, Montañez-Barrera, Moyard, Niu, O'Riordan,
  Oud, Panigrahi, Park, Polatajko, Quesada, Roberts, Sá, Schoch, Shi, Shu,
  Sim, Singh, Strandberg, Soni, Száva, Thabet, Vargas-Hernández, Vincent,
  Vitucci, Weber, Wierichs, Wiersema, Willmann, Wong, Zhang, and
  Killoran]{bergholm2022pennylaneautomaticdifferentiationhybrid}
Ville Bergholm, Josh Izaac, Maria Schuld, Christian Gogolin, Shahnawaz Ahmed,
  Vishnu Ajith, M.~Sohaib Alam, Guillermo Alonso-Linaje, B.~AkashNarayanan, Ali
  Asadi, Juan~Miguel Arrazola, Utkarsh Azad, Sam Banning, Carsten Blank,
  Thomas~R Bromley, Benjamin~A. Cordier, Jack Ceroni, Alain Delgado, Olivia~Di
  Matteo, Amintor Dusko, Tanya Garg, Diego Guala, Anthony Hayes, Ryan Hill,
  Aroosa Ijaz, Theodor Isacsson, David Ittah, Soran Jahangiri, Prateek Jain,
  Edward Jiang, Ankit Khandelwal, Korbinian Kottmann, Robert~A. Lang, Christina
  Lee, Thomas Loke, Angus Lowe, Keri McKiernan, Johannes~Jakob Meyer, J.~A.
  Montañez-Barrera, Romain Moyard, Zeyue Niu, Lee~James O'Riordan, Steven Oud,
  Ashish Panigrahi, Chae-Yeun Park, Daniel Polatajko, Nicolás Quesada, Chase
  Roberts, Nahum Sá, Isidor Schoch, Borun Shi, Shuli Shu, Sukin Sim, Arshpreet
  Singh, Ingrid Strandberg, Jay Soni, Antal Száva, Slimane Thabet, Rodrigo~A.
  Vargas-Hernández, Trevor Vincent, Nicola Vitucci, Maurice Weber, David
  Wierichs, Roeland Wiersema, Moritz Willmann, Vincent Wong, Shaoming Zhang,
  and Nathan Killoran.
\newblock Pennylane: Automatic differentiation of hybrid quantum-classical
  computations.
\newblock \emph{arXiv:1811.04968}, July 2022.
\newblock \doi{10.48550/arXiv.1811.04968}.
\newblock Code available at \url{https://github.com/PennyLaneAI/pennylane}.

\bibitem[Lubasch et~al.(2018)Lubasch, Moinier, and Jaksch]{Lubasch2018}
Michael Lubasch, Pierre Moinier, and Dieter Jaksch.
\newblock Multigrid renormalization.
\newblock \emph{Journal of Computational Physics}, 372:\penalty0 587--602, June
  2018.
\newblock ISSN 0021-9991.
\newblock \doi{10.1016/j.jcp.2018.06.065}.

\bibitem[Lubasch et~al.(2020)Lubasch, Joo, Moinier, Kiffner, and
  Jaksch]{Lubasch2020}
Michael Lubasch, Jaewoo Joo, Pierre Moinier, Martin Kiffner, and Dieter Jaksch.
\newblock Variational quantum algorithms for nonlinear problems.
\newblock \emph{Physical Review A}, 101:\penalty0 010301, January 2020.
\newblock \doi{10.1103/PhysRevA.101.010301}.

\bibitem[Gourianov et~al.(2022)Gourianov, Lubasch, Dolgov, van~den Berg,
  Babaee, Givi, Kiffner, and Jaksch]{Gourianov2022}
Nikita Gourianov, Michael Lubasch, Sergey Dolgov, Quincy~Y. van~den Berg,
  Hessam Babaee, Peyman Givi, Martin Kiffner, and Dieter Jaksch.
\newblock A quantum-inspired approach to exploit turbulence structures.
\newblock \emph{Nature Computational Science}, 2:\penalty0 30--37, January
  2022.
\newblock \doi{10.1038/s43588-021-00181-1}.

\bibitem[Hölscher et~al.(2025)Hölscher, Rao, Müller, Klepsch, Luckow,
  Stollenwerk, and Wilhelm]{Hoelscher2025}
Leonhard Hölscher, Pooja Rao, Lukas Müller, Johannes Klepsch, Andre Luckow,
  Tobias Stollenwerk, and Frank~K. Wilhelm.
\newblock Quantum-inspired fluid simulation of two-dimensional turbulence with
  {GPU} acceleration.
\newblock \emph{Physical Review Research}, 7:\penalty0 013112, January 2025.
\newblock \doi{10.1103/physrevresearch.7.013112}.

\bibitem[Ritter et~al.(2024)Ritter, N\'u\~nez Fern\'andez, Wallerberger, von
  Delft, Shinaoka, and Waintal]{Ritter2024}
Marc~K. Ritter, Yuriel N\'u\~nez Fern\'andez, Markus Wallerberger, Jan von
  Delft, Hiroshi Shinaoka, and Xavier Waintal.
\newblock Quantics tensor cross interpolation for high-resolution parsimonious
  representations of multivariate functions.
\newblock \emph{Physical Review Letters}, 132:\penalty0 056501, January 2024.
\newblock \doi{10.1103/PhysRevLett.132.056501}.

\bibitem[Amankwah et~al.(2022)Amankwah, Camps, Bethel, Beeumen, and
  Perciano]{Amankwah2022}
Mercy~G. Amankwah, Daan Camps, E.~Wes Bethel, Roel~Van Beeumen, and Talita
  Perciano.
\newblock Quantum pixel representations and compression for ${N}$-dimensional
  images.
\newblock \emph{Scientific Reports}, 12:\penalty0 7712, May 2022.
\newblock \doi{10.1038/s41598-022-11024-y}.

\bibitem[Le et~al.(2011{\natexlab{a}})Le, Dong, and Hirota]{Le2011}
Phuc~Q. Le, Fangyan Dong, and Kaoru Hirota.
\newblock A flexible representation of quantum images for polynomial
  preparation, image compression, and processing operations.
\newblock \emph{Quantum Information Processing}, 10:\penalty0 63--84, February
  2011{\natexlab{a}}.
\newblock \doi{10.1007/s11128-010-0177-y}.

\bibitem[Le et~al.(2011{\natexlab{b}})Le, Iliyasu, Dong, and Hirota]{Le2011_2}
Phuc~Q. Le, Abdullahi~M. Iliyasu, Fangyan Dong, and Kaoru Hirota.
\newblock \emph{A Flexible Representation and Invertible Transformations for
  Images on Quantum Computers}, volume 372, pages 179--202.
\newblock Springer Berlin Heidelberg, Berlin, Heidelberg, 2011{\natexlab{b}}.
\newblock ISBN 978-3-642-11739-8.
\newblock \doi{10.1007/978-3-642-11739-8_9}.

\bibitem[Savostyanov and Oseledets(2011)]{Savostyanov2011}
Dmitry~V. Savostyanov and Ivan Oseledets.
\newblock Fast adaptive interpolation of multi-dimensional arrays in tensor
  train format.
\newblock In \emph{The 2011 International Workshop on Multidimensional (nD)
  Systems}, pages 1--8, September 2011.
\newblock \doi{10.1109/nDS.2011.6076873}.

\bibitem[Savostyanov(2014)]{Savostyanov2014}
Dmitry~V. Savostyanov.
\newblock Quasioptimality of maximum-volume cross interpolation of tensors.
\newblock \emph{Linear Algebra and its Applications}, 458:\penalty0 217--244,
  June 2014.
\newblock ISSN 0024-3795.
\newblock \doi{10.1016/j.laa.2014.06.006}.

\bibitem[Dolgov and Savostyanov(2020)]{Dolgov2020}
Sergey Dolgov and Dmitry~V. Savostyanov.
\newblock Parallel cross interpolation for high-precision calculation of
  high-dimensional integrals.
\newblock \emph{Computer Physics Communications}, 246:\penalty0 106869, January
  2020.
\newblock ISSN 0010-4655.
\newblock \doi{10.1016/j.cpc.2019.106869}.

\bibitem[N\'u\~nez Fern\'andez et~al.(2022)N\'u\~nez Fern\'andez, Jeannin,
  Dumitrescu, Kloss, Kaye, Parcollet, and Waintal]{Nunez-Fernandez2022}
Yuriel N\'u\~nez Fern\'andez, Matthieu Jeannin, Philipp~T. Dumitrescu, Thomas
  Kloss, Jason Kaye, Olivier Parcollet, and Xavier Waintal.
\newblock Learning {F}eynman diagrams with tensor trains.
\newblock \emph{Physical Review X}, 12:\penalty0 041018, November 2022.
\newblock \doi{10.1103/PhysRevX.12.041018}.

\bibitem[Fernández et~al.(2024)Fernández, Ritter, Jeannin, Li, Kloss, Louvet,
  Terasaki, Parcollet, von Delft, Shinaoka, and Waintal]{Nunez-Fernandez2024}
Yuriel~Núñez Fernández, Marc~K. Ritter, Matthieu Jeannin, Jheng-Wei Li,
  Thomas Kloss, Thibaud Louvet, Satoshi Terasaki, Olivier Parcollet, Jan von
  Delft, Hiroshi Shinaoka, and Xavier Waintal.
\newblock Learning tensor networks with tensor cross interpolation: new
  algorithms and libraries.
\newblock \emph{arXiv:2407.02454}, July 2024.
\newblock \doi{https://arxiv.org/abs/2407.02454}.

\bibitem[Sun et~al.(2011)Sun, Le, Iliyasu, Yan, Garcia, Dong, and
  Hirota]{Sun2011}
Bo~Sun, Phuc~Q. Le, Abdullah~M. Iliyasu, Fei Yan, J.~Adrian Garcia, Fangyan
  Dong, and Kaoru Hirota.
\newblock A multi-channel representation for images on quantum computers using
  the {RGB}$\alpha$ color space.
\newblock In \emph{2011 IEEE 7th International Symposium on Intelligent Signal
  Processing}, pages 1--6, October 2011.
\newblock \doi{10.1109/WISP.2011.6051718}.

\bibitem[Sun et~al.(2013)Sun, Iliyasu, Yan, Dong, and Hirota]{Sun2013}
Bo~Sun, Abdullah~M. Iliyasu, Fei Yan, Fangyan Dong, and Kaoru Hirota.
\newblock An {RGB} multi-channel representation for images on quantum
  computers.
\newblock \emph{Journal of Advanced Computational Intelligence and Intelligent
  Informatics}, 17\penalty0 (3):\penalty0 404--417, March 2013.
\newblock \doi{10.20965/jaciii.2013.p0404}.

\bibitem[Schuld(2021)]{schuld2021supervisedquantummachinelearning}
Maria Schuld.
\newblock Supervised quantum machine learning models are kernel methods.
\newblock \emph{arXiv:2101.11020}, January 2021.
\newblock \doi{10.48550/arXiv.2101.11020}.

\bibitem[Holmes et~al.(2023)Holmes, Coble, Sornborger, and
  Suba\ifmmode\mbox{\c{s}}\else\c{s}\fi{}\ifmmode\imath\else\i\fi{}]{Holmes2023}
Zo\"e Holmes, Nolan~J. Coble, Andrew~T. Sornborger, and
  Yi\ifmmode\breve{g}\else\u{g}\fi{}it
  Suba\ifmmode\mbox{\c{s}}\else\c{s}\fi{}\ifmmode\imath\else\i\fi{}.
\newblock Nonlinear transformations in quantum computation.
\newblock \emph{Physical Review Research}, 5:\penalty0 013105, February 2023.
\newblock \doi{10.1103/PhysRevResearch.5.013105}.

\bibitem[Sch\"on et~al.(2005)Sch\"on, Solano, Verstraete, Cirac, and
  Wolf]{Schoen2005}
Christian Sch\"on, Enrique Solano, Frank Verstraete, J.~Ignacio Cirac, and
  Michael~M. Wolf.
\newblock Sequential generation of entangled multiqubit states.
\newblock \emph{Physical Review Letters}, 95:\penalty0 110503, September 2005.
\newblock \doi{10.1103/PhysRevLett.95.110503}.

\bibitem[Sch\"on et~al.(2007)Sch\"on, Hammerer, Wolf, Cirac, and
  Solano]{Schoen2007}
Christian Sch\"on, Klemens Hammerer, Michael~M. Wolf, J.~Ignacio Cirac, and
  Enrique Solano.
\newblock Sequential generation of matrix-product states in cavity {QED}.
\newblock \emph{Physical Review A}, 75:\penalty0 032311, March 2007.
\newblock \doi{10.1103/PhysRevA.75.032311}.

\bibitem[Smith et~al.(2022)Smith, Jobst, Green, and Pollmann]{Smith2022}
Adam Smith, Bernhard Jobst, Andrew~G. Green, and Frank Pollmann.
\newblock Crossing a topological phase transition with a quantum computer.
\newblock \emph{Physical Review Research}, 4:\penalty0 L022020, April 2022.
\newblock \doi{10.1103/PhysRevResearch.4.L022020}.

\bibitem[Lin et~al.(2021)Lin, Dilip, Green, Smith, and Pollmann]{Lin2021}
Sheng-Hsuan Lin, Rohit Dilip, Andrew~G. Green, Adam Smith, and Frank Pollmann.
\newblock Real- and imaginary-time evolution with compressed quantum circuits.
\newblock \emph{PRX Quantum}, 2:\penalty0 010342, March 2021.
\newblock \doi{10.1103/PRXQuantum.2.010342}.

\bibitem[Barratt et~al.(2021)Barratt, Dborin, Bal, Stojevic, Pollmann, and
  Green]{Barratt2021}
Fergus Barratt, James Dborin, Matthias Bal, Vid Stojevic, Frank Pollmann, and
  Andrew~G. Green.
\newblock Parallel quantum simulation of large systems on small {NISQ}
  computers.
\newblock \emph{npj Quantum Information}, 7\penalty0 (1):\penalty0 79, May
  2021.
\newblock ISSN 2056-6387.
\newblock \doi{10.1038/s41534-021-00420-3}.

\bibitem[Ran(2020)]{PhysRevA.101.032310}
Shi-Ju Ran.
\newblock Encoding of matrix product states into quantum circuits of one- and
  two-qubit gates.
\newblock \emph{Physical Review A}, 101:\penalty0 032310, Mar 2020.
\newblock \doi{10.1103/PhysRevA.101.032310}.

\bibitem[Rudolph et~al.(2023{\natexlab{a}})Rudolph, Chen, Miller, Acharya, and
  Perdomo-Ortiz]{Rudolph2023_2}
Manuel~S. Rudolph, Jing Chen, Jacob Miller, Atithi Acharya, and Alejandro
  Perdomo-Ortiz.
\newblock Decomposition of matrix product states into shallow quantum circuits.
\newblock \emph{Quantum Science and Technology}, 9\penalty0 (1):\penalty0
  015012, nov 2023{\natexlab{a}}.
\newblock \doi{10.1088/2058-9565/ad04e6}.

\bibitem[Ben-Dov et~al.(2024)Ben-Dov, Shnaiderov, Makmal, and
  Dalla~Torre]{Ben-Dov2024}
Matan Ben-Dov, David Shnaiderov, Adi Makmal, and Emanuele~G. Dalla~Torre.
\newblock Approximate encoding of quantum states using shallow circuits.
\newblock \emph{npj Quantum Information}, 10\penalty0 (1), July 2024.
\newblock ISSN 2056-6387.
\newblock \doi{10.1038/s41534-024-00858-1}.

\bibitem[Schollwöck(2011)]{SCHOLLWOCK201196}
Ulrich Schollwöck.
\newblock The density-matrix renormalization group in the age of matrix product
  states.
\newblock \emph{Annals of Physics}, 326\penalty0 (1):\penalty0 96--192, 2011.
\newblock ISSN 0003-4916.
\newblock \doi{https://doi.org/10.1016/j.aop.2010.09.012}.
\newblock January 2011 Special Issue.

\bibitem[Orús(2014)]{Or_s_2014}
Román Orús.
\newblock A practical introduction to tensor networks: Matrix product states
  and projected entangled pair states.
\newblock \emph{Annals of Physics}, 349:\penalty0 117--158, October 2014.
\newblock ISSN 0003-4916.
\newblock \doi{10.1016/j.aop.2014.06.013}.

\bibitem[Wei et~al.(2022)Wei, Malz, and Cirac]{Wei2022}
Zhi-Yuan Wei, Daniel Malz, and J.~Ignacio Cirac.
\newblock Sequential generation of projected entangled-pair states.
\newblock \emph{Physical Review Letters}, 128:\penalty0 010607, January 2022.
\newblock \doi{10.1103/PhysRevLett.128.010607}.

\bibitem[Bohun et~al.(2024)Bohun, Lukin, Luhanko, Korpas, Brouwer, Maksymenko,
  and Koch-Janusz]{Bohun2024}
Vladyslav Bohun, Illia Lukin, Mykola Luhanko, Georgios Korpas, Philippe J.
  S.~De Brouwer, Mykola Maksymenko, and Maciej Koch-Janusz.
\newblock Scalable and shallow quantum circuits encoding probability
  distributions informed by asymptotic entanglement analysis.
\newblock \emph{arXiv:2412.05202}, December 2024.
\newblock \doi{10.48550/arXiv.2412.05202}.

\bibitem[Shende et~al.(2004{\natexlab{a}})Shende, Bullock, and
  Markov]{PhysRevA.70.012310}
Vivek~V. Shende, Stephen~S. Bullock, and Igor~L. Markov.
\newblock Recognizing small-circuit structure in two-qubit operators.
\newblock \emph{Physical Review A}, 70:\penalty0 012310, Jul
  2004{\natexlab{a}}.
\newblock \doi{10.1103/PhysRevA.70.012310}.

\bibitem[Shende et~al.(2004{\natexlab{b}})Shende, Markov, and
  Bullock]{PhysRevA.69.062321}
Vivek~V. Shende, Igor~L. Markov, and Stephen~S. Bullock.
\newblock Minimal universal two-qubit controlled-{NOT}-based circuits.
\newblock \emph{Physical Review A}, 69:\penalty0 062321, Jun
  2004{\natexlab{b}}.
\newblock \doi{10.1103/PhysRevA.69.062321}.

\bibitem[Wei and Di(2012)]{Wei:2012zgm}
Hai-Rui Wei and Yao-Min Di.
\newblock {Decomposition of orthogonal matrix and synthesis of two-qubit and
  three-qubit orthogonal gates}.
\newblock \emph{Quant. Inf. Comput.}, 12\penalty0 (3-4):\penalty0 0262--0270,
  2012.
\newblock \doi{10.26421/QIC12.3-4-6}.

\bibitem[Evenbly and Vidal(2009)]{Evenbly2009}
G.~Evenbly and G.~Vidal.
\newblock Algorithms for entanglement renormalization.
\newblock \emph{Physical Review B}, 79:\penalty0 144108, Apr 2009.
\newblock \doi{10.1103/PhysRevB.79.144108}.

\bibitem[Nocedal and Wright(2006)]{NoceWrig06}
Jorge Nocedal and Stephen~J. Wright.
\newblock \emph{Numerical Optimization}.
\newblock Springer, New York, NY, USA, 2nd edition, July 2006.
\newblock ISBN 978-0-387-30303-1.
\newblock \doi{10.1007/978-0-387-40065-5}.

\bibitem[Hauru et~al.(2021)Hauru, Damme, and Haegeman]{Hauru_2021}
Markus Hauru, Maarten~Van Damme, and Jutho Haegeman.
\newblock {Riemannian optimization of isometric tensor networks}.
\newblock \emph{SciPost Physics}, 10:\penalty0 040, 2021.
\newblock \doi{10.21468/SciPostPhys.10.2.040}.

\bibitem[Dborin et~al.(2022)Dborin, Barratt, Wimalaweera, Wright, and
  Green]{Dborin2022}
James Dborin, Fergus Barratt, Vinul Wimalaweera, Lewis Wright, and Andrew~G.
  Green.
\newblock Matrix product state pre-training for quantum machine learning.
\newblock \emph{Quantum Science and Technology}, 7\penalty0 (3):\penalty0
  035014, May 2022.
\newblock \doi{10.1088/2058-9565/ac7073}.

\bibitem[Zhang et~al.(2022)Zhang, Liu, Hsieh, and Tao]{Zhang2022}
Kaining Zhang, Liu Liu, Min-Hsiu Hsieh, and Dacheng Tao.
\newblock Escaping from the barren plateau via gaussian initializations in deep
  variational quantum circuits.
\newblock In Alice~H. Oh, Alekh Agarwal, Danielle Belgrave, and Kyunghyun Cho,
  editors, \emph{Advances in Neural Information Processing Systems}, October
  2022.
\newblock URL \url{https://openreview.net/forum?id=jXgbJdQ2YIy}.

\bibitem[Rudolph et~al.(2023{\natexlab{b}})Rudolph, Miller, Motlagh, Chen,
  Acharya, and Perdomo-Ortiz]{Rudolph2023}
Manuel~S. Rudolph, Jacob Miller, Danial Motlagh, Jing Chen, Atithi Acharya, and
  Alejandro Perdomo-Ortiz.
\newblock Synergistic pretraining of parametrized quantum circuits via tensor
  networks.
\newblock \emph{Nature Communications}, 14\penalty0 (1), December
  2023{\natexlab{b}}.
\newblock ISSN 2041-1723.
\newblock \doi{10.1038/s41467-023-43908-6}.

\bibitem[Park and Killoran(2024)]{Park2024}
Chae-Yeun Park and Nathan Killoran.
\newblock Hamiltonian variational ansatz without barren plateaus.
\newblock \emph{{Quantum}}, 8:\penalty0 1239, February 2024.
\newblock ISSN 2521-327X.
\newblock \doi{10.22331/q-2024-02-01-1239}.

\bibitem[Wang et~al.(2024)Wang, Qi, Ferrie, and Dong]{Wang2024}
Yabo Wang, Bo~Qi, Chris Ferrie, and Daoyi Dong.
\newblock Trainability enhancement of parameterized quantum circuits via
  reduced-domain parameter initialization.
\newblock \emph{Physical Review Applied}, 22:\penalty0 054005, November 2024.
\newblock \doi{10.1103/PhysRevApplied.22.054005}.

\bibitem[Puig et~al.(2025)Puig, Drudis, Thanasilp, and Holmes]{Puig2025}
Ricard Puig, Marc Drudis, Supanut Thanasilp, and Zo\"e Holmes.
\newblock Variational quantum simulation: A case study for understanding warm
  starts.
\newblock \emph{PRX Quantum}, 6:\penalty0 010317, January 2025.
\newblock \doi{10.1103/PRXQuantum.6.010317}.

\bibitem[Mhiri et~al.(2025)Mhiri, Puig, Lerch, Rudolph, Chotibut, Thanasilp,
  and Holmes]{Mhiri2025}
Hela Mhiri, Ricard Puig, Sacha Lerch, Manuel~S. Rudolph, Thiparat Chotibut,
  Supanut Thanasilp, and Zoë Holmes.
\newblock A unifying account of warm start guarantees for patches of quantum
  landscapes.
\newblock \emph{arXiv:2502.07889}, February 2025.
\newblock \doi{10.48550/arXiv.2502.07889}.

\bibitem[Higham(1989)]{Higham1989}
Nicholas~J. Higham.
\newblock Matrix nearness problems and applications.
\newblock In M.~J.~C. Gover and S.~Barnett, editors, \emph{Applications of
  Matrix Theory}, pages 1--27. Oxford University Press, October 1989.

\bibitem[Myronenko and
  Song(2009)]{myronenko2009closedformsolutionrotationmatrix}
Andriy Myronenko and Xubo Song.
\newblock On the closed-form solution of the rotation matrix arising in
  computer vision problems.
\newblock \emph{arXiv:0904.1613}, April 2009.
\newblock \doi{10.48550/arXiv.0904.1613}.

\bibitem[Moritz et~al.(2018)Moritz, Nishihara, Wang, Tumanov, Liaw, Liang,
  Elibol, Yang, Paul, Jordan, and Stoica]{moritz_ray2018}
Philipp Moritz, Robert Nishihara, Stephanie Wang, Alexey Tumanov, Richard Liaw,
  Eric Liang, Melih Elibol, Zongheng Yang, William Paul, Michael~I. Jordan, and
  Ion Stoica.
\newblock Ray: a distributed framework for emerging ai applications.
\newblock In \emph{Proceedings of the 13th USENIX Conference on Operating
  Systems Design and Implementation}, OSDI'18, pages 561--577, USA, October
  2018. USENIX Association.
\newblock ISBN 9781931971478.
\newblock URL
  \url{https://www.usenix.org/conference/osdi18/presentation/moritz}.
\newblock Code available at \url{https://docs.ray.io/en/latest/index.html}.

\bibitem[Bradbury et~al.(2025)Bradbury, Frostig, Hawkins, Johnson, Leary,
  Maclaurin, Necula, Paszke, Vander{P}las, Wanderman-{M}ilne, and Zhang]{jax}
James Bradbury, Roy Frostig, Peter Hawkins, Matthew~James Johnson, Chris Leary,
  Dougal Maclaurin, George Necula, Adam Paszke, Jake Vander{P}las, Skye
  Wanderman-{M}ilne, and Qiao Zhang.
\newblock {JAX}: composable transformations of {P}ython+{N}um{P}y programs,
  March 2025.
\newblock Code available at \url{https://github.com/jax-ml/jax}.

\bibitem[Virtanen et~al.(2020)Virtanen, Gommers, Oliphant, Haberland, Reddy,
  Cournapeau, Burovski, Peterson, Weckesser, Bright, {van der Walt}, Brett,
  Wilson, Millman, Mayorov, Nelson, Jones, Kern, Larson, Carey, Polat, Feng,
  Moore, {VanderPlas}, Laxalde, Perktold, Cimrman, Henriksen, Quintero, Harris,
  Archibald, Ribeiro, Pedregosa, {van Mulbregt}, and {SciPy 1.0
  Contributors}]{scipy2020}
Pauli Virtanen, Ralf Gommers, Travis~E. Oliphant, Matt Haberland, Tyler Reddy,
  David Cournapeau, Evgeni Burovski, Pearu Peterson, Warren Weckesser, Jonathan
  Bright, St{\'e}fan~J. {van der Walt}, Matthew Brett, Joshua Wilson, K.~Jarrod
  Millman, Nikolay Mayorov, Andrew R.~J. Nelson, Eric Jones, Robert Kern, Eric
  Larson, C~J Carey, {\.I}lhan Polat, Yu~Feng, Eric~W. Moore, Jake
  {VanderPlas}, Denis Laxalde, Josef Perktold, Robert Cimrman, Ian Henriksen,
  E.~A. Quintero, Charles~R. Harris, Anne~M. Archibald, Ant{\^o}nio~H. Ribeiro,
  Fabian Pedregosa, Paul {van Mulbregt}, and {SciPy 1.0 Contributors}.
\newblock {{SciPy} 1.0: Fundamental Algorithms for Scientific Computing in
  Python}.
\newblock \emph{Nature Methods}, 17:\penalty0 261--272, February 2020.
\newblock \doi{10.1038/s41592-019-0686-2}.
\newblock Code available at \url{https://scipy.org/}.

\bibitem[Ciregan et~al.(2012)Ciregan, Meier, and Schmidhuber]{Ciregan2012}
Dan Ciregan, Ueli Meier, and Jürgen Schmidhuber.
\newblock Multi-column deep neural networks for image classification.
\newblock In \emph{2012 IEEE Conference on Computer Vision and Pattern
  Recognition}, pages 3642--3649, June 2012.
\newblock \doi{10.1109/CVPR.2012.6248110}.

\bibitem[Wan et~al.(2013)Wan, Zeiler, Zhang, Le~Cun, and Fergus]{Wan2013}
Li~Wan, Matthew Zeiler, Sixin Zhang, Yann Le~Cun, and Rob Fergus.
\newblock Regularization of neural networks using dropconnect.
\newblock In Sanjoy Dasgupta and David McAllester, editors, \emph{Proceedings
  of the 30th International Conference on Machine Learning}, volume 28\,(3) of
  \emph{Proceedings of Machine Learning Research}, pages 1058--1066, Atlanta,
  Georgia, USA, June 2013. PMLR.
\newblock URL \url{https://proceedings.mlr.press/v28/wan13.html}.

\bibitem[Hasanpour et~al.(2016)Hasanpour, Rouhani, Fayyaz, and
  Sabokrou]{Hasanpour2016}
Seyyed~Hossein Hasanpour, Mohammad Rouhani, Mohsen Fayyaz, and Mohammad
  Sabokrou.
\newblock Lets keep it simple, using simple architectures to outperform deeper
  and more complex architectures.
\newblock \emph{arXiv:1608.06037}, August 2016.
\newblock \doi{10.48550/arXiv.1608.06037}.

\bibitem[Rajasegaran et~al.(2019)Rajasegaran, Jayasundara, Jayasekara,
  Jayasekara, Seneviratne, and Rodrigo]{Rajasegaran2019}
Jathushan Rajasegaran, Vinoj Jayasundara, Sandaru Jayasekara, Hirunima
  Jayasekara, Suranga Seneviratne, and Ranga Rodrigo.
\newblock {DeepCaps}: Going deeper with capsule networks.
\newblock In \emph{2019 IEEE/CVF Conference on Computer Vision and Pattern
  Recognition (CVPR)}, pages 10717--10725, June 2019.
\newblock \doi{10.1109/CVPR.2019.01098}.

\bibitem[Meshkini et~al.(2020)Meshkini, Platos, and Ghassemain]{Meshkini2020}
Khatereh Meshkini, Jan Platos, and Hassan Ghassemain.
\newblock An analysis of convolutional neural network for fashion images
  classification ({Fashion-MNIST}).
\newblock In Sergey Kovalev, Valery Tarassov, Vaclav Snasel, and Andrey
  Sukhanov, editors, \emph{Proceedings of the Fourth International Scientific
  Conference ``Intelligent Information Technologies for Industry'' (IITI'19)},
  pages 85--95, Cham, June 2020. Springer International Publishing.
\newblock ISBN 978-3-030-50097-9.
\newblock \doi{10.1007/978-3-030-50097-9_10}.

\bibitem[Kayed et~al.(2020)Kayed, Anter, and Mohamed]{Kayed2020}
Mohammed Kayed, Ahmed Anter, and Hadeer Mohamed.
\newblock Classification of garments from {Fashion MNIST} dataset using {CNN}
  {LeNet}-5 architecture.
\newblock In \emph{2020 International Conference on Innovative Trends in
  Communication and Computer Engineering (ITCE)}, pages 238--243, February
  2020.
\newblock \doi{10.1109/ITCE48509.2020.9047776}.

\bibitem[Huang et~al.(2017)Huang, Liu, Van Der~Maaten, and
  Weinberger]{Huang2017}
Gao Huang, Zhuang Liu, Laurens Van Der~Maaten, and Kilian~Q. Weinberger.
\newblock Densely connected convolutional networks.
\newblock In \emph{2017 IEEE Conference on Computer Vision and Pattern
  Recognition (CVPR)}, pages 2261--2269, July 2017.
\newblock \doi{10.1109/CVPR.2017.243}.

\bibitem[Deng et~al.(2009)Deng, Dong, Socher, Li, Li, and Fei-Fei]{ImageNet}
Jia Deng, Wei Dong, Richard Socher, Li-Jia Li, Kai Li, and Li~Fei-Fei.
\newblock {ImageNet}: A large-scale hierarchical image database.
\newblock In \emph{2009 IEEE Conference on Computer Vision and Pattern
  Recognition}, pages 248--255, 2009.
\newblock \doi{10.1109/CVPR.2009.5206848}.
\newblock Dataset available at \url{https://www.image-net.org}.

\bibitem[Tange(2024)]{tange_2024_13826092}
Ole Tange.
\newblock {GNU Parallel 20240922 ('Gold Apollo AR924')}, September 2024.
\newblock Code available at \url{https://doi.org/10.5281/zenodo.13826092}.

\bibitem[Mitarai et~al.(2018)Mitarai, Negoro, Kitagawa, and
  Fujii]{PhysRevA.98.032309}
K.~Mitarai, M.~Negoro, M.~Kitagawa, and K.~Fujii.
\newblock Quantum circuit learning.
\newblock \emph{Physical Review A}, 98:\penalty0 032309, Sep 2018.
\newblock \doi{10.1103/PhysRevA.98.032309}.

\bibitem[Farhi and Neven(2018)]{farhi2018classificationquantumneuralnetworks}
Edward Farhi and Hartmut Neven.
\newblock Classification with quantum neural networks on near term processors.
\newblock \emph{arXiv:1802.06002}, February 2018.
\newblock \doi{10.48550/arXiv.1802.06002}.

\bibitem[Liu et~al.(2022)Liu, Yu, Duan, and Deng]{Liu2022}
Zidu Liu, Li-Wei Yu, L.-M. Duan, and Dong-Ling Deng.
\newblock Presence and absence of barren plateaus in tensor-network based
  machine learning.
\newblock \emph{Physical Review Letters}, 129:\penalty0 270501, December 2022.
\newblock \doi{10.1103/PhysRevLett.129.270501}.

\bibitem[Zhang et~al.(2024)Zhang, Liu, and Zhang]{Zhang2024}
Hao-Kai Zhang, Shuo Liu, and Shi-Xin Zhang.
\newblock Absence of barren plateaus in finite local-depth circuits with
  long-range entanglement.
\newblock \emph{Physical Review Letters}, 132:\penalty0 150603, April 2024.
\newblock \doi{10.1103/PhysRevLett.132.150603}.

\bibitem[Barthel and Miao(2025)]{Barthel2025}
Thomas Barthel and Qiang Miao.
\newblock Absence of barren plateaus and scaling of gradients in the energy
  optimization of isometric tensor network states.
\newblock \emph{Communications in Mathematical Physics}, 406\penalty0 (86),
  March 2025.
\newblock ISSN 1432-0916.
\newblock \doi{10.1007/s00220-024-05217-x}.

\bibitem[Kingma and Ba(2015)]{kingma2015adam}
Diederik~P. Kingma and Jimmy Ba.
\newblock Adam: A method for stochastic optimization.
\newblock In \emph{3rd International Conference on Learning Representations
  (ICLR)}, San Diego, CA, USA, May 2015.
\newblock \doi{10.48550/arXiv.1412.6980}.

\bibitem[Hornik(1991)]{hornik_1991}
Kurt Hornik.
\newblock Approximation capabilities of multilayer feedforward networks.
\newblock \emph{Neural Networks}, 4\penalty0 (2):\penalty0 251--257, 1991.
\newblock ISSN 0893-6080.
\newblock \doi{10.1016/0893-6080(91)90009-T}.

\bibitem[Jerbi et~al.(2023)Jerbi, Fiderer, Nautrup, Kübler, Briegel, and
  Dunjko]{Jerbi2023}
Sofiene Jerbi, Lukas~J. Fiderer, Hendrik~Poulsen Nautrup, Jonas~M. Kübler,
  Hans~J. Briegel, and Vedran Dunjko.
\newblock Quantum machine learning beyond kernel methods.
\newblock \emph{Nature Communications}, 14\penalty0 (1), January 2023.
\newblock \doi{10.1038/s41467-023-36159-y}.

\bibitem[Schölkopf and Smola(2001)]{Schoelkopf2001}
Bernhard Schölkopf and Alexander~J. Smola.
\newblock \emph{Learning with Kernels: Support Vector Machines, Regularization,
  Optimization, and Beyond}.
\newblock MIT Press, Cambridge, MA, December 2001.
\newblock ISBN 978-0262194754.
\newblock \doi{10.7551/mitpress/4175.001.0001}.

\bibitem[Pedregosa et~al.(2011)Pedregosa, Varoquaux, Gramfort, Michel, Thirion,
  Grisel, Blondel, Prettenhofer, Weiss, Dubourg, Vanderplas, Passos,
  Cournapeau, Brucher, Perrot, and Duchesnay]{scikit-learn}
F.~Pedregosa, G.~Varoquaux, A.~Gramfort, V.~Michel, B.~Thirion, O.~Grisel,
  M.~Blondel, P.~Prettenhofer, R.~Weiss, V.~Dubourg, J.~Vanderplas, A.~Passos,
  D.~Cournapeau, M.~Brucher, M.~Perrot, and E.~Duchesnay.
\newblock Scikit-learn: Machine learning in {P}ython.
\newblock \emph{Journal of Machine Learning Research}, 12:\penalty0 2825--2830,
  October 2011.
\newblock URL \url{https://jmlr.csail.mit.edu/papers/v12/pedregosa11a.html}.
\newblock Code available at \url{https://scikit-learn.org/stable/}.

\bibitem[Novikov et~al.(2018)Novikov, Trofimov, and Oseledets]{Novikov2018}
Alexander Novikov, Mikhail Trofimov, and Ivan Oseledets.
\newblock Exponential machines.
\newblock \emph{Bulletin of the Polish Academy of Sciences Technical Sciences},
  66, No 6 (Special Section on Deep Learning: Theory and Practice):\penalty0
  789--797, December 2018.
\newblock \doi{10.24425/bpas.2018.125926}.

\bibitem[Huggins et~al.(2019)Huggins, Patil, Mitchell, Whaley, and
  Stoudenmire]{Huggins2019}
William Huggins, Piyush Patil, Bradley Mitchell, K~Birgitta Whaley, and
  E.~Miles Stoudenmire.
\newblock Towards quantum machine learning with tensor networks.
\newblock \emph{Quantum Science and Technology}, 4\penalty0 (2):\penalty0
  024001, January 2019.
\newblock \doi{10.1088/2058-9565/aaea94}.

\bibitem[Efthymiou et~al.(2019)Efthymiou, Hidary, and
  Leichenauer]{Efthymiou2019}
Stavros Efthymiou, Jack Hidary, and Stefan Leichenauer.
\newblock {TensorNetwork} for machine learning.
\newblock \emph{arXiv:1906.06329}, June 2019.
\newblock \doi{10.48550/arXiv.1906.06329}.

\bibitem[Barratt et~al.(2022)Barratt, Dborin, and Wright]{Barratt2022}
Fergus Barratt, James Dborin, and Lewis Wright.
\newblock Improvements to gradient descent methods for quantum tensor network
  machine learning.
\newblock In \emph{Second Workshop on Quantum Tensor Networks in Machine
  Learning, 35th Conference on Neural Information Processing Systems (NeurIPS
  2021)}, March 2022.
\newblock \doi{10.48550/arXiv.2203.03366}.

\bibitem[Lin et~al.(2023)Lin, Kuijpers, Peterhansl, and Pollmann]{Lin:2023put}
Sheng-Hsuan Lin, Olivier Kuijpers, Sebastian Peterhansl, and Frank Pollmann.
\newblock Distributive pre-training of generative modeling using matrix-product
  states.
\newblock In \emph{35th Conference on Neural Information Processing Systems},
  June 2023.
\newblock \doi{10.48550/arXiv.2306.14787}.

\bibitem[Nemkov et~al.(2023)Nemkov, Kiktenko, and Fedorov]{Nemkov2023}
Nikita~A. Nemkov, Evgeniy~O. Kiktenko, and Aleksey~K. Fedorov.
\newblock Fourier expansion in variational quantum algorithms.
\newblock \emph{Physical Review A}, 108:\penalty0 032406, September 2023.
\newblock \doi{10.1103/PhysRevA.108.032406}.

\bibitem[Okumura and Ohzeki(2023)]{Okumura2023}
Shun Okumura and Masayuki Ohzeki.
\newblock Fourier coefficient of parameterized quantum circuits and barren
  plateau problem.
\newblock \emph{arXiv:2309.06740}, September 2023.
\newblock \doi{10.48550/arXiv.2309.06740}.

\bibitem[Barthe and P{\'{e}}rez-Salinas(2024)]{Barthe2024}
Alice Barthe and Adri{\'{a}}n P{\'{e}}rez-Salinas.
\newblock Gradients and frequency profiles of quantum re-uploading models.
\newblock \emph{{Quantum}}, 8:\penalty0 1523, November 2024.
\newblock ISSN 2521-327X.
\newblock \doi{10.22331/q-2024-11-14-1523}.

\bibitem[LeCun et~al.(1989)LeCun, Boser, Denker, Henderson, Howard, Hubbard,
  and Jackel]{NIPS1989_53c3bce6}
Yann LeCun, Bernhard Boser, John Denker, Donnie Henderson, R.~Howard, Wayne
  Hubbard, and Lawrence Jackel.
\newblock Handwritten digit recognition with a back-propagation network.
\newblock In D.~Touretzky, editor, \emph{Advances in Neural Information
  Processing Systems}, volume~2. Morgan-Kaufmann, 1989.
\newblock URL
  \url{https://proceedings.neurips.cc/paper_files/paper/1989/file/53c3bce66e43be4f209556518c2fcb54-Paper.pdf}.

\bibitem[Dosovitskiy et~al.(2021)Dosovitskiy, Beyer, Kolesnikov, Weissenborn,
  Zhai, Unterthiner, Dehghani, Minderer, Heigold, Gelly, Uszkoreit, and
  Houlsby]{visiontransormer2023}
Alexey Dosovitskiy, Lucas Beyer, Alexander Kolesnikov, Dirk Weissenborn,
  Xiaohua Zhai, Thomas Unterthiner, Mostafa Dehghani, Matthias Minderer, Georg
  Heigold, Sylvain Gelly, Jakob Uszkoreit, and Neil Houlsby.
\newblock An image is worth 16x16 words: Transformers for image recognition at
  scale.
\newblock In \emph{International Conference on Learning Representations}, 2021.
\newblock URL \url{https://openreview.net/forum?id=YicbFdNTTy}.

\bibitem[Zeiler(2012)]{zeiler2012adadeltaadaptivelearningrate}
Matthew~D. Zeiler.
\newblock Adadelta: An adaptive learning rate method.
\newblock \emph{arXiv:1212.5701}, December 2012.
\newblock \doi{10.48550/arXiv.1212.5701}.

\bibitem[Glorot and Bengio(2010)]{pmlr-v9-glorot10a}
Xavier Glorot and Yoshua Bengio.
\newblock Understanding the difficulty of training deep feedforward neural
  networks.
\newblock In Yee~Whye Teh and Mike Titterington, editors, \emph{Proceedings of
  the Thirteenth International Conference on Artificial Intelligence and
  Statistics}, volume~9 of \emph{Proceedings of Machine Learning Research},
  pages 249--256, Chia Laguna Resort, Sardinia, Italy, 13--15 May 2010. PMLR.
\newblock URL \url{https://proceedings.mlr.press/v9/glorot10a.html}.

\bibitem[Li et~al.(2020)Li, Jamieson, Rostamizadeh, Gonina, Ben-tzur, Hardt,
  Recht, and Talwalkar]{MLSYS2020_a06f20b3}
Liam Li, Kevin Jamieson, Afshin Rostamizadeh, Ekaterina Gonina, Jonathan
  Ben-tzur, Moritz Hardt, Benjamin Recht, and Ameet Talwalkar.
\newblock A system for massively parallel hyperparameter tuning.
\newblock In I.~Dhillon, D.~Papailiopoulos, and V.~Sze, editors,
  \emph{Proceedings of Machine Learning and Systems}, volume~2, pages 230--246,
  2020.
\newblock URL
  \url{https://proceedings.mlsys.org/paper_files/paper/2020/file/a06f20b349c6cf09a6b171c71b88bbfc-Paper.pdf}.

\bibitem[Bergstra et~al.(2013)Bergstra, Yamins, and Cox]{pmlr-v28-bergstra13}
James Bergstra, Daniel Yamins, and David Cox.
\newblock Making a science of model search: Hyperparameter optimization in
  hundreds of dimensions for vision architectures.
\newblock In Sanjoy Dasgupta and David McAllester, editors, \emph{Proceedings
  of the 30th International Conference on Machine Learning}, volume 28\,(1) of
  \emph{Proceedings of Machine Learning Research}, pages 115--123, Atlanta,
  Georgia, USA, 17--19 Jun 2013. PMLR.
\newblock URL \url{https://proceedings.mlr.press/v28/bergstra13.html}.

\bibitem[Page(1993)]{Page1993}
Don~N. Page.
\newblock Average entropy of a subsystem.
\newblock \emph{Physical Review Letters}, 71:\penalty0 1291--1294, August 1993.
\newblock \doi{10.1103/PhysRevLett.71.1291}.

\bibitem[Bravyi and Kitaev(2005)]{Bravyi_2005}
Sergey Bravyi and Alexei Kitaev.
\newblock Universal quantum computation with ideal {C}lifford gates and noisy
  ancillas.
\newblock \emph{Phys. Rev. A}, 71:\penalty0 022316, Feb 2005.
\newblock \doi{10.1103/PhysRevA.71.022316}.

\bibitem[Ross and Selinger(2016)]{Ross_2014}
Neil~J. Ross and Peter Selinger.
\newblock {Optimal ancilla-free Clifford+$T$ approximation of $Z$-rotations}.
\newblock \emph{Quant. Inf. Comput.}, 16\penalty0 (11-12):\penalty0 0901--0953,
  2016.
\newblock \doi{10.26421/QIC16.11-12-1}.

\bibitem[Yamamoto and Yoshioka(2024)]{pygridsynth}
Shuntaro Yamamoto and Nobuyuki Yoshioka.
\newblock Pygridsynth: A native python implementation of the gridsynth
  algorithm, November 2024.
\newblock Code available at
  \url{https://github.com/quantum-programming/pygridsynth}.

\end{thebibliography}

\end{document}